\documentclass{egpubl}
\usepackage{eg2022}
\usepackage{soul}

\ConferencePaper        
\usepackage[T1]{fontenc}
\usepackage{dfadobe}  

\biberVersion
\BibtexOrBiblatex
\usepackage[backend=biber,bibstyle=EG,citestyle=alphabetic,backref=true]{biblatex} 
\electronicVersion
\PrintedOrElectronic

\ifpdf \usepackage[pdftex]{graphicx} \pdfcompresslevel=9
\else \usepackage[dvips]{graphicx} \fi

\usepackage{egweblnk} 

\title[Learning Spectral Unions of Partial Deformable 3D Shapes]%
      {Learning Spectral Unions of Partial Deformable 3D Shapes}

\author[L. Moschella et al.]
{\parbox{\textwidth}{\centering Luca Moschella$^{1}$,
	  Simone Melzi$^{1,2}$,
      Luca Cosmo$^{3}$,
      Filippo Maggioli$^{1}$,
      Or Litany$^{4}$,\\
      Maks Ovsjanikov$^{5}$,
      Leonidas Guibas$^{6}$,
      Emanuele Rodol\`{a}$^{1}$
      }
      \\
{\parbox{\textwidth}{\centering $^1$ Sapienza University of Rome \quad $^2$ University of Milano-Bicocca \quad $^3$ Ca' Foscari University of Venice \quad  $^4$ NVIDIA \quad $^5$ LIX, \'Ecole Polytechnique, CNRS \quad $^6$ Stanford University
}
}
}
\usepackage{amscd} 
\usepackage{amsfonts} 
\usepackage{subfigure} 
\usepackage{caption} 
\usepackage{mathtools}
\usepackage[utf8]{inputenc}

\usepackage{placeins}
\usepackage{algorithm}
\usepackage{algorithmic}
\usepackage{array}
\usepackage{overpic}
\usepackage{bm}
\usepackage{wrapfig}
\usepackage{pgfplots}
\usepackage{pgf,tikz}
\usepackage{multirow}
\usepackage[normalem]{ulem}

\newcommand{\M}{\mathcal{M}}

\definecolor{darkblue}{rgb}{0,0.447,0.741}
\definecolor{darkred}{rgb}{0.54,0,0}

\definecolor{changecolor}{rgb}{0.0, 0.5, 1.0}

\newcommand{\changed}[1]{}  
\begin{document}


\maketitle
\begin{abstract}

Spectral geometric methods have brought revolutionary changes to the field of geometry processing. Of particular interest is the study of the Laplacian spectrum as a compact, isometry and permutation-invariant representation of a shape. Some recent works show how the intrinsic geometry of a full shape can be recovered from its spectrum, but there are approaches that consider the more challenging 
problem of recovering the geometry from the spectral information of partial shapes.
In this paper, we propose a possible way to fill this gap. We introduce a learning-based method to estimate the Laplacian spectrum of the union of partial non-rigid 3D shapes, without actually computing the 3D geometry of the union or any correspondence between those partial shapes. We do so by operating purely in the spectral domain and by defining the union operation between short sequences of eigenvalues.
We show that the approximated union spectrum can be used as-is to reconstruct the complete geometry~\cite{Instant2020}, perform region localization on a template~\cite{rampini2019correspondence} and retrieve shapes from a database, generalizing ShapeDNA~\cite{shapeDNA2006} to work with partialities.
Working with eigenvalues allows us to deal with unknown correspondence, different sampling, and different discretizations (point clouds and meshes alike), making this operation especially robust and general. 
Our approach is data-driven and can generalize to isometric and non-isometric deformations of the surface, as long as these stay within the same semantic class (e.g., human bodies or horses), as well as to partiality artifacts not seen at training time.

\ccsdesc[500]{Computing methodologies~Shape analysis}
\ccsdesc[300]{Theory of computation~Computational geometry}

    \printccsdesc   
\end{abstract}  

\section{Introduction}\label{sec:introduction}
%
Recent progress in spectral geometry processing has brought to significant qualitative leaps {that lead to better results} in a range of challenging tasks
such as deformable shape matching~\cite{ovsjanikov2012functional,litany2017fully}, retrieval~\cite{shapeDNA2006,bronstein2011shape}, 
style~\cite{Instant2020} and pose transfer~\cite{kovnatsky2013coupled,yin2015spectral} among others. 
More recently, a great deal of attention has been put on the study of the eigenvalues of the Laplace-Beltrami operator (i.e., the \textit{Laplacian spectrum}) as a compact, isometry and permutation-invariant representation of the input shape. It has been shown that, with the appropriate knowledge on the input domain, this representation contains enough information for localize shape's regions \cite{rampini2019correspondence} and even to reconstruct the geometry of the shape \cite{cosmo2019isospectralization, marin2019high}. 
However, these methods typically operate in a controlled scenario requiring to have access to the \textit{full} geometry of the shape, ignoring the fact that real-world data are riddled with partiality artifacts.

{
In this paper, we propose a learning-based framework to predict the Laplacian spectrum of the union of two shapes directly from the spectrum of the individual parts. This enables the  aforementioned spectral methods to be applied directly in case of partial views of the same shape, without resorting to methods that explicitly fuse the 3D geometry of partial shapes.
}
{Indeed, a typical pipeline to combine partial shapes can be very cumbersome, and requires to match the corresponding regions, extract a set of (non-rigid) transformations from the matches, and merge the partial views into a consistent discretization.}
%
%
%
%
Each of these steps can be error-prone and difficult to solve, as testified by a wealth of literature on non-rigid shape matching and reconstruction, {especially in the case of partial shapes}. 
For example, the mere presence of inconsistent surface sampling can cause problems in most matching pipelines \cite{melzi2019matching}.
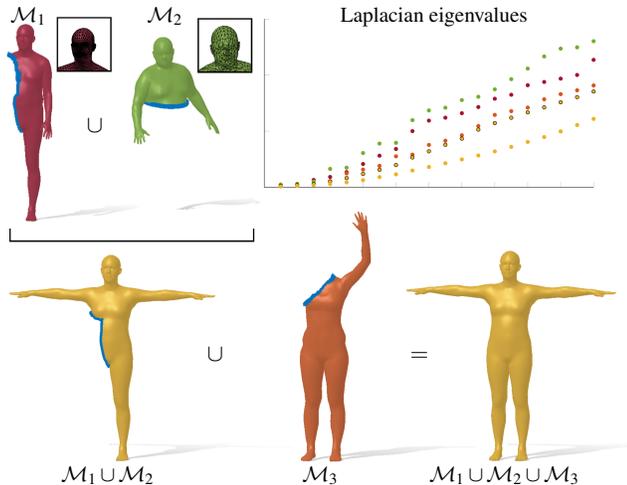
\begin{figure}[t]
\centering
     \begin{overpic}[trim=0cm 0cm 0cm 0cm,clip,width=\linewidth]{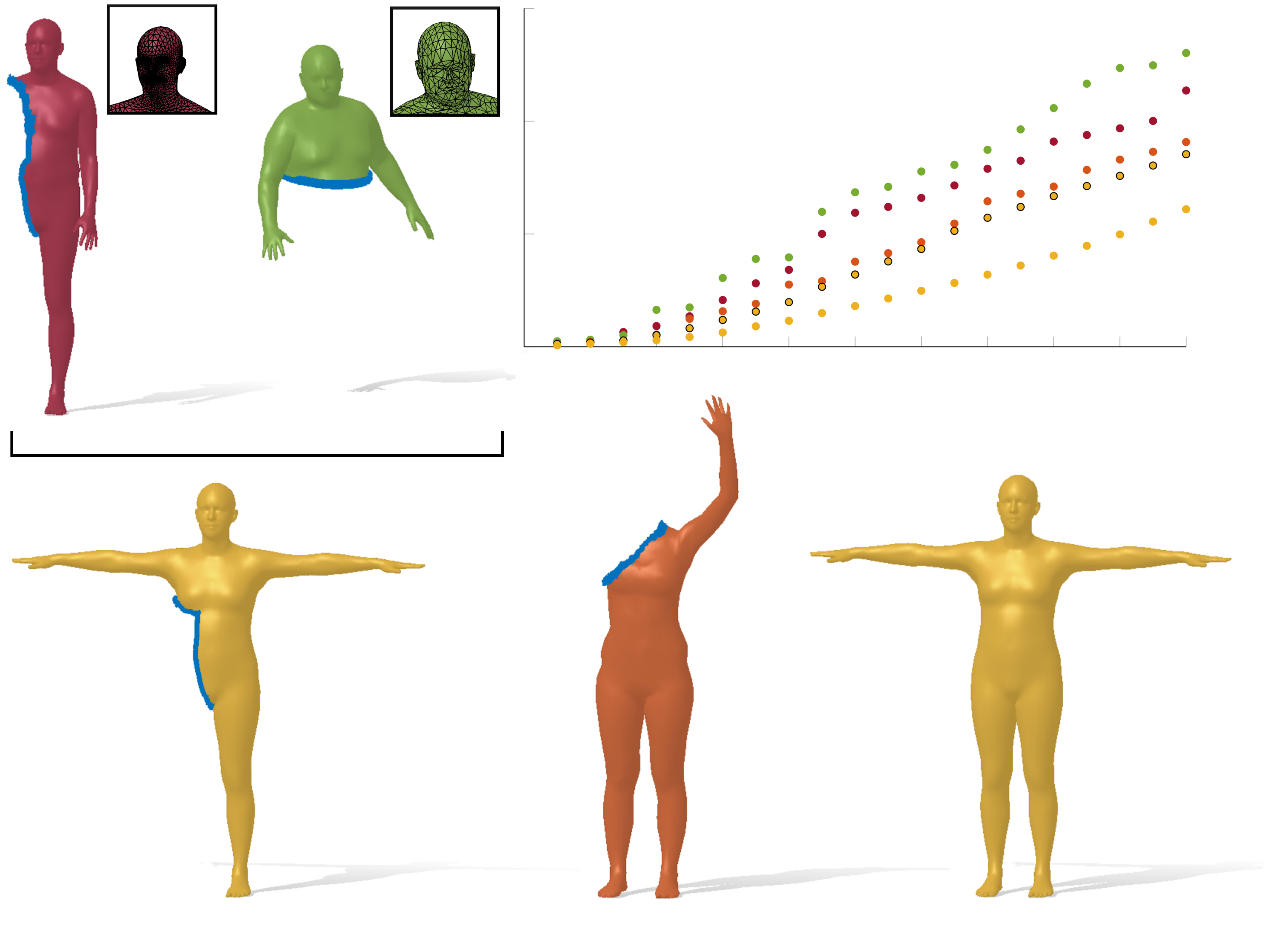}
     \put(1, 73.5){\footnotesize $\M_1$ }
     \put(13, 56){$\cup$ }
      \put(22.5, 73.5){\footnotesize $\M_2$ }
     \put(9, 1){\footnotesize $\M_1 \cup \M_2$}
     \put(32, 20){$\cup$ }
     \put(47, 1){\footnotesize $\M_3$}
     \put(53, 73.5){\footnotesize Laplacian eigenvalues}
     \put(64, 20){$=$ }
     \put(67, 1){\footnotesize $\M_1 \cup \M_2 \cup \M_3$}
     
   \end{overpic}
\caption{\label{fig:teaser}
Given a collection of partial deformable shapes $\{\M_1,\M_2,\M_3\}$ as input, our method predicts the Laplacian eigenvalues of their union without first having to compute a correspondence or a transformation between the input shapes. The resulting eigenvalues (top right plots, colors correspond to each surface) can be used to reconstruct the final shape if needed, up to isometry/pose (bottom right). In this example, the input shapes have different poses, varying overlap, and different mesh connectivity.} 


\end{figure}

     
  
Motivated by the excellent results {achieved} by the methods that exploit the Laplacian spectrum representation, we propose a different perspective. 
We claim that, in many cases, it is not necessary to have the extrinsic geometry of a target full shape and propose to directly estimate the intrinsic properties of the sought full shape {\em without} having to materialize its surface geometry.
This is done by translating the {objective} of merging partial shapes from the spatial to a purely {\em spectral} domain.
%
%
For each partial surface, our method takes as input the truncated sequence of its Laplacian eigenvalues, which act as a surrogate of the shape geometry, and {predicts} as output the eigenvalue sequence of the {3D model obtained from the union of the partial surfaces (or an isometric deformation thereof)} -- but not the 3D model itself, as visualized in Figure~\ref{fig:teaser}. This {eigenvalues} prediction task is in general ill-posed, but can be resolved by means of a data prior, namely by training a deep net on a few hundred examples. The advantages of this approach are numerous, and include associativity (i.e. $A \cup B \cup C = (A \cup B) \cup C$), invariance to deformations and sampling, and generalization to different discrete representations for the input geometry. In a way, this recalls the notion of ``homomorphic encryption'' in secure computation~\cite{rivest1979data}, where the task is to perform calculations on encrypted data without decrypting it first.

\vspace{1ex}\noindent\textbf{Contribution.}
In this paper, we introduce a learning-based method to estimate the Laplacian spectrum of the union of partial non-rigid 3D shapes, without actually computing the 3D geometry of the union. 
%
%
Sidestepping the reconstruction means that we do not have to commit to one specific 3D embedding in the output (e.g. a specific pose for a human body), but leave this choice to task-specific blocks. 
{Moreover, our method takes advantage of the geometric insight that the spectra can be used not only for single shape recovery and processing (as done in prior works) but also to enable multi-shape operations such as unions.}
Once a spectrum is predicted, it can be fed {\em as-is} to any existing spectral pipeline operating with eigenvalues. 
For example, we can reconstruct the full 3D geometry by using the method in~\cite{Instant2020} as an output module. If the geometry is not needed, e.g., for tasks of shape retrieval~\cite{shapeDNA2006} and region localization~\cite{rampini2019correspondence},  we achieve the same accuracy that can be obtained in the case where the full shape is given.

%

%
\section{Related work}
\label{sec:related}
We discuss two lines of research that are most closely related to our spectral aggregation task: partial non-rigid aggregation of shapes in their extrinsic form (e.g, mesh or point cloud), and spectral analysis of partial shapes. 

%
\subsection{Nonrigid shape aggregation} 
Recovering deformable 3D shapes from partial scans has numerous applications in AR/VR, manufacturing, and robot manipulation. A common setting for this problem is non-rigid registration, where the scans are captured sequentially and exhibit mild inter-frame deformations, and significant overlap. In such cases, template-less methods have been shown to perform well by using general deformation models such as thin-plate splines~\cite{brown2004non,brown2007global} or as-rigid-as-possible energies~\cite{li2008global}.
%
%
Wand et al.~\cite{wand2007reconstruction,wand2009efficient} used dynamic ``surfels'' to represent the input surfaces, and proposed a statistical model to recover the underlying template shape. Temporal coherence has been used in~\cite{tevs2012animation} to generate dense correspondences from robust landmarks, and in~\cite{mitra2007dynamic} to reconstruct a space-time surface embedded in 4D. Sharf et al.~\cite{sharf2008space} incorporated a mass conservation prior 
 to control the plausibility of the reconstructed surface.  The ``Dynamic Fusion'' method of Newcomb et al.~\cite{newcombe2015dynamicfusion} and follow-up work~\cite{slavcheva2017killingfusion}, demonstrated real-time, template-free non-rigid reconstruction allowing both the object and the camera to move. 

 More related to our setting are cases where the input set is sparser, and the deformations between the scans can change significantly. In fact, we do {\em not} assume temporal coherence or an initial alignment. Similar to us, methods designed for these settings usually assume a strong prior on the shape category or even rely on a parametric model.  In~\cite{zhang2014quality}, a generic human template was used for building a personalized parametric human body model similar to SCAPE~\cite{anguelov2005scape}. Chang and Zwicker~\cite{chang2009range,chang2011global} assumed an articulated model and solved for joints and skinning weights. More recently, advances in geometric deep learning for processing point clouds and meshes were used to leverage data-driven priors (e.g. from ~\cite{Bogo_2017_CVPR}) for deformable shape completion and fusion~\cite{litany2018deformable,halimi2020greater}.

\subsection{Eigenvalues and partiality}
Spectral representations based on the Laplacian are widely used in the analysis of deformable shapes, mainly due to their isometric invariance. Much less attention has been given to the effect of partiality on the spectrum. 

\vspace{1ex}\noindent\textbf{Shape correspondence.} A first attempt at utilizing the Laplacian eigen{\em functions} to recover dense correspondences between a partial and a full shape was shown in~\cite{rodola2017partial}, building upon the seminal functional maps framework~\cite{ovsjanikov2012functional}. 
This was further extended to matching shapes in the presence of clutter~\cite{cosmo2016matching}, and to a more efficient fully spectral variant in~\cite{litany2017fully}. In the context of partial shape aggregation, most relevant is an extension to the multi-part matching algorithm (a.k.a ``non-rigid puzzle'') proposed in~\cite{litany2016non}. 
Recently, deep learning techniques have also been utilizing Laplacian eigenfunctions for matching~\cite{litany2017deep, halimi2019unsupervised,ginzburg2020cyclic,aygun2020unsupervised,roufosse2019unsupervised,attaiki2021dpfm}. Replacing eigenfunctions with a basis learned from data was recently proven more robust and therefore applicable to challenging settings including point clouds and partiality~\cite{marin2020correspondence}. 
%

    

\vspace{1ex}\noindent\textbf{Reconstruction.}
Aside from matching, other works have investigated spectral methods for non-rigid completion and registration. In FARM~\cite{marin2020farm} and its high-resolution variant~\cite{marin2019high}, a functional maps representation is incorporated into a parametric model-based regression pipeline. 
%

\vspace{1ex}\noindent\textbf{Shape from Spectrum.}
Most closely related to ours are works that aim directly to recover the shape from its underlying spectrum, also known as the problem of ``hearing the shape of a drum''~\cite{kac1966can}. This procedure was recently studied by Cosmo et al.~\cite{cosmo2019isospectralization} in practical rather than purely theoretical settings. Their pipeline, dubbed ``isospectralization'', was proven useful in multiple application scenarios, and extended in~\cite{Instant2020} by replacing the regularizers of~\cite{cosmo2019isospectralization} with a data-driven prior. 

In this work, we aim to perform the union of partial deformable shapes from their spectral representation.
Our method differs from the ones listed above for two main reasons:
i) we consider the shape from spectrum problem in the more challenging setting of partiality, and ii) rather than recovering the geometry of the full shape, we aim to recover its Laplacian spectrum, given the spectrum of two parts. 
In other words, we introduce the problem of spectral unions of partial shapes and propose an effective solution.
In this light our work is related to \cite{sung2018learning}, which devised a framework to learn fuzzy representations that enable set operations on man-made objects.
    



\section{Proposed method}\label{sec:method}
%
Let us be given two partial shapes $\mathcal{M}_1$ and $\mathcal{M}_2$, and let $\mathcal{M}_1\cup \mathcal{M}_2$
denote their non-rigid alignment, as depicted on the left side of Figure~\ref{fig:teaser}.  We seek an answer to the following question: 
what can we say about $\mathcal{M}_1 \cup \mathcal{M}_2$, {without} actually computing this union?
More specifically: can we predict the spectrum of $\mathcal{M}_1 \cup \mathcal{M}_2$ without having to solve for the point-to-point correspondence between them?

With no additional priors, the question is ill-posed; for example, there are infinitely many ways in which two sheets of paper can be glued together.
%
In the sequel, we claim that the spectrum of the union can be predicted by coupling Laplacian eigenvalues with a data prior without solving a correspondence or reconstruction problem in the process.
%
%

\setlength{\columnsep}{3pt}
\begin{wrapfigure}[]{r}{0.4\linewidth}
\vspace{-0.4cm}
\begin{center}
\begin{overpic}
[trim=0cm 0cm 0cm 0cm,clip,width=1.0\linewidth]{{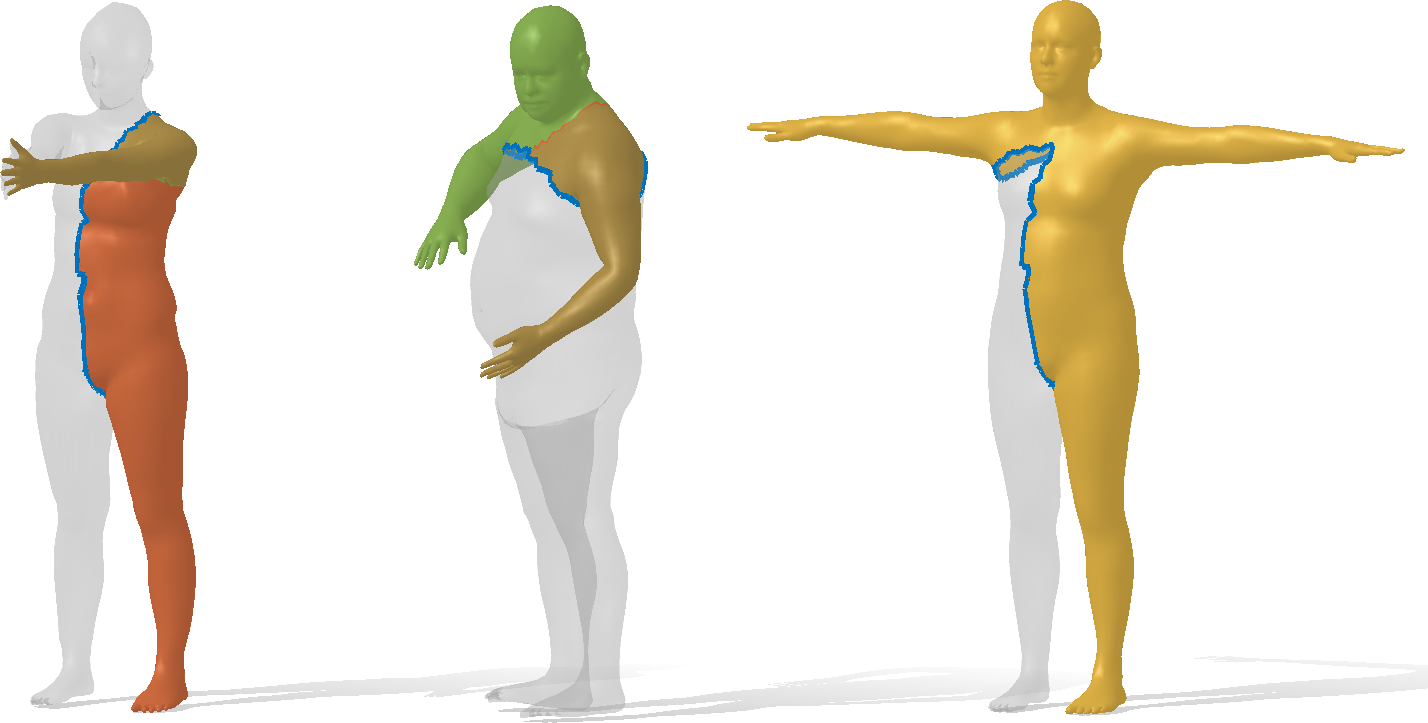}}
        \put(2, -7){\footnotesize $\M_1$}
        \put(33, -7){\footnotesize $\M_2$}
        \put(12.5, 42){\footnotesize $\mathcal{R}_1$}
        \put(45.2, 32){\footnotesize $\mathcal{R}_2$}
        \put(60, -7){\footnotesize $\M_1 \cup \M_2$}        
\end{overpic}
\end{center}
\end{wrapfigure}
\vspace{1ex}\noindent\textbf{Mathematical preliminaries.}
We model shapes as Riemannian manifolds $\M$ with boundary $\partial\M$. Each manifold identifies an equivalence class of isometries, and thus has infinitely many embeddings in $\mathbb{R}^3$ (e.g. changes in pose). Let us be given two manifolds $\M_1$ and $\M_2$, together with a diffeomorphism $\pi:\mathcal{R}_1\to\mathcal{R}_2$ between regions $\mathcal{R}_1\subseteq\M_1$ and $\mathcal{R}_2\subseteq\M_2$. 
A third manifold $\M_1 \cup \M_2$ can be obtained by attaching $\M_1$
to $\M_2$ over the common region via the map $\pi$ (as depicted in the inset figure). We refer to $\M_1
\cup \M_2$ as the {\em union shape}\footnote{We keep the mathematical
  description simple for the sake of clarity. Formally, this operation
  is called {\em connected sum}, denoted by $\M_1\#\M_2$, and is part
  of the surgery theory of manifolds, see, e.g.,
  \cite{rolfsen1976knots}.}. 
%

On each $\M$ we consider the Laplace-Beltrami operator $\Delta$, extending the notion of Laplace operator from Euclidean geometry to surfaces. This operator admits a spectral decomposition:
\begin{align}
    \Delta \phi_i(x) &= \lambda_i \phi_i(x) & x\in\mathrm{int}(\mathcal{M})\\
    \phi_i(x) &= 0   &x\in\partial\mathcal{M} \label{eq:bc}
\end{align}
into eigenvalues $\lambda_1\le\lambda_2\le\lambda_3\le\cdots$ and associated eigenfunctions $\phi_1,\phi_2,\phi_3,\dots$; we adopt homogeneous Dirichlet boundary conditions~\eqref{eq:bc}. The set of eigenvalues forms a discrete {\em spectrum}, which we assume to be ordered non-decreasingly. In this paper, we consider truncated spectra of length $k$, and introduce the vector-valued function:
\begin{align}
    \bm{\lambda}:\M\mapsto (\lambda_1,\dots,\lambda_k)\,.
\end{align}
In particular, we completely discard the eigenfunctions $\phi_1(x),\phi_2(x),\dots$, which are point-based quantities and thus highly dependent on shape discretization.

\vspace{1ex}\noindent\textbf{\textit{Remark.}}
Since the Laplacian $\Delta$ is invariant to isometries, so is its truncated spectrum encoded in $\bm{\lambda}$. This means that eigenvalues capture shape information {\em up to pose}, a fundamental property that is at the basis of our method.

\begin{figure}[t]
\centering

   
    \begin{overpic}[trim=0cm 0cm 0cm -0.5cm,clip,width=0.8\linewidth]{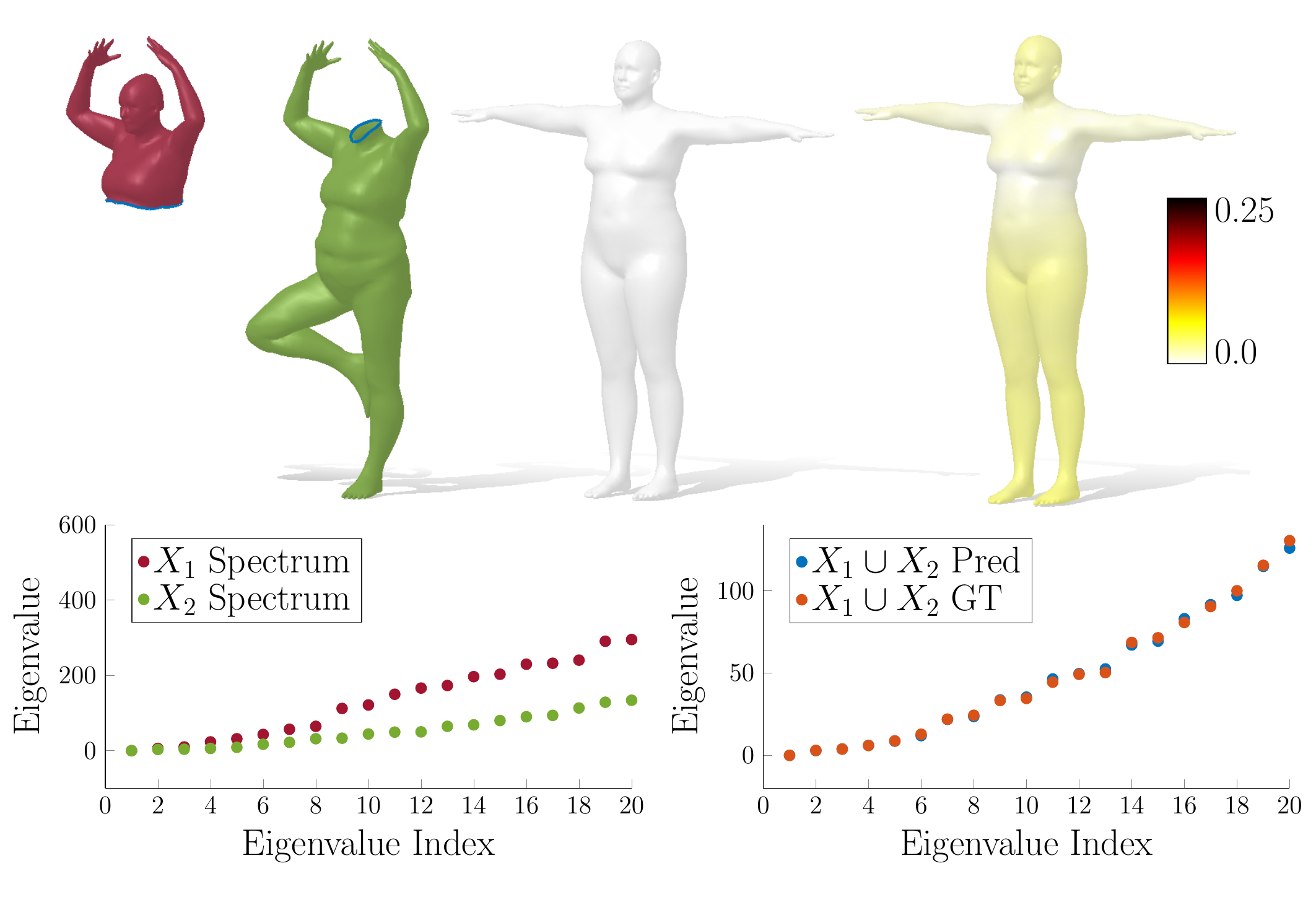}
        \put(6.5, 69){\footnotesize $\M_1$ }
        \put(16.5, 69){$\cup$ }
        \put(23,69){\footnotesize $\M_2$ }
        \put(37.5, 69){\footnotesize Ground Truth}
        \put(75.25, 69){\footnotesize Ours}
    \end{overpic}

    \caption{\label{fig:instant_recovery}Example of partial shapes whose union entirely covers the full shape. This is the simplest setting that we consider in this paper. Given the spectra of the partial shapes (red and green), we recover the spectrum of their union, and from the spectrum we recover the geometry in standard T-pose using a shape-from-spectrum reconstruction method~\cite{Instant2020}. The white shape is recovered from ground-truth eigenvalues; ours is colored with a heatmap, which encodes reconstruction error.
    }
\end{figure}
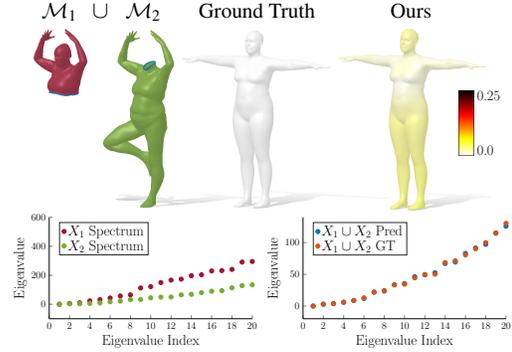

\vspace{1ex}\noindent\textbf{Problem statement.}
In non-rigid alignment, one is given 3D embeddings (e.g. point clouds) for $\M_1$ and $\M_2$, and must recover a 3D embedding of their union $\M_1\cup\M_2$.
Since, in this setting, $\M_1$ and $\M_2$ may undergo wildly different deformations, there is no guarantee that they have the same 3D coordinates on the common region. Therefore, it is not clear how a 3D embedding for $\M_1\cup\M_2$ should look like.



In our work, we propose to mitigate this problem by switching from a discrete representation of the 3D embedding of $\M_1\cup\M_2$ to a discrete representation of the entire isometry class, given by $\bm{\lambda}(\M_1\cup\M_2)$. Then, we translate the problem of recovering an alignment between 3D embeddings to the estimation of a parametric nonlinear operator $\mathcal{U}_\Theta:\mathbb{R}^k\times\mathbb{R}^k\to\mathbb{R}^k$, 
 such that:
\begin{align}\label{eq:utheta}
    \bm{\lambda}(\M_1 \cup \M_2) = \mathcal{U}_\Theta( \bm{\lambda}(\M_1),  \bm{\lambda}(\M_2))\,.
\end{align}
%
%
%

We call $\mathcal{U}_\Theta$ the {\em spectral union} operator, and model it as a deep neural network with learnable parameters $\Theta$. A specific definition for the architecture and the loss are given in Section~\ref{sec:dl}.

\vspace{1ex}\noindent\textbf{\textit{Remark.}}
In general, the spectrum of the union shape $\M_1\cup\M_2$ is {\em not} simply the union of the spectra of $\M_1,\M_2$. This is only true if $\M_1$ and $\M_2$ correspond to disjoint regions of the complete shape (see e.g. \cite[Sec.~3.1]{rodola2017partial}), while in this paper we consider the case in which $\M_1$ and $\M_2$ partially overlap, thus making the interactions between the two spectra much more complex.

\vspace{1ex}\noindent\textbf{Difficulty settings and associativity.}
%
%

Estimating an operator $\mathcal{U}_\Theta$ that makes
Eq.~\eqref{eq:utheta} hold for many different pairs $(\M_1,\M_2)$ is
not a simple problem, even with short sequences (in this paper we use $k=20$).
In fact, it is known that Laplacian spectra can vary wildly under partiality perturbations~\cite{filoche2009strong}, and predicting these variations can be difficult.





\begin{figure}[t]
\centering
    \begin{overpic}[trim=0cm 0cm 0cm 0cm, clip, tics=10, width=0.9\linewidth]{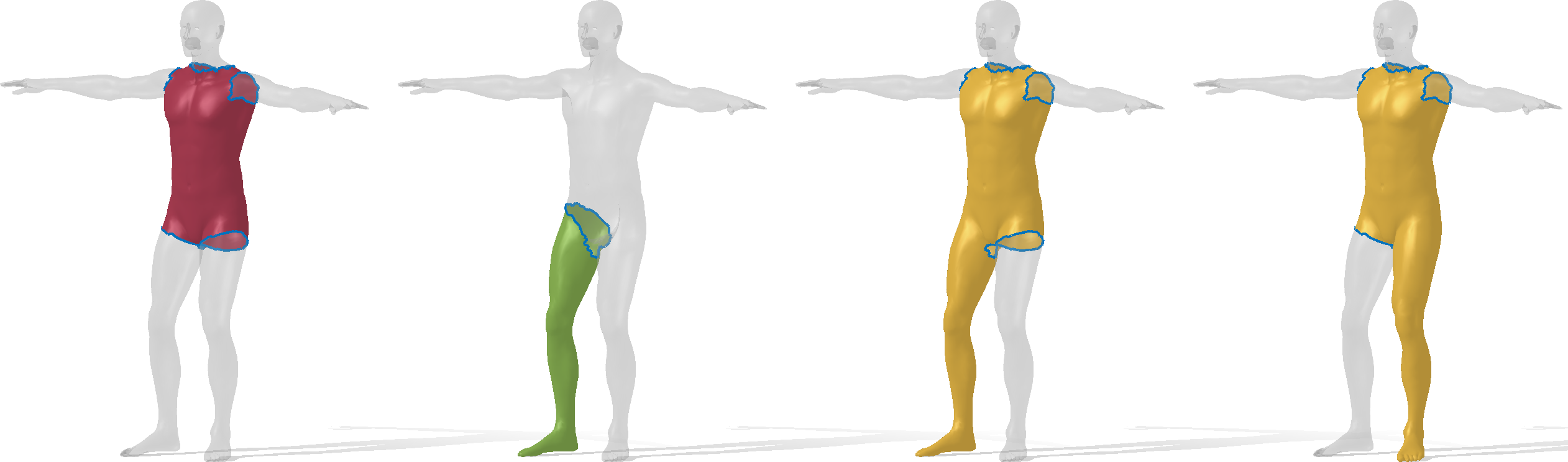}
        \put(4.6, 27){\footnotesize $\M_1$}
        \put(29.8, 27){\footnotesize $\M_2$}
        \put(24, 13){$\cup$}
        \put(48, 13){$=$}
        \put(49, 13){$=$}
        \put(74, 13){or}
    \end{overpic}
    \vspace{0.2cm}
    
    \begin{overpic}[trim=0cm 0cm 0cm 0cm, clip, tics=10, width=0.9\linewidth]{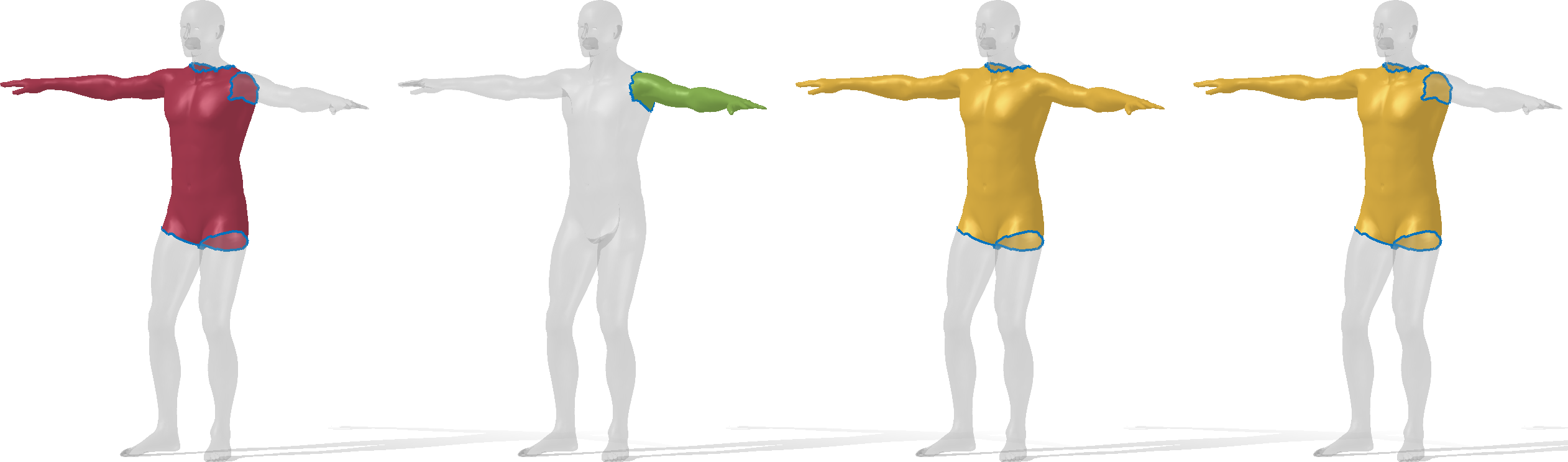}
    \put(4.6, 27){\footnotesize $\M_1$}
        \put(29.8, 27){\footnotesize $\M_2$}
            \put(24, 13){$\cup$}
        \put(48, 13){$=$}
        \put(49, 13){$=$}
        \put(74, 13){or}
    \end{overpic}

    \caption{\label{fig:ambiguities}Spectra capture isometry classes, thus there exist ambiguous cases where unions have multiple valid solutions. {\em Top}: The two solutions are isometric, hence intrinsically equivalent. {\em Bottom}: Since each part is isospectral to its symmetric version, the union of the two spectra can result in three possible solutions (we only show two for simplicity). The semi-transparent full shape is for reference.}
\end{figure}

Based on these observations, we consider two different scenarios with different characteristics:
\begin{enumerate}
    \item $\M_1\cup\M_2$ is a complete, watertight shape;
    \item $\M_1\cup\M_2$ is a partial shape itself.
\end{enumerate}
%
As we demonstrate below, Scenario 1 is simple enough to be solved with a feed-forward network, and generalizes well to unseen data, as shown in Figure~\ref{fig:instant_recovery}.
%
Scenario 2 is more difficult, since allowing partiality on the union shape introduces another dimension of variability, as well as more ambiguity on the possible output; see Figure~\ref{fig:ambiguities} for examples.

Despite being more difficult to solve, the latter scenario lends itself to modeling more complex interactions. In particular, exploiting the associative property of the union, we can compose  $m>2$ partial shapes simply by aggregating pairwise unions:
\begin{align}
    \bm{\lambda}(\M_1  \cup & \M_2 \cup \cdots \cup \M_m)=\\ &\mathcal{U}_\Theta(\cdots(\mathcal{U}_\Theta(\bm{\lambda}(\M_1),\bm{\lambda}(\M_2)),\cdots),\bm{\lambda}(\M_m))\nonumber
\end{align}
\setlength{\columnsep}{7pt}
\begin{wrapfigure}[]{r}{0.4\linewidth}
\vspace{-0.3cm}
\begin{center}
\begin{overpic}
[trim=0cm 0cm 0cm 0cm,clip,width=1.0\linewidth]{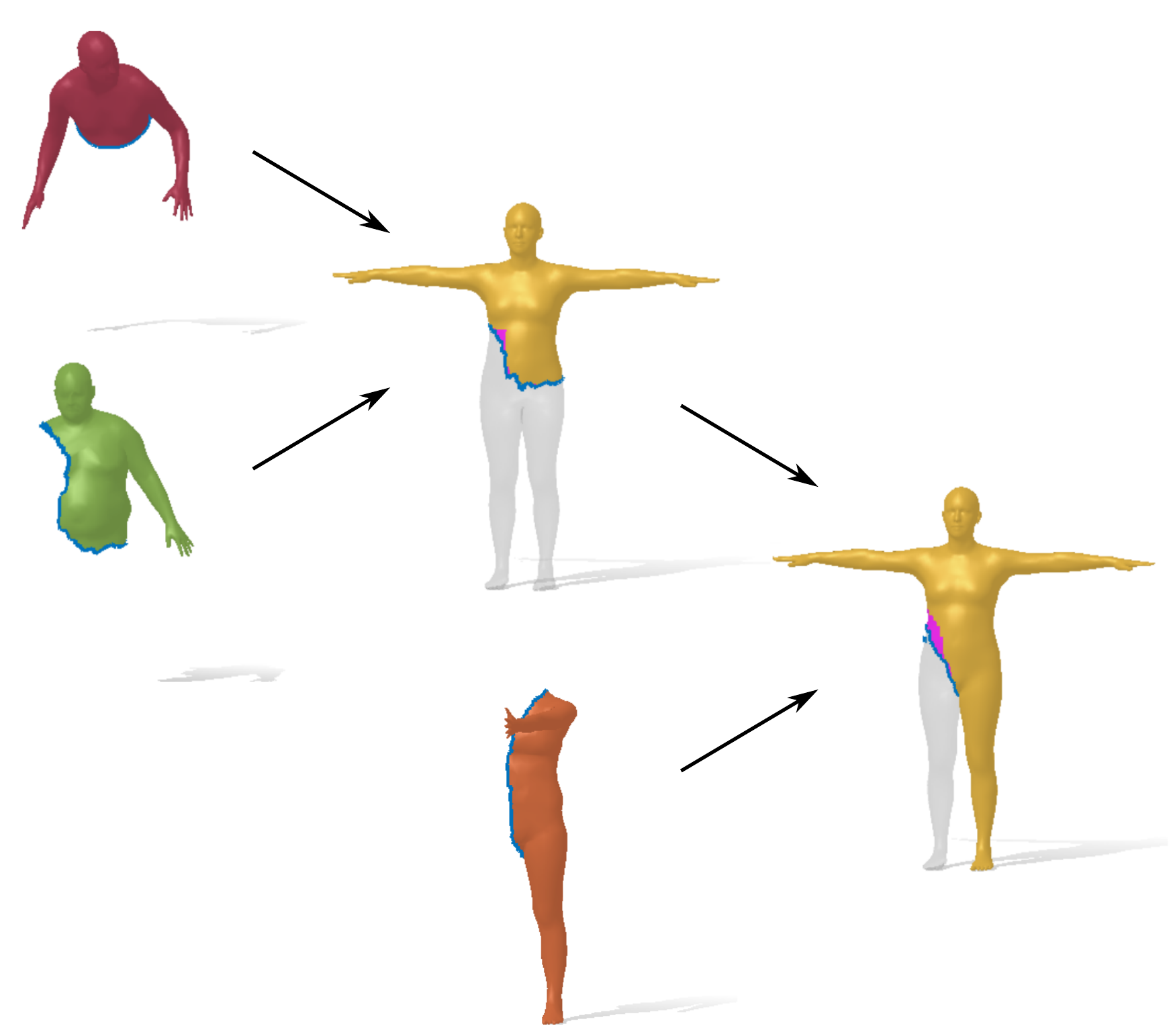}
        \put(14, 82){\footnotesize $\M_1$}
        \put(14, 34){\footnotesize $\M_2$}
        \put(50, 7){\footnotesize $\M_3$}
\end{overpic}
\end{center}
\end{wrapfigure}
See Figure~\ref{fig:teaser} and the inset on the right for an illustration.
Note that composing $m$ partial shapes resembles the `non-rigid puzzle' setting seen in~\cite{litany2016non}, although with a crucial difference: the method of~\cite{litany2016non}  has access to the complete shape, which is instead unknown to us.


\section{Network architecture}\label{sec:dl}
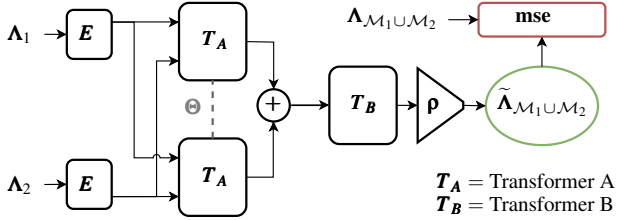
\begin{figure}[t]
\centering
    \begin{overpic}[trim=0cm 0cm 0cm 0cm,clip, tics=10, width=0.9\linewidth]{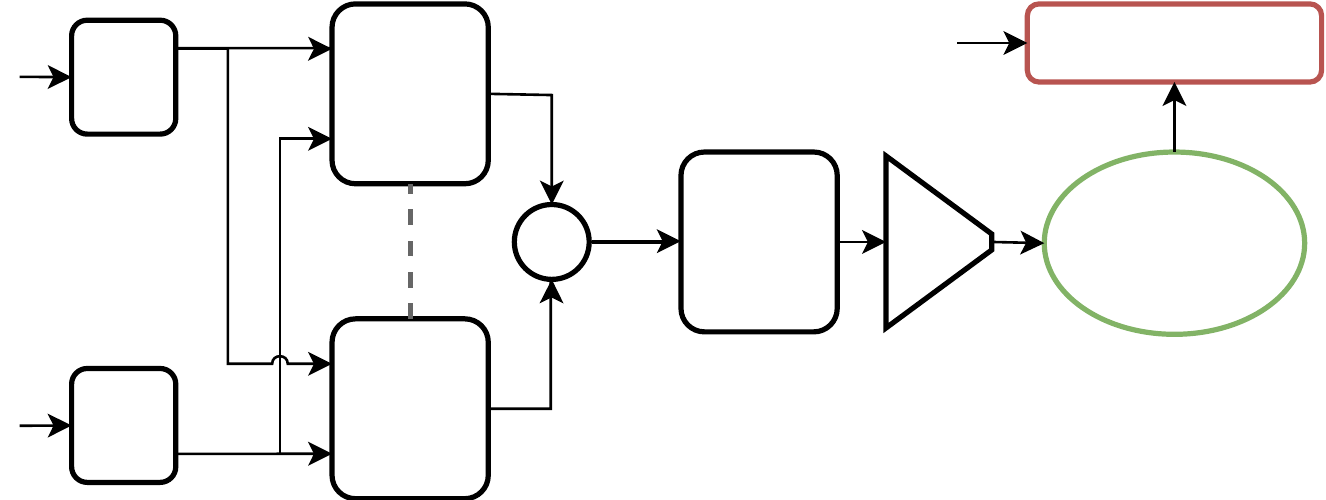}
    \put(-5, 31){\footnotesize $\bm{\Lambda}_{1}$}
    \put(-5, 4.5){\footnotesize $\bm{\Lambda}_{2}$}
    
    \put(7.5, 31){\footnotesize $\bm{E}$}
    \put(7.5, 4.5){\footnotesize $\bm{E}$}
    
    
    \put(29, 6){\footnotesize $\bm{T_A}$}
    \put(26, 18){\footnotesize \color{gray}{$\bm{\Theta}$}}
    \put(29, 30){\footnotesize $\bm{T_A}$}
    \put(40, 18.75){\footnotesize $\bm{+}$}
    \put(55, 18.75){\footnotesize $\bm{T_B}$}
    \put(68.5, 18.75){\footnotesize $\bm{\rho}$}
    
    \put(80.75, 18.75){\footnotesize $ \bm{\widetilde{\Lambda}}_{\M_1 \cup \M_2}$}
    \put(54.5, 34){\footnotesize $\bm{\Lambda}_{\M_1 \cup \M_2}$}
    \put(84, 34){\footnotesize $\mathbf{mse}$}
    
    \put(70, 5){\footnotesize $\bm{T_A} = $ Transformer A}
    \put(70, 1){\footnotesize $\bm{T_B} = $ Transformer B}
    \end{overpic}

    \caption{\label{fig:encoder}{Our neural architecture.
    $\bm{\Lambda}_{1}$ and $\bm{\Lambda}_{2}$ are the input eigenvalues of the partialities,
     $\bm{E}$ embeds the eigenvalues into a high dimensional space,
     the transformer $\bm{T_A}$ produces a latent representation of the inputs that are summed up to obtain a commutative latent representation of the union,
    the transformer $\bm{T_B}$ plus the linear dimensionality reduction $\bm{\rho}$ decodes this latent representation to obtain the predicted eigenvalues $ \bm{\widetilde{\Lambda}}_{\M_1 \cup \M_2}$.}}
\end{figure}

Our network takes as input two sequences of $k$ eigenvalues, each associated with a partial shape, and outputs a sequence of $k$ eigenvalues, as a prediction of the spectral union. 
Figure~\ref{fig:encoder} illustrates the neural architecture. It is composed of three main blocks: (1) the projection of the input eigenvalues into a high dimensional space; (2) two transformers, forced to be commutative, to learn the union operation; (3) a dimensionality reduction to decode the spectral union.


\vspace{1ex}\noindent\textbf{Eigenvalue embeddings.}
The Laplacian eigenvalues of surfaces form a non-decreasing sequence that approximately grows linearly with rate inversely proportional to surface area, a behavior described by Weyl's asymptotic law~\cite{weyl11}.
{This results in the input eigenvalues hugely varying depending on the area of the partial shape. To guard against network instability we encode the spectra via the offset representation:}
%
%
%
\begin{equation*}
    \mathrm{off}(\lambda_i)  =
    \lambda_i - \lambda_{i-1}\,,
\end{equation*}
with $\mathrm{off}(\lambda_1)=\lambda_1$. {This representation has the further advantage of imposing the increasing order constraint on the predicted eigenvalues, by just requiring the non-negativity of the predicted offset sequence.}




In practice, the network sees each spectrum as a sequence of length $k$ offsets $\bm{\Lambda}=(\mathrm{off}(\lambda_1), \dots, \mathrm{off}(\lambda_k)) \in \mathbb{R}^k$, 
each one is
then embedded into a higher-dimensional representation of length $2\ell+1$, constructed as follows:
%
%
\begin{equation*}
    \mathrm{off}(\lambda_i)
    \mapsto
    \left(
    {\color{darkblue}\vec{\theta}_a^i},
    \quad
    \mathrm{off}(\lambda_i)
    {\color{darkblue}\vec{\theta}_b},
    \quad
    \mathrm{off}(\lambda_i)
    \right)
\end{equation*}
%
  

where ${\color{darkblue}\vec{\theta_a^i}}$ is a $\ell$-dimensional
vector acting as a positional encoding for the \textit{i}-th offset,
and ${\color{darkblue}\vec{\theta_b}}$ is a linear mapping of the
offset to a $\ell$-dimensional space. The learnable vectors
${\color{darkblue}\vec{\theta_a^i}}$ and
${\color{darkblue}\vec{\theta_b}}$, once learned, are independent from the input shapes and eigenvalues. 

This representation encodes both the eigenvalue quantity and its position in the sequence, which is a fundamental information for recovering the geometry area.

\vspace{1ex}\noindent\textbf{Symmetric architecture.}
%
Given the eigenvalue sequences of $\bm{\Lambda}_1$ and $\bm{\Lambda}_2$ (associated to $\M_1$ and $\M_2$ respectively), our neural architecture learns how to perform their union without ever leaving the spectral domain.
%
%
We further require our model to be commutative, i.e., the result should {\em not} depend on which pair between $(\bm{\Lambda}_1,\bm{\Lambda}_2)$ or $(\bm{\Lambda}_2,\bm{\Lambda}_1)$ is given as input.



{We gain this invariance by using a single transformer $\bm{T_A}$ on the embedded eigenvalues, performing two symmetric operations to obtain a {representation} of $\bm{\Lambda}_{1}$ {informed about} $\bm{\Lambda}_{2}$ and vice-versa. The two transformed {representations} are summed together to obtain a commutative representation of the union.
}
We then feed the result to the second transformer $\bm{T_B}$, whose task is to decode the union into a representation that can be easily reduced, via a simple linear layer $\bm{\rho}$, from the high-dimensional {representation} back to a sequence of eigenvalues. The whole architecture is illustrated in Figure~\ref{fig:encoder}. 
 
{The transformers are position-aware neural networks, where the output for each eigenvalue depends on its value and position together with all the other eigenvalues and their positions. It employs an \emph{attention}  mechanism to learn relation among eigenvalues.}
 
{In the {network,} the dimensionality of each representation is $32$,
$\bm{T_A}$ has $8$ heads and $6$ layers meanwhile $\bm{T_B}$ has $8$ heads and $3$ layers.
Thus, $\bm{\rho}$ reduces the representation dimensionality from $32$ to $1$.
Refer to the supplementary materials for further details.}


\vspace{1ex}\noindent\textbf{Training.}
Our model is trained with a mean squared error loss between the predicted and ground truth spectra.  Before entering the loss, the offset representation for the eigenvalues is decoded with a cumulative sum.
{Experimental results show that penalizing the loss according to the linear increase of the eigenvalues does not yield significant improvements.}
%
%
{In the training phase, we augmented the partial regions with small random changes in their surface area.}
The optimizer used is Adam with a learning rate of 2e-4 and weight decay of 1e-5. We use a learning rate scheduler to escape local minima and stabilize the training, in particular the cosine annealing with warm restarts scheduler \cite{loshchilov2017sgdr}, doubling at each restart the number of epochs between restarts.
{We trained the model for 6741 epochs for a total of 1d 13h 46m on a GeForce RTX 2080 TI, tracking the experiments with \cite{wandb}.} 

%
%

\section{Data and evaluation.}
%
In our experiments we use 3D data from the FAUST~\cite{bogo2014faust} and SURREAL~\cite{varol17_surreal} datasets of deformable human shapes with different identities. This provides us with a total of 50 different identities, each in 10 different poses.
To produce partial data, we first extract surface patches of various sizes from the full shapes, and then combine the patches randomly to form two datasets: 
\begin{itemize}
    \item A dataset of $\sim$150 partial pairs, where each union
      covers the entire surface. We test in three different settings
      depending on the information given at training time: (i) known
      identity, unknown partiality; (ii) unknown identity, known
      partiality; (iii) both identity and partiality are
      unknown.  {We define an identity as \emph{known} if the training set contains any partiality in any pose of the same shape, and we} {consider a partiality \emph{known} if the two input partiality types together with their corresponding union are in the training dataset in any shape identity or pose.}
    \item A dataset of $\sim$100 partial pairs, whose union does not cover the entire surface. For training, the partial shapes are augmented by enlarging/shrinking the patches randomly. We consider the same three settings as above.
\end{itemize}
As shown in Figure~\ref{fig:ambiguities}, {there are cases in which more than one region on the template is a valid solution to the union problem}{. Two different ambiguities arise: 
(a) symmetric counterparts of one or both input partial regions may produce different union regions with different spectra;
(b) symmetric union regions are described by the same spectra even though {they} are localized in different parts of the shape.
We remove these ambiguities in the training data by following a minimum union area principle and privileging ``left-sided'' symmetries exploiting a ground truth symmetry map and labels of the template left side.
{By this choice, associativity is promoted as we show empirically in the results.}
}

%

We define two test sets. 
In \textsc{Test A}, the pose or the type of partiality have never been seen, but the predicted union may be seen in a different pose or identity at training time. \textsc{Test B} is more challenging, since the union of the two parts has never been seen at training (neither in a different pose or identity, nor as a union of different partialities). 
%
{The number of samples in the test datasets is about 15\% of all the data available, the remaining data is used for training.}

\begin{wrapfigure}[4]{r}{0.54\linewidth}
\vspace{-0.3cm}
\begin{center}
\begin{overpic}
[trim=0cm 0cm 0cm 0cm,clip,width=1.0\linewidth]{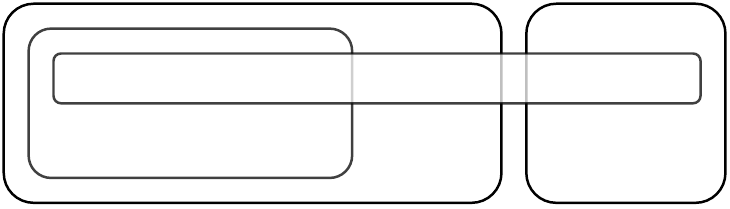}
        \put(50,9){\scriptsize unknown}
        \put(55, 4){\scriptsize man}
        \put(77, 9){\scriptsize unknown}
        \put(79, 4){\scriptsize woman}
        %
        \put(7.5, 7){\scriptsize known (in train data)}

        \put(20, 16.5){\scriptsize re-meshed (dropping $30\%$ vertices)}
\end{overpic}
\end{center}
\end{wrapfigure}
We analyze both \textsc{Test A} and \textsc{Test B} scenarios in different settings summarized in the inset Venn diagram.



In Table~\ref{table:encoder_evaluation} we report a quantitative analysis of the predictive power of our learning model, according to the mean squared error (mse) and mean absolute error (mae) metrics. 

Further, we perform qualitative experiments on different classes, on
horses from TOSCA \cite{TOSCA} and earphones from PartNet
\cite{Mo_2019_CVPR}, {some examples are in Figure~\ref{fig:results:mask_prediction_tosca},~\ref{fig:results:mask_prediction:pc} and~\ref{fig:results:compositionality}. 
Through these experiments we prove our method generalization ability to any shape category. Additionally, we successfully trained the neural network on humans and fine-tuned it to work with horses, where the data is scarce, demonstrating that it is possible to perform transfer learning between different shape classes.} 
Details about the data generation and training process for these classes is described in the supplementary material.
%
%
%
{Sample code and data are available online\footnote[1]{Link to the repository \href{https://github.com/lucmos/spectral-unions}{https://github.com/lucmos/spectral-unions}}.}

\begin{table}[t]
    \centering
    \begin{tabular}{ cl|cl}
        \multicolumn{2}{c}{} 
                &   mse   &   mae                    \\[.5ex]
        \multirow{6}{*}{\rotatebox[origin=c]{90}{\textsc{Test A}}}
            & known man & \textbf{{\color{darkblue}11.14}}  & \textbf{{\color{darkblue}2.09}}   \\ 
            & unknown man & 13.25  & 2.59    \\ 
            & unknown woman  & 36.92  &  3.93   \\[.5ex]
            & known man re-meshed & 29.61  & 3.31   \\ 
            & unknown man re-meshed & 32.67  & 3.60    \\ 
            & unknown woman re-meshed & 62.33  & 5.23    \\ 
        \hline
        \multirow{6}{*}{\rotatebox[origin=c]{90}{\textsc{Test B}}}
            & known man & 15.41  & 2.59    \\ 
            & unknown man & 24.05  & 3.60    \\ 
            & unknown woman  & 64.47  & 4.99    \\[.5ex]
            & known man re-meshed & 51.20  & 4.54    \\ 
            & unknown man re-meshed & 75.91  & 5.90    \\ 
            & unknown woman re-meshed & \textbf{{\color{darkred}110.17}} & \textbf{{\color{darkred}6.78}}    \\  
    \end{tabular}
    \caption{Error between the predicted and ground truth eigenvalues in different experimental settings. 
    In each row, ``known'' denotes an identity included in the
    training set, ``unknown'' one not included, and ``re-meshed''
    indicates that the shapes were re-meshed {by removing $30\%$ of their vertices}     before computing their
    spectrum . 
    }
    \label{table:encoder_evaluation}
\end{table}

\section{Applications}\label{sec:results}
%







We can easily plug our method into existing pipelines that take as input Laplacian eigenvalues. Unique to our approach is that it addresses the scenario in which only partial views of the complete shape are available. 
We also refer to the Supplementary for further details and results.

\subsection{Geometry reconstruction}\label{sec:instant_recovery}
To recover the shape geometry from its predicted union eigenvalues, we use the data-driven method of~\cite{Instant2020}, which takes eigenvalues as input and directly yields a 3D mesh embedding as output. 
%
An example is given in Figure~\ref{fig:instant_recovery}, where we compare the geometry recovered from our estimated spectra with the one obtained from the ground truth eigenvalues.
%
{%
For human meshes where the correspondence between their T-Pose and the connectivity adopted in \cite{Instant2020} is known, 
we can compute the point-wise reconstruction error as the L2 distances between correspondent points.}
We plot this Euclidean error on the reconstructed surface. White color corresponds to zero error and dark red encodes a larger error. 
%
%
Our spectrum prediction is accurate enough to retain the core geometric information of the original partial shapes, as it can be seen in these examples.
{For these experiments we used the pre-trained network provided by the authors of~\cite{Instant2020} {and~\cite{marin2021spectral}. We sampled the test shapes outside the training set adopted in these papers.} Thus the network is not specifically trained to handle spectra predicted by our pipeline. Moreover, since the spectrum encodes just intrinsic properties (i.e. appearance) of the shape, all the reconstructions of ~\cite{Instant2020} are in the T-pose.}

%



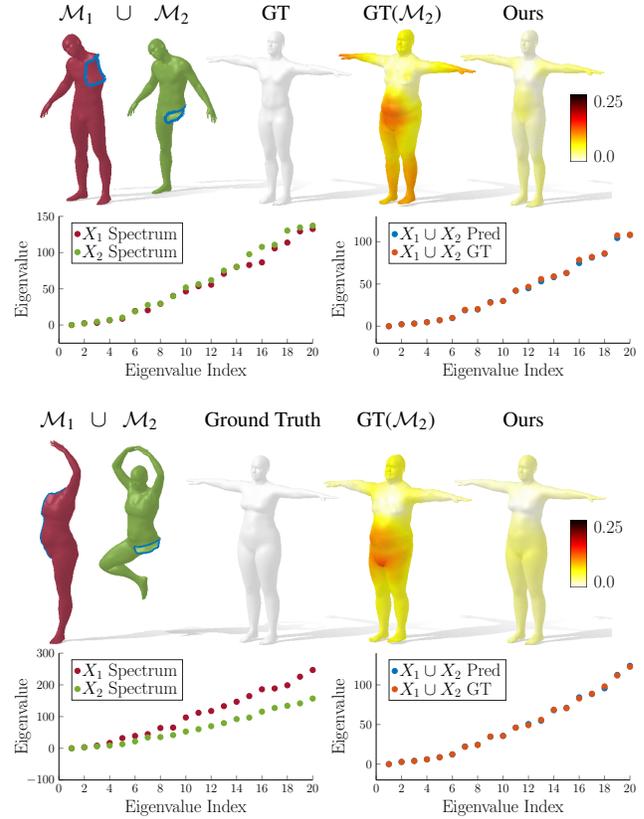
\begin{figure}[t]
\centering
    \begin{overpic}[trim=0cm 0cm 0cm -0.4cm,clip,width=\linewidth]{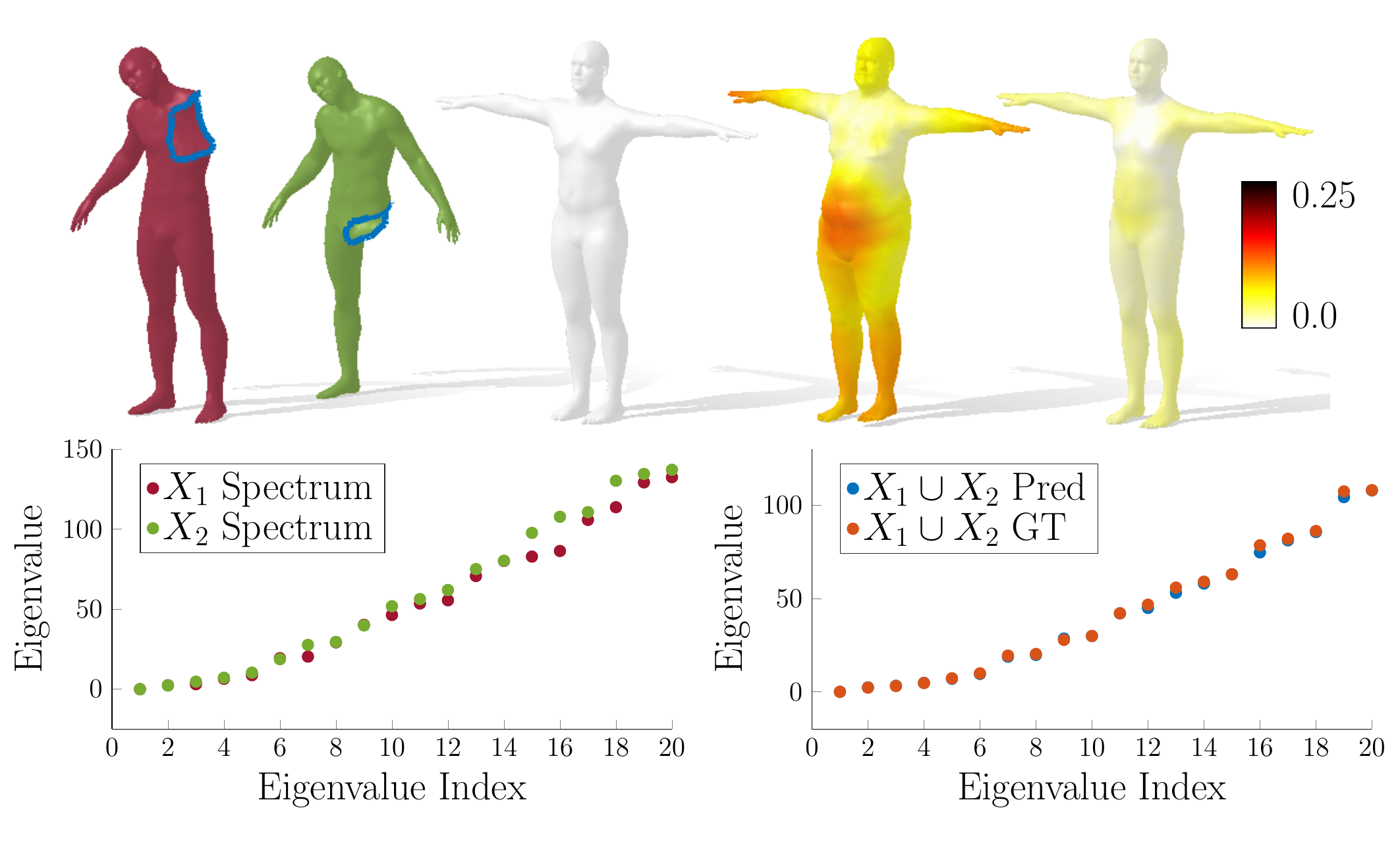}
        \put(8,60){\footnotesize $\M_1$ }
        \put(17,60){$\cup$ }
        \put(23,60){\footnotesize $\M_2$ }
        \put(40,60){\footnotesize GT}
        \put(56,60){\footnotesize GT($\M_2$)}
        \put(78,60){\footnotesize Ours}
    \end{overpic}\\[1ex]
    \begin{overpic}[trim=0cm 0cm 0cm -1cm,clip,width=\linewidth]{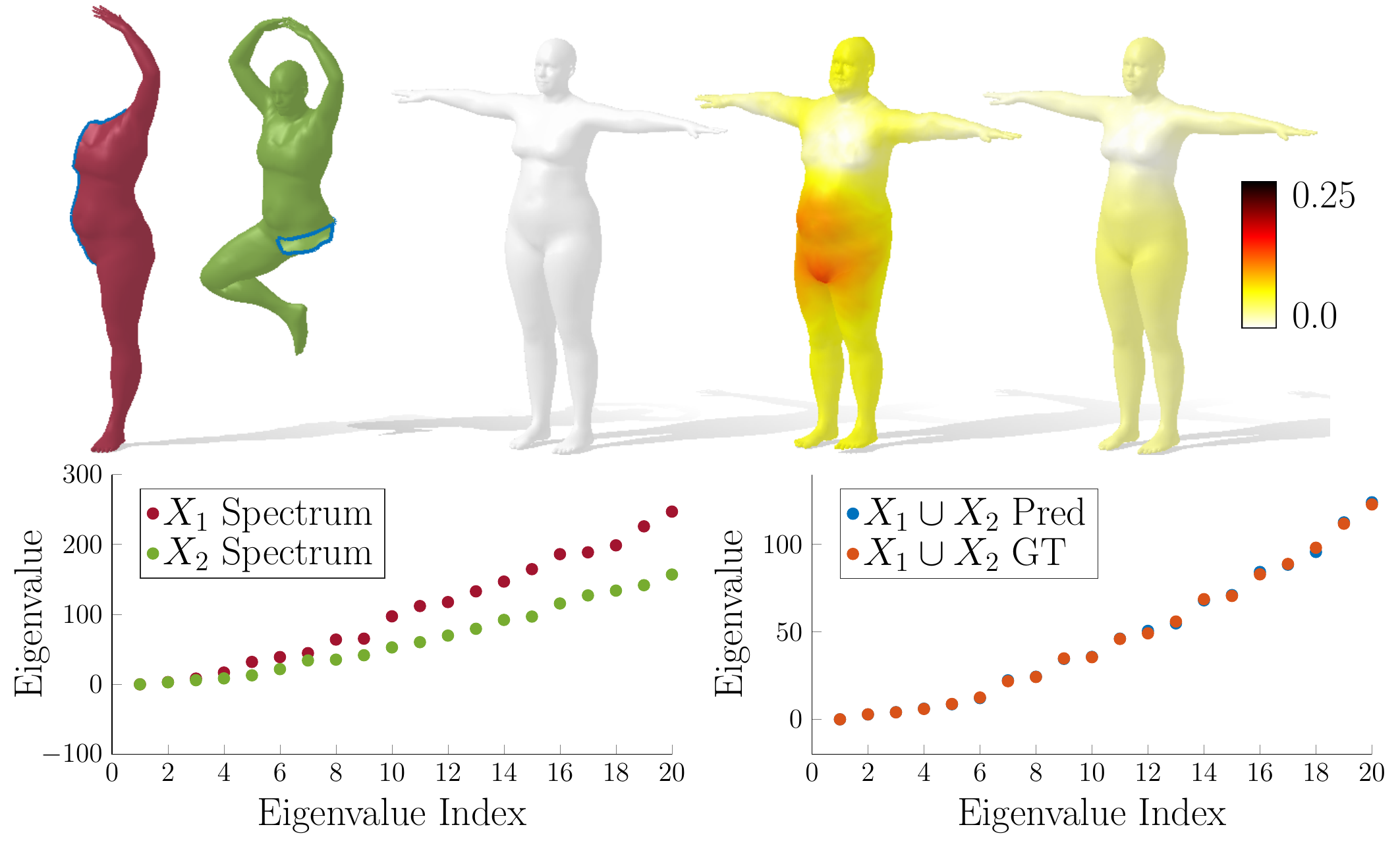}
        \put(5,63){\footnotesize $\M_1$ }
        \put(13,63){$\cup$ }
        \put(18,63){\footnotesize $\M_2$ }
        \put(31,63){\footnotesize Ground Truth}
        \put(55,63){\footnotesize GT($\M_2$)}
        \put(78, 63){\footnotesize Ours}
    \end{overpic}
    \caption{\label{fig:results:instant_recovery_combined}
    Given two partial shapes as input, we compare the reconstruction obtained by running the method of~\cite{Instant2020} only on a partial input (the green shape), yielding the fourth shape, with the reconstruction obtained from our predicted full spectrum, yielding the last shape.
    }
\end{figure}

To emphasize the importance of having an aggregated spectrum, as predicted by our model, in Figure~\ref{fig:results:instant_recovery_combined} we show the reconstructions obtained with the method of~\cite{Instant2020} when using the spectrum of just one of the two partial shapes as input. 
The result in this case is quite different from what is expected, showing that existing state-of-the-art pipelines are not able to handle partial shapes correctly. 

\begin{table}[t!]
    \centering
    \begin{tabular}{ cl|cl}
        \multicolumn{2}{c}{} 
                &   IoU   &   Acc.                    \\[.5ex]
        \multirow{6}{*}{\rotatebox[origin=c]{90}{\textsc{Test A}}}
            & known man & \textbf{{\color{darkblue}99.28\%}} &  \textbf{{\color{darkblue}99.61\% }}  \\
            & unknown man & 93.78\% & 95.83\%  \\
            & unknown woman  & 94.19\% & 96.32\%   \\[.5ex]
            & known man re-meshed & 98.54\% & 99.06\%  \\
            & unknown man re-meshed & 91.44\% & 94.08\%   \\
            & unknown woman re-meshed & 93.47\% & 95.56\%  \\
        \hline
        \multirow{6}{*}{\rotatebox[origin=c]{90}{\textsc{Test B}}}
            & known man & 97.96\% & 98.55\%  \\
            & unknown man & 87.58\% & 92.52\%  \\
            & unknown woman  & 96.05\% & 98.46\%  \\[.5ex]
            & known man re-meshed & 93.04\% & 97.33\%  \\
            & unknown man re-meshed & \textbf{{\color{darkred}83.69\%}} & \textbf{{\color{darkred}91.08\%}} \\
            & unknown woman re-meshed & 95.59\% & 98.43\%  \\
    \end{tabular}
    \caption{Intersection over union (IoU) and accuracy in the region localization task, in different experimental settings. Model trained on a single identity, to show generalization. }
    \label{table:mask_prediction:single_identity:performance}
\end{table}


\begin{table}[t!]
    \centering
    \begin{tabular}{ cl|cl}
        \multicolumn{2}{c}{} 
                &   IoU   &   Acc.                    \\[.5ex]
        \multirow{6}{*}{\rotatebox[origin=c]{90}{\textsc{Test A}}}
          & known man & \textbf{{\color{darkblue}98.24\%}} &\textbf{{\color{darkblue} 99.09\%}} \\
            & unknown man & 96.26\% & 97.64\% \\
            & unknown woman  & 96.17\% & 98.04\% \\[.5ex]
            & known man re-meshed & 97.70\% & 98.74\% \\
            & unknown man re-meshed & 95.88 \% & 97.78\% \\
            & unknown woman re-meshed & 96.04\% & 97.66\% \\
        \hline
        \multirow{6}{*}{\rotatebox[origin=c]{90}{\textsc{Test B}}}
            & known man & 97.43\% & 99.14\% \\
            & unknown man & 93.31\% & 98.23\% \\
            & unknown woman  & 95.74\% & 98.59\% \\[.5ex]
            & known man re-meshed & 97.61\% & 99.11\% \\
            & unknown man re-meshed & \textbf{{\color{darkred}90.85 \%}} & \textbf{{\color{darkred}97.63\%}} \\
            & unknown woman re-meshed & 96.81\% & 98.98\% \\
    \end{tabular}
    \caption{Performance when training on six different identities instead of a single identity (compare with Table~\ref{table:mask_prediction:single_identity:performance}).
    }
    \label{table:mask_prediction:multi_identity:performance}
\end{table}

\begin{figure}[t]
\centering
    \begin{overpic}[trim=0cm 0cm 0cm -0.5cm,clip,width=\linewidth]{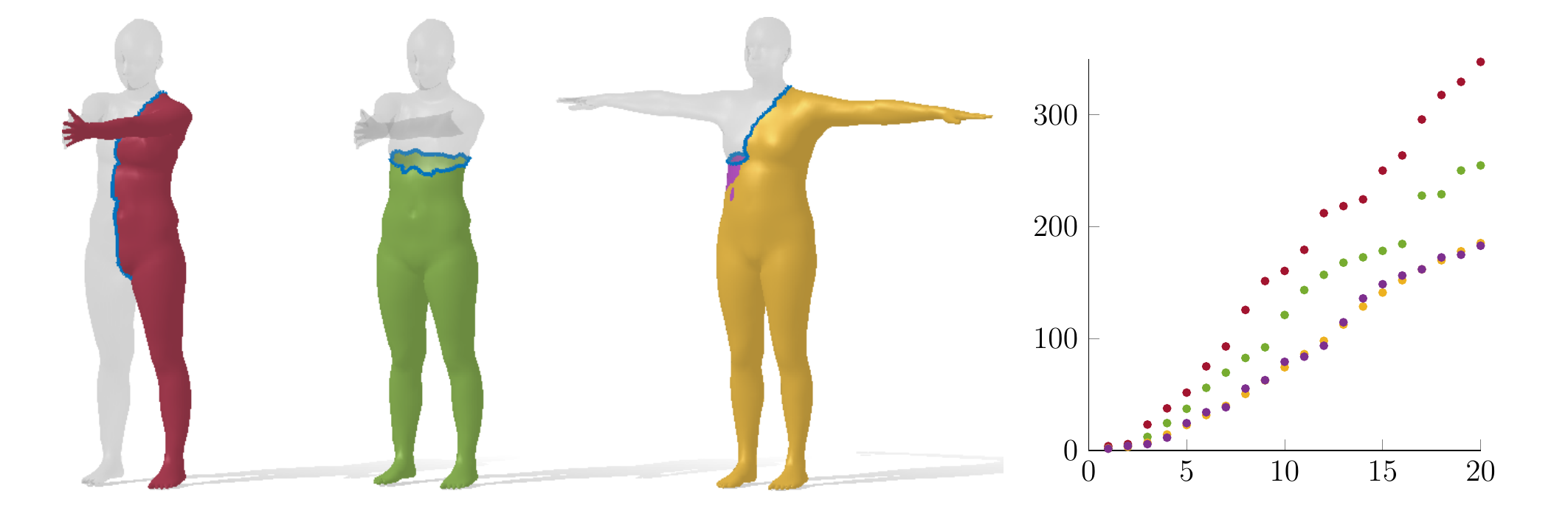}
        \put(6,33){\footnotesize $\M_1$ }
        \put(25,33){\footnotesize $\M_2$ }
        \put(45,33){\footnotesize Mask}
        \put(66,33){\footnotesize Laplacian eigenvalues}
        \put(17,33){$\cup$}
        \put(36.5,33){$=$}
    \end{overpic}
    \includegraphics[width=\linewidth]{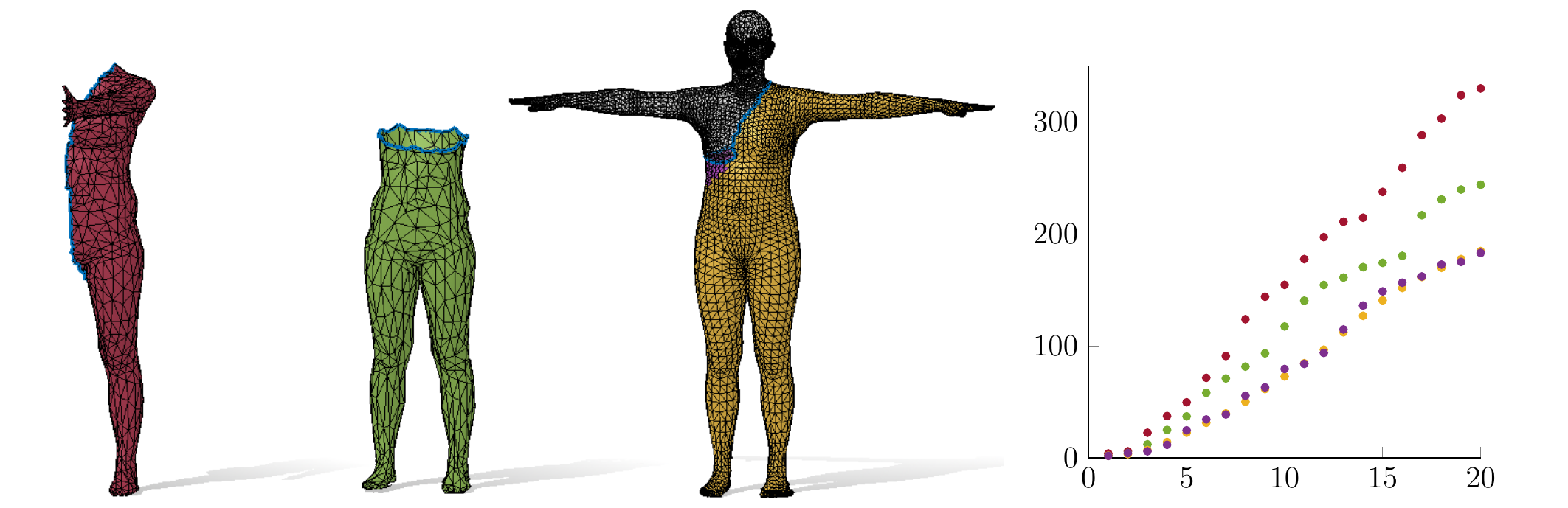} \\
    \includegraphics[width=\linewidth]{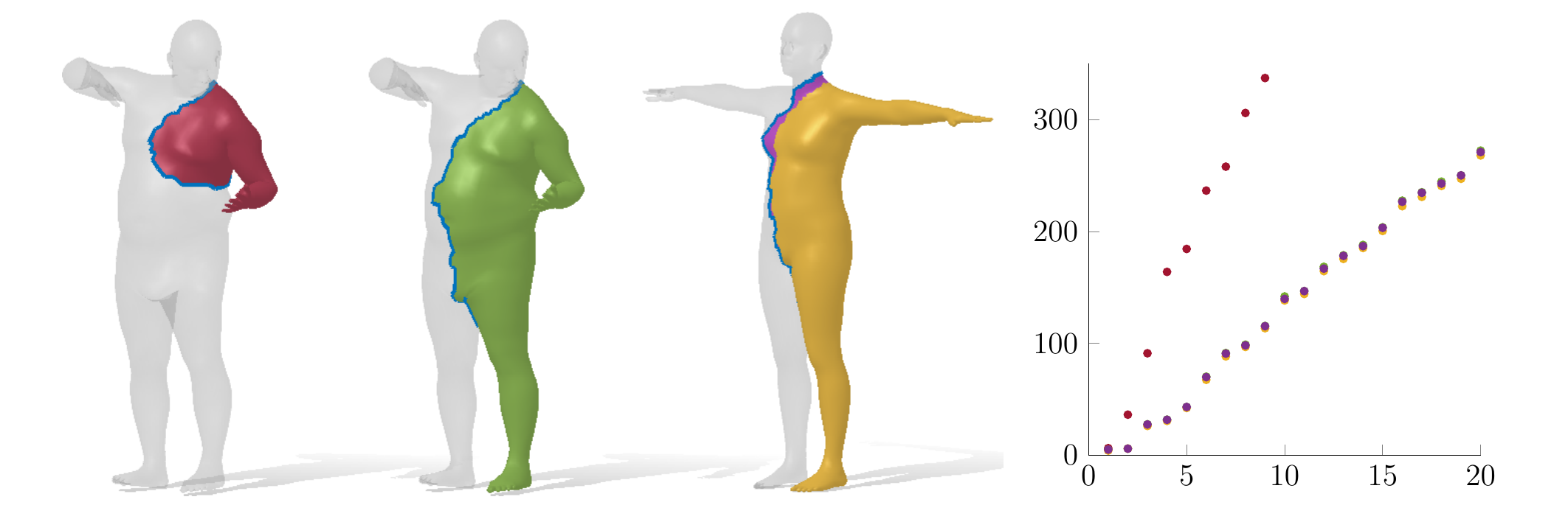} \\
    \includegraphics[width=\linewidth]{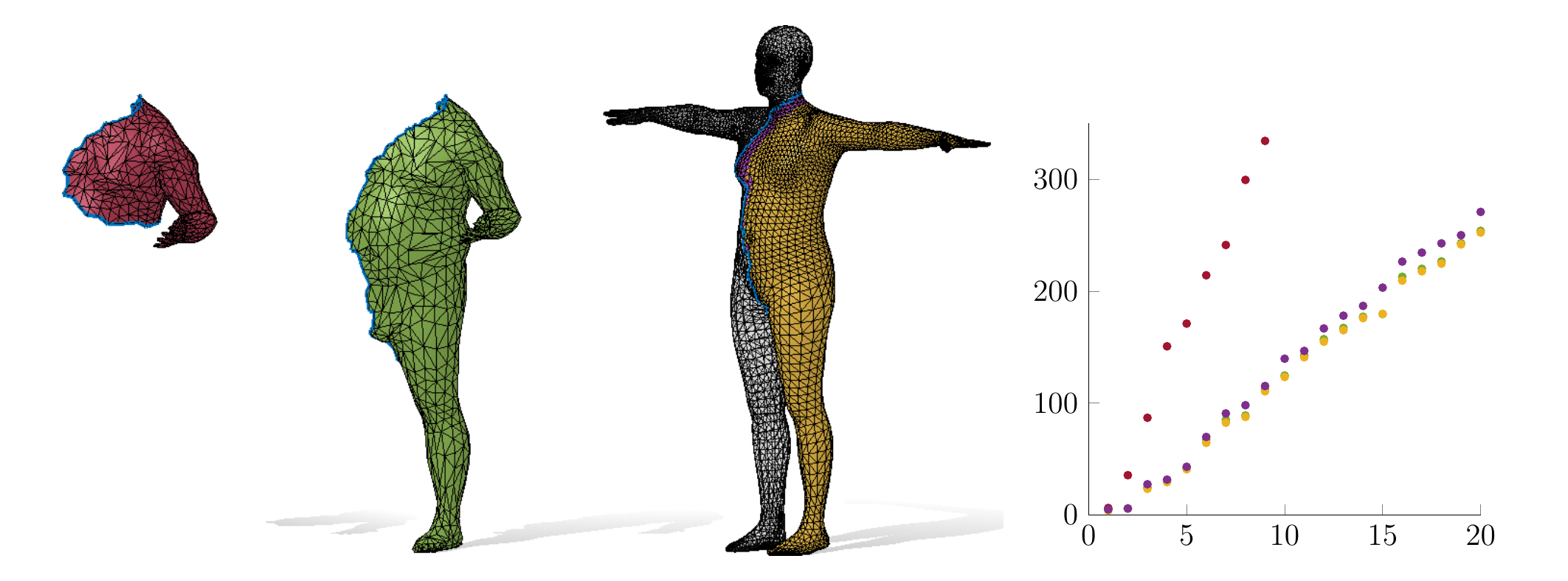} 
    \caption{\label{fig:resuls:mask_prediction:comparison_with_gt} Region localization task, under the effect of different mesh connectivity. Given the eigenvalues of two partial shapes, we correctly predict an indicator function that represents the union of the two over a fixed template.}
\end{figure}

\subsection{Region localization}\label{sec:mask_prediction}
%
This task, introduced in~\cite{rampini2019correspondence}, consists in locating, on a fixed template, the region corresponding to a given partial shape.
%
%
%
{To solve this problem we combine the spectral union model introduced in Section~\ref{sec:dl} with a simple MLP, described in detail in the supplementary materials. The MLP takes as input the predicted eigenvalues of the union, and outputs an indicator function over the vertices of the template. The spectral union operator is not trained to solve the region localization but is used as-is with frozen weights.}
 
{In principle, substituting $\bm{T_B}$ in Figure~\ref{fig:encoder} with the region localization MLP would work if the whole system is trained end-to-end. However, the goal of this work is to perform the union operation in the spectral space. Moreover, if we do not impose the union to be a spectrum, we would not be able to compose the predicted union spectra with another partiality.}

In the loss definition, one must take care of the potential ambiguities exemplified in Figure~\ref{fig:ambiguities}; we do so by implementing a symmetry-invariant loss, that does not penalize symmetric solutions.
The MLP is trained using the train/test splits described in Section~\ref{sec:dl}, with the difference that we used just 6 different identities in the training phase. 

To analyze the prediction quality on this task we adopt two metrics: intersection over union (IoU) of the predicted mask with the ground truth mask, and accuracy, i.e. the ratio of correctly predicted vertices over the full template.
We show several qualitative results in  Figure~\ref{fig:resuls:mask_prediction:comparison_with_gt} {and attach an interactive demo in the supplementary materials}. 

\begin{figure}[t]
\centering
    \begin{overpic}[trim=0cm -0.5cm 0cm -0cm,clip,width=0.85\linewidth]{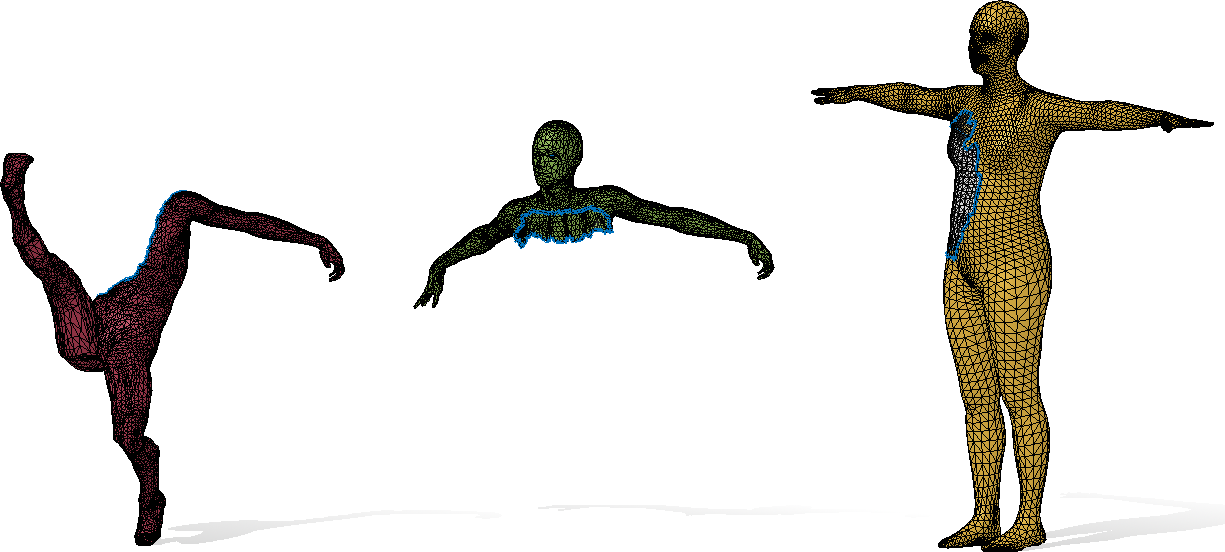}
        \put(7.5,42){\footnotesize $\M_1$ }
        \put(28,42){\footnotesize $\cup$ }
        \put(43,42){\footnotesize $\M_2$ }
        \put(57,42){$=$}
        \put(68,42){\footnotesize Mask}
    \end{overpic}
     \vspace{0.1cm}
    \includegraphics[width=0.84\linewidth]{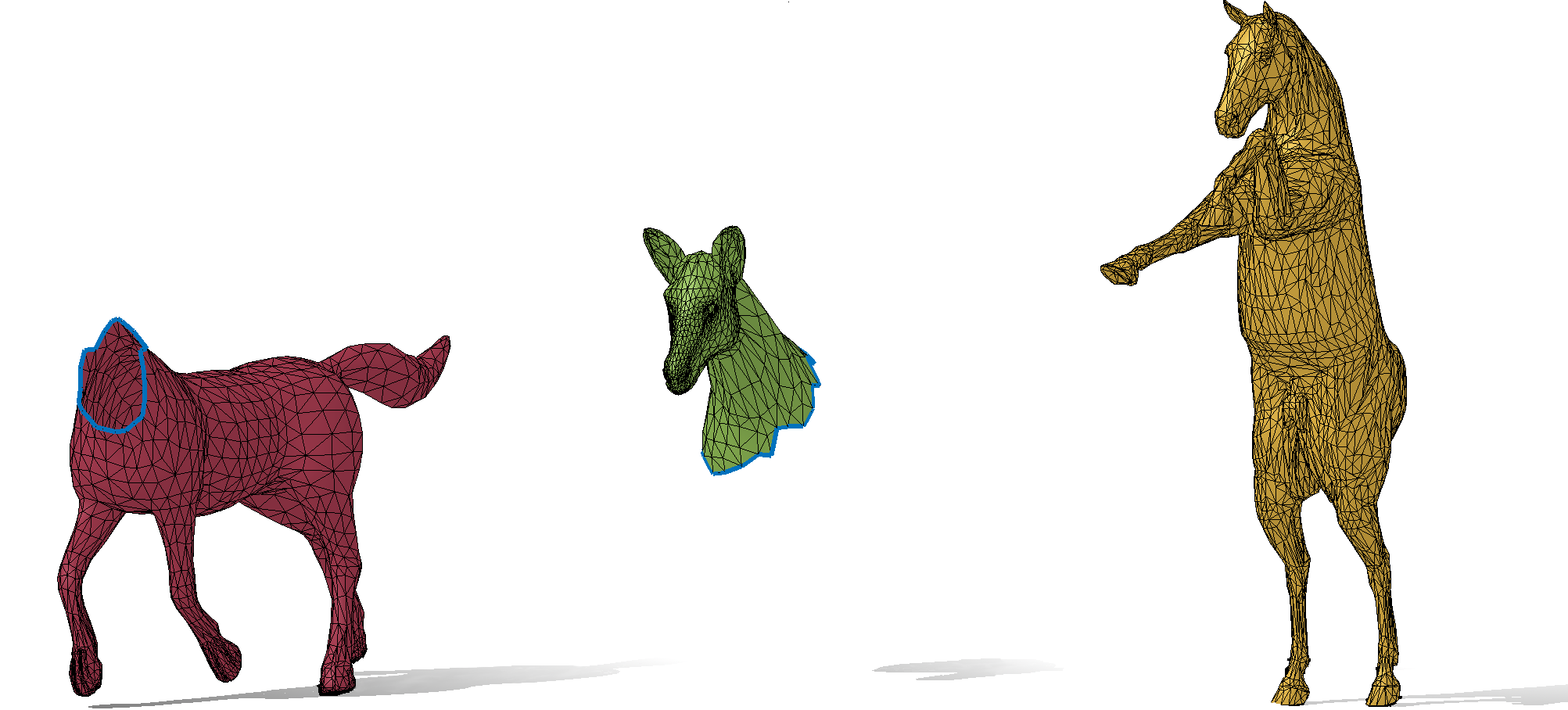}
    \vspace{0.1cm}
    \includegraphics[width=0.85\linewidth]{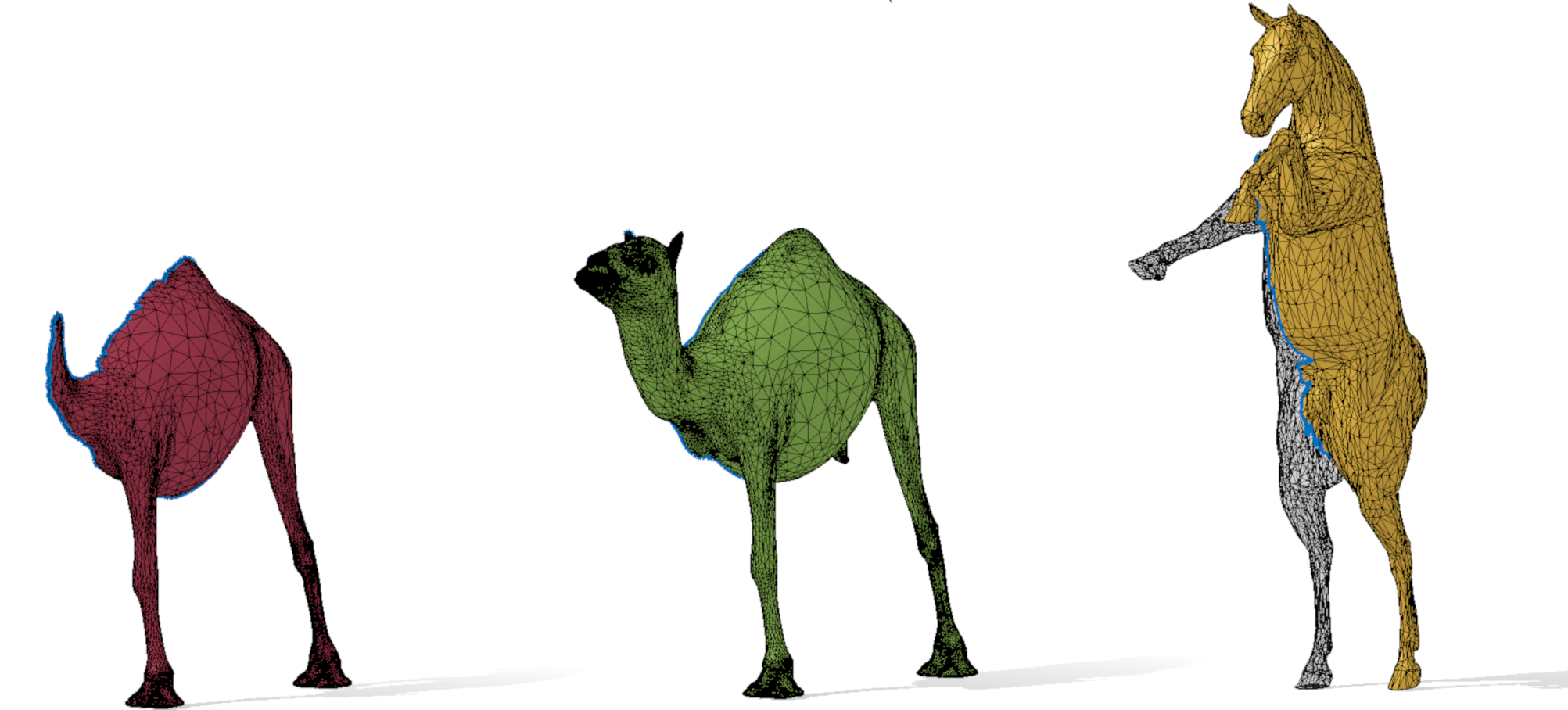}
    \caption{\label{fig:results:mask_prediction_tosca} Region localization across different datasets. Partial shapes come from datasets not involved in the training.}
\end{figure}

\vspace{1ex}\noindent\textbf{Robustness to {re-meshing}.}
One key aspect of Laplacian eigenvalues is that they are robust to shape discretization and mesh connectivity. Our model inherits this robustness; see  Figure~\ref{fig:resuls:mask_prediction:comparison_with_gt}, where we highlight the {re-meshed} inputs by visualizing their surface triangulation. This is supported also by Tables~\ref{table:mask_prediction:single_identity:performance} and \ref{table:mask_prediction:multi_identity:performance}, where the performance on the {re-meshed} shapes is comparable with the original ones. In these experiments, we test our network with the eigenvalues computed from noisy, re-meshed partial shapes obtained by removing $30\%$ of their vertices with an edge collapse algorithm~\cite{garland1997surface}.

\vspace{1ex}\noindent\textbf{Generalization to new identities.}
Our approach generalizes to identities unseen at training time as can be noted in Table~\ref{table:mask_prediction:multi_identity:performance}. To further stress this aspect, we devised an experimental setup in which we used as training set just a single identity. The results of this setup are shown in Table~\ref{table:mask_prediction:single_identity:performance}. 

\vspace{1ex}\noindent\textbf{Generalization to different datasets.}
In Figure~\ref{fig:results:mask_prediction_tosca} we use partial shapes from other datasets to localize regions on the fixed template. These shapes have different triangulation, vertex density and style, confirming generalization across datasets. More specifically: a shape from TOSCA~\cite{TOSCA} for humans (first row), one from SMAL~\cite{SMAL} for the horses (second row) and a camel shape that has a different triangulation and comes from a different class (third row).

\begin{figure}[t]
\centering
    \begin{overpic}[trim=0cm 0cm -3cm 0cm,clip,width=0.82\linewidth]{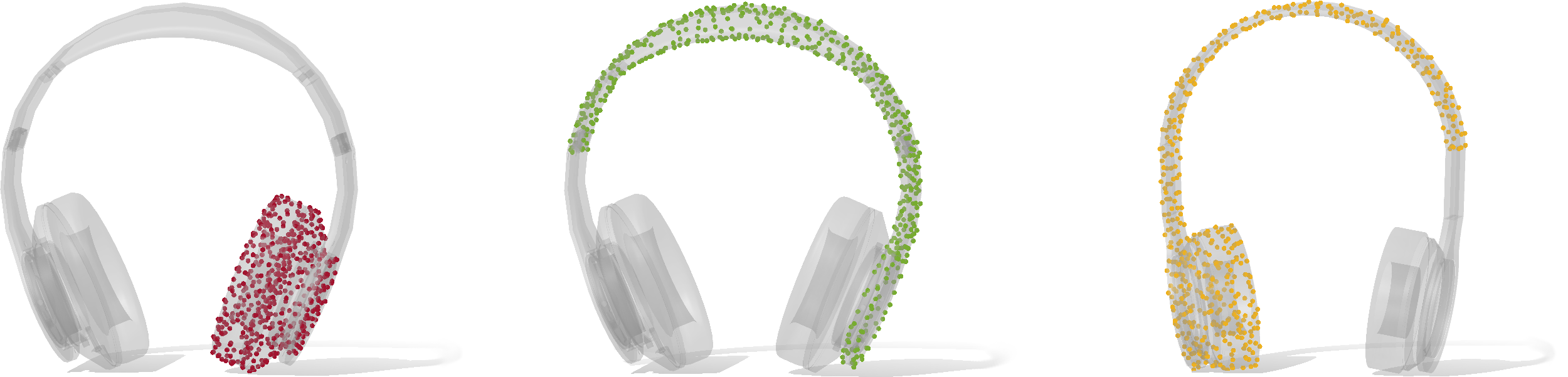}
        \put(25,10){$\cup$}
        \put(60,10){$=$}
    \end{overpic}
    \vspace{0.1cm}
    \begin{overpic}[trim=0cm 0cm 0cm 0cm,clip,width=0.88\linewidth]{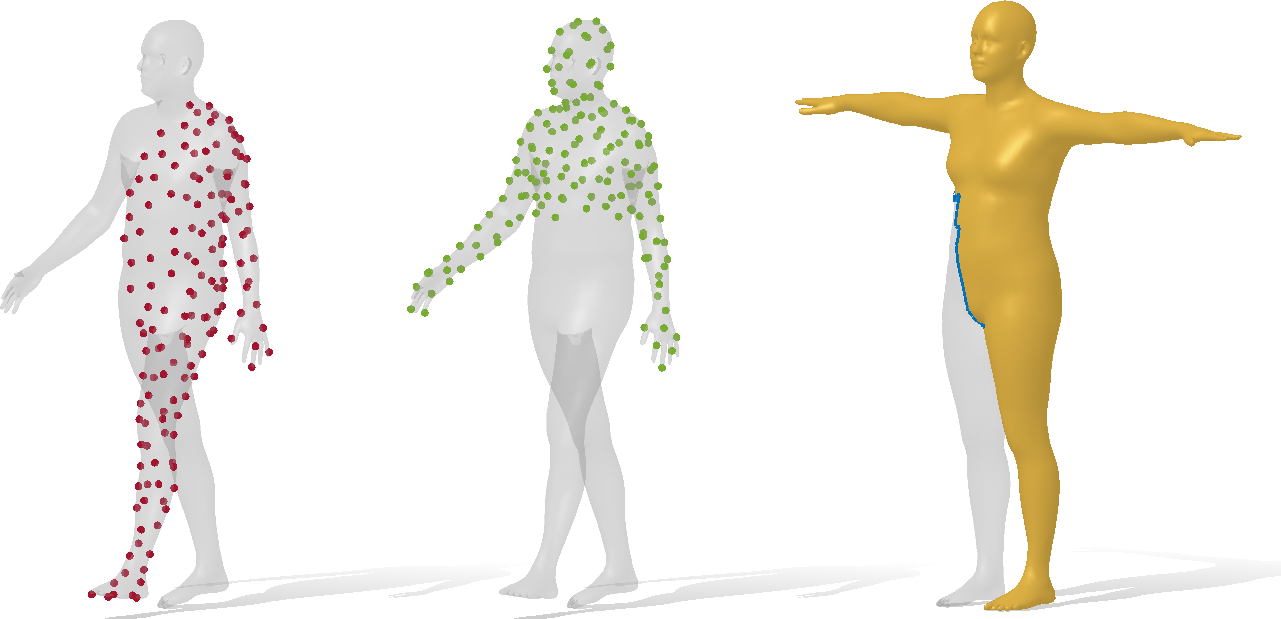}
        \put(27,28){$\cup$}
        \put(59,28){$=$}
    \end{overpic}
    \caption{\label{fig:results:mask_prediction:pc} Region localization from partial point cloud spectra. The white mesh is just shown as a visual reference.}
\end{figure}

\vspace{1ex}\noindent\textbf{Generalization to point clouds.}
We obtain good results also on point clouds, as shown in Figure~\ref{fig:results:mask_prediction:pc}. 
For earphones, in the top row, we perform both training and testing on point clouds.
In the bottom row, we show that our model trained on human meshes generalizes to point clouds.
We compute the Laplacian for point clouds with the method of~\cite{Sharp:2020:LNT}. 

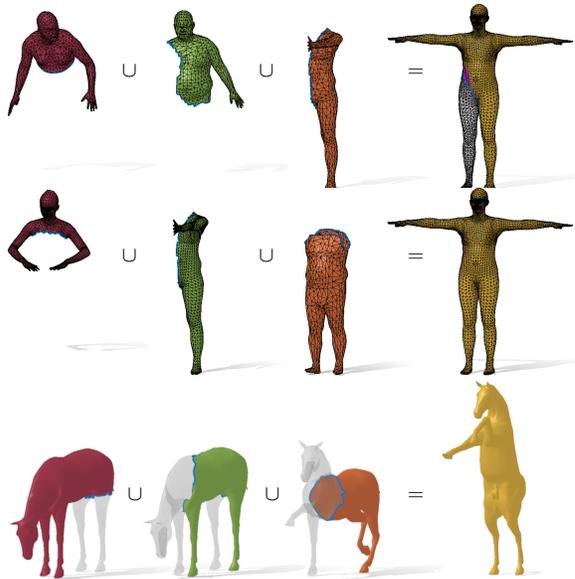
\begin{figure}[t]
\centering
    \begin{overpic}[trim=0cm 0cm 0cm 0cm,clip,width=0.9\linewidth]{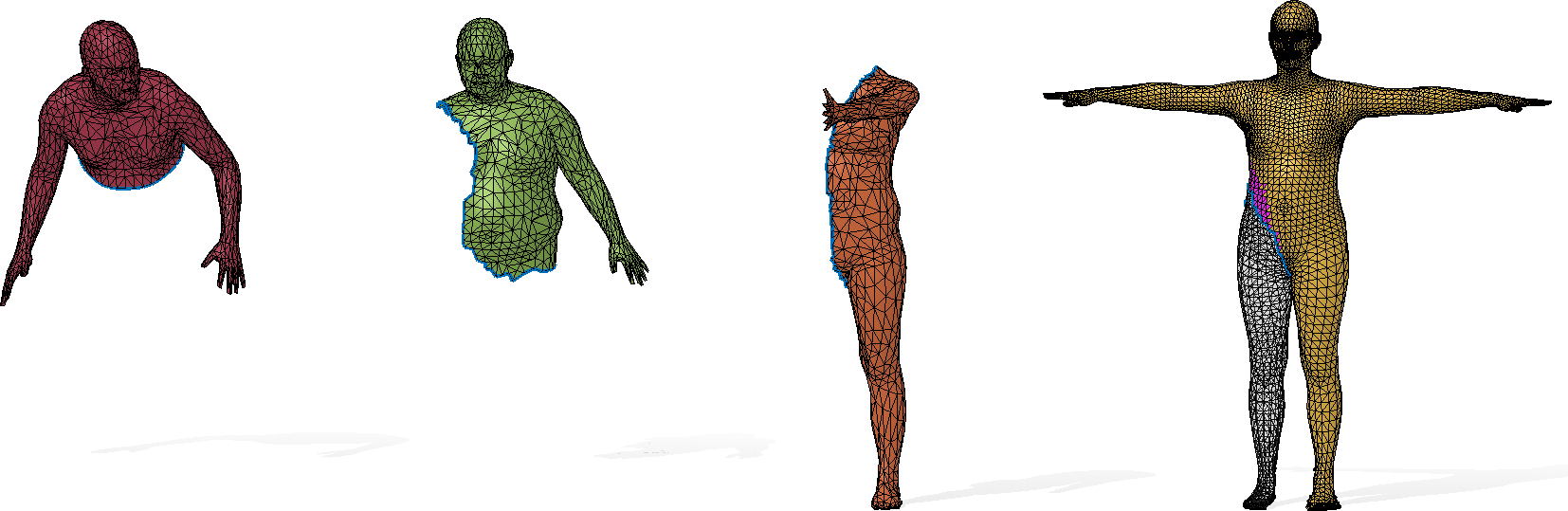}
        \put(20, 20){\footnotesize$\cup$}
        \put(44, 20){\footnotesize$\cup$}
        \put(70, 20){\footnotesize$=$}
    \end{overpic}
    \vspace{0.1cm}
    \begin{overpic}[trim=0cm 0cm 0cm 0cm,clip,width=0.9\linewidth]{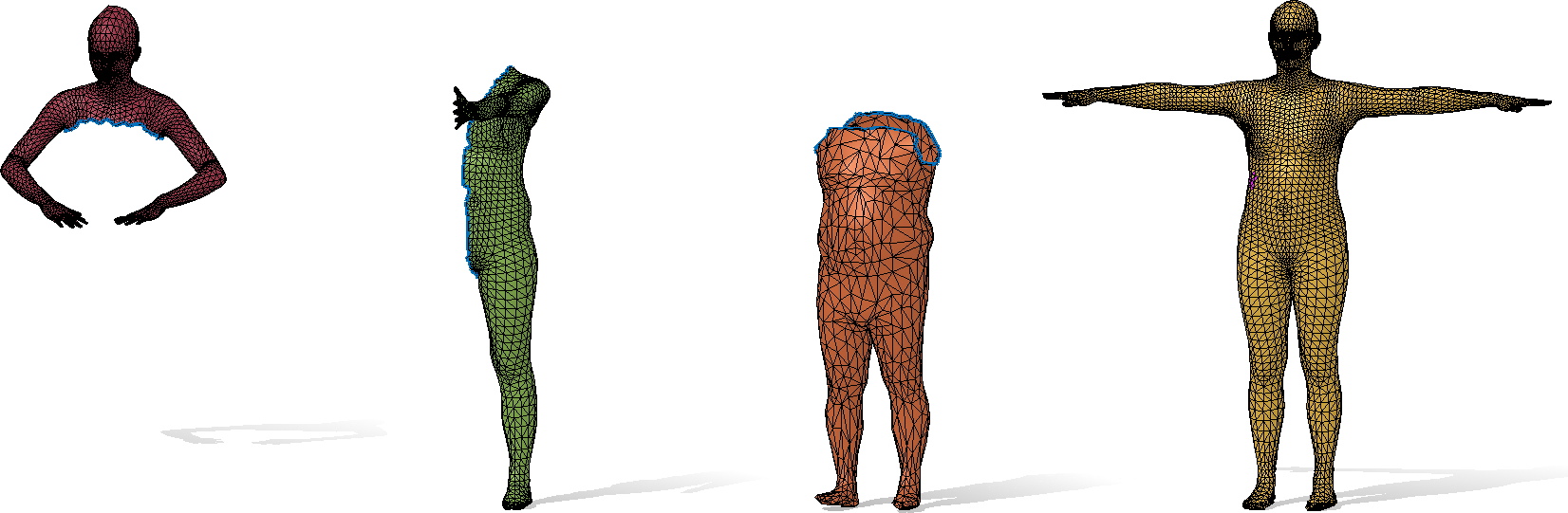}
        \put(20, 20){\footnotesize$\cup$}
        \put(44, 20){\footnotesize$\cup$}
        \put(70, 20){\footnotesize$=$}
    \end{overpic}
    \vspace{0.1cm}
    \begin{overpic}[trim=0cm 0cm 0cm 0cm,clip,width=0.9\linewidth]{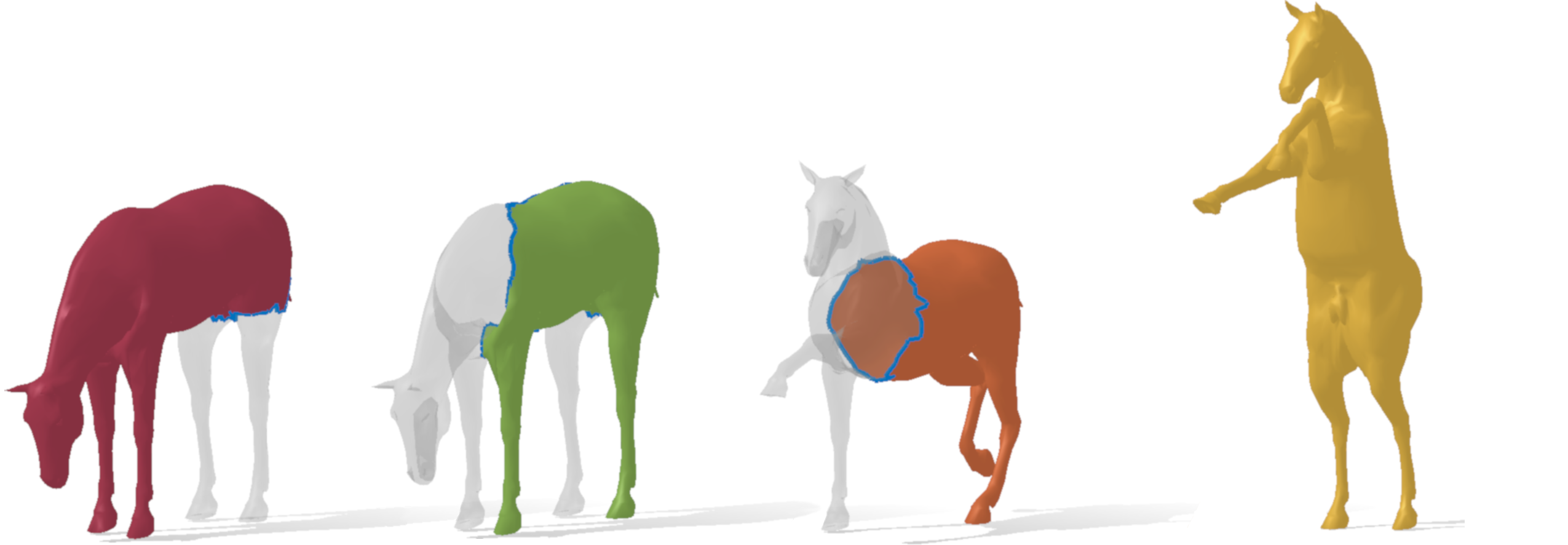}
        \put(21, 14){\footnotesize$\cup$}
        \put(45, 14){\footnotesize$\cup$}
        \put(70, 14){\footnotesize$=$}
    \end{overpic}
    \caption{\label{fig:results:compositionality} Example of associativity. Note that all the human meshes involved have different connectivity.
    }
\end{figure}

\vspace{1ex}\noindent\textbf{Associativity.}
We can compute spectral unions of $>2$ partial shapes {iteratively} 
as described in Section~\ref{sec:method}. 
%
In Figure~\ref{fig:results:compositionality} we show qualitative results over three parts. 

\begin{figure}[t!]
\centering
    \begin{overpic}[trim=0cm 0cm 0cm 0cm,clip,width=\linewidth]{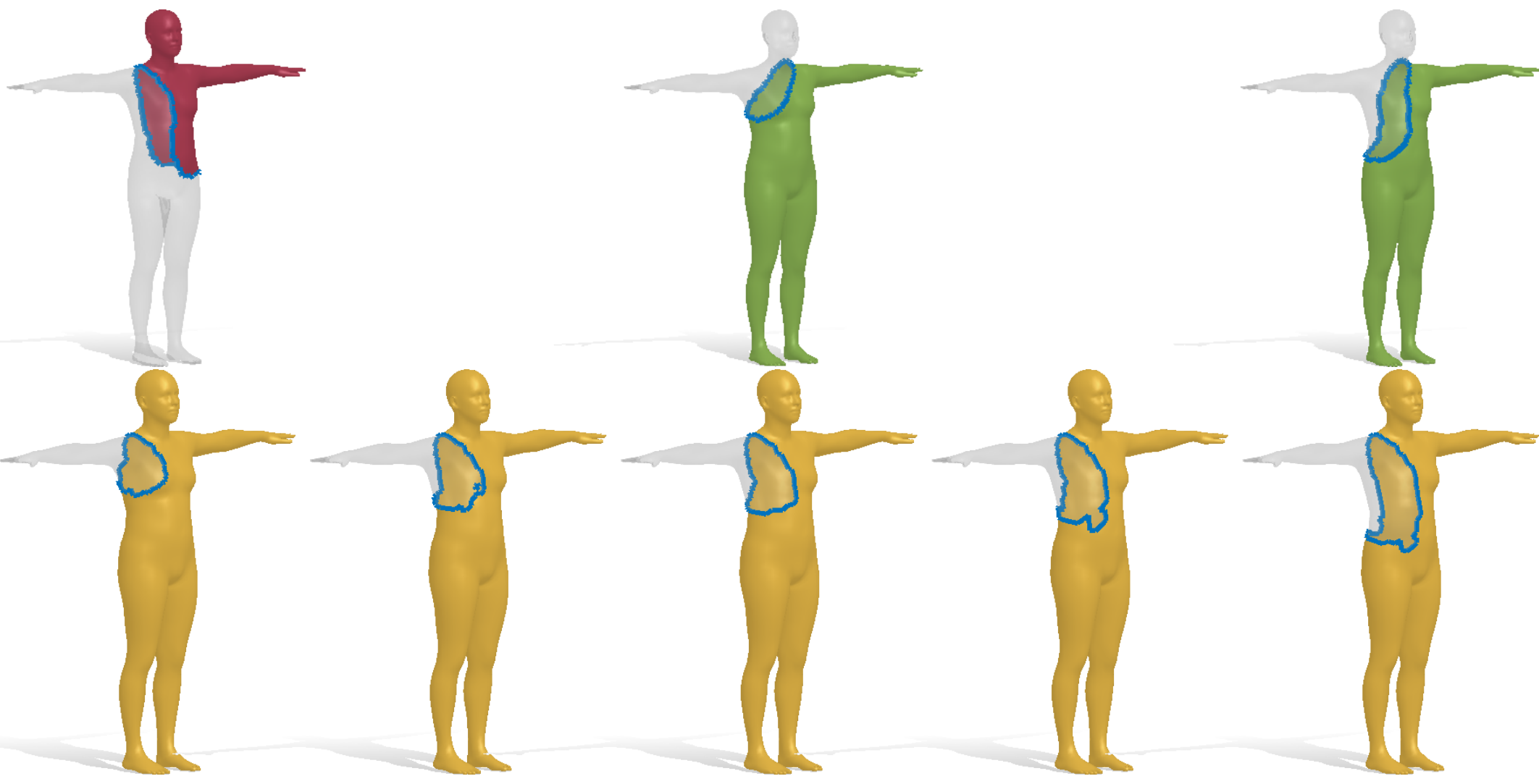}
        \put(28,40){$\cup$}
        \put(68,40){$\dots$}
    \end{overpic}\\[2ex]
    \begin{overpic}[trim=0cm 0cm 0cm 0cm,clip,width=\linewidth]{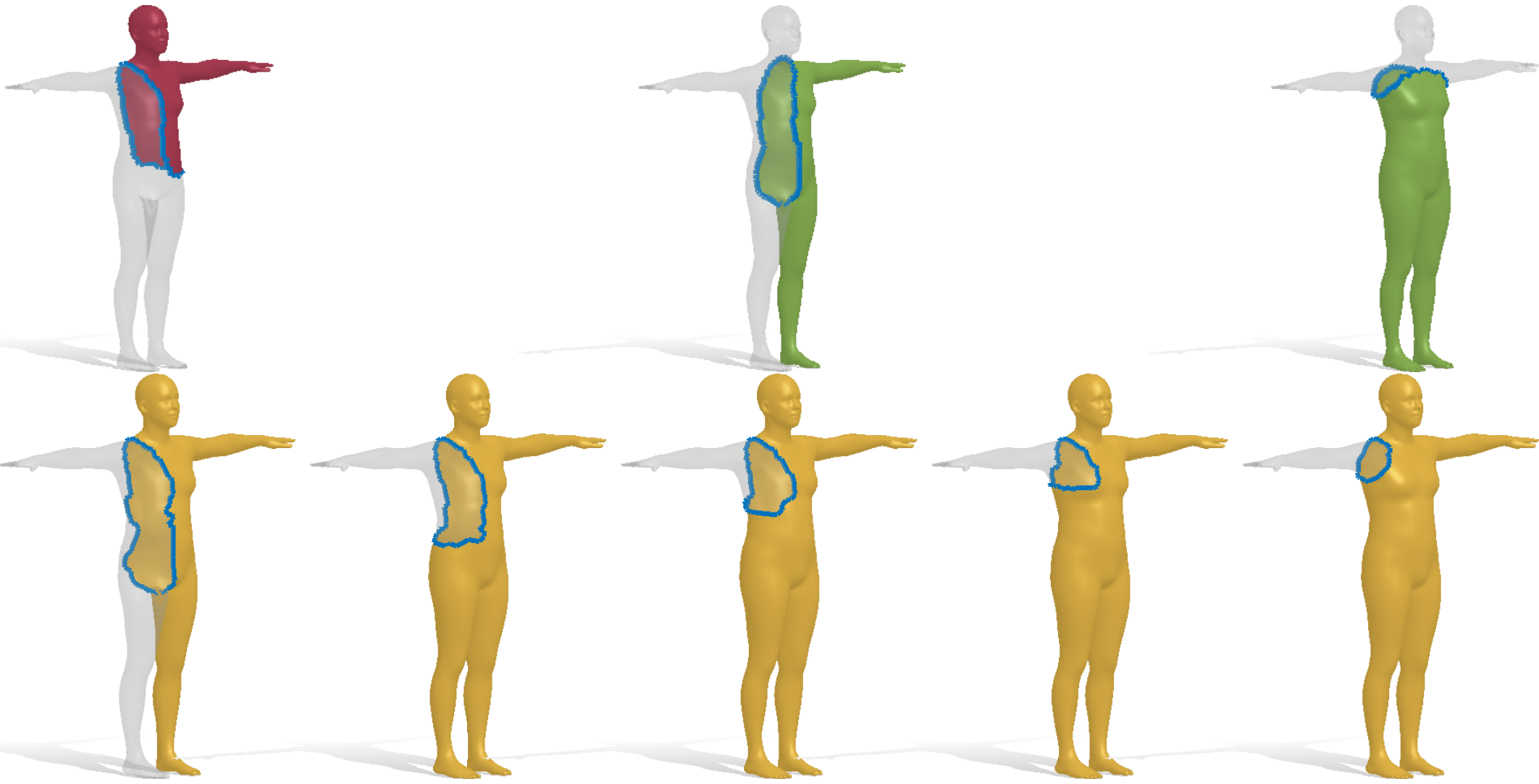}
        \put(28,40){$\cup$}
        \put(68,40){$\dots$}
    \end{overpic}
    \caption{\label{fig:interpolation} 
    Two examples of linear interpolation of eigenvalues (green shapes), and the resulting predicted masks (in yellow). Please refer to the main text for details.}
\end{figure}

\vspace{1ex}\noindent\textbf{Interpolation.}
Finally, in Figure~\ref{fig:interpolation} we first interpolate the spectra of two partial shapes (in green), and then compute the union of the interpolated spectra with the spectrum of a fixed shape (in red). From each of these unions, we predict a mask on the given template (in yellow). We can see how in the first example (top row) the mask changes smoothly. On the other hand, in the second example it is less obvious how to interpolate the completely missing leg, resulting in an abrupt discontinuity in the predicted mask. 

\begin{table}[b!]
    \centering
    \begin{tabular}{ c|c|c|c } 
                & top-1     & top-5 & top-10 \\ 
        \hline
        Ours       & $86.14\%$ &  $\mathbf{97.75\%}$     &  $\mathbf{99.20\%}$      \\ 
        ShapeDNA & $\mathbf{86.59\%}$ &  $96.81\%$     &  $97.72\%$
    \end{tabular}
    \caption{Comparisons on the shape retrieval task.
    }
    \label{table:identity_retrieval}
\end{table}

\subsection{Shape retrieval}\label{sec:identity_retrieval}
This task consists in retrieving a query from a database of shapes that could undergo several deformations.
A well-known spectral method to tackle this problem, ShapeDNA~\cite{shapeDNA2006}, adopts the Laplacian spectrum as a shape signature. In the space of these signatures, nearest-neighbor search yields the desired result. However, in order to work correctly, ShapeDNA needs the spectrum of a complete shape; extensions of this signature to the partial case have proven unsuccessful to date~\cite{rodolashrec}.
Our method applies directly to this case, since we can estimate the eigenvalues of the unknown complete shape whenever the input query is just a collection of its partial views.

We run our tests on a dataset of 440 complete shapes (44 identities in 10 poses each). For our method, we evaluate 4400 pairs of partial shapes; for each pair we predict {the ShapeDNA signature of their union} 
and use it to query the database.  {We compare it with the accuracy obtained by standard ShapeDNA on each of the 440 complete shapes in the database, which assumes exact knowledge of the union spectra; nevertheless, the identity it retrieves may be wrong due to spectra variations caused by deformations between different poses.}
We measure the performance using top-$k$ metrics, which count the number of times a shape with the correct identity is in the first $k$ retrieved shapes; we use $k=1,5,10$. The results are reported in Table~\ref{table:identity_retrieval}, and show that our predicted eigenvalues are accurate enough to compete with, and even surpass, ShapeDNA for this task. 
The better performance is due to the robustness of our method to the noise induced by the pose change.













\section{Conclusion}
\label{sec:conclusion}
We introduced a method to recover the aggregated Laplacian spectrum of a collection of partial deformable shapes, while avoiding the computational burden of computing correspondences or extrinsic alignments. Our method involves a deep net that, given two eigenvalue sequences as input, simply produces another eigenvalue sequence as output. In spite of its apparent simplicity, this method allows to address a number of applications that traditionally require solving for a correspondence, and retains a comparable quality (in some cases, even higher) to methods that have direct access to the 3D geometry of the full shape.

\vspace{1ex}\noindent\textbf{{Limitations and future directions.}}
Perhaps the main limitation of our method lies in the missing mathematical guarantee that our predicted sequences are actual Laplacian eigenvalues, despite our positive empirical results. 
We consider enforcing this constraint as an interesting direction of further research.
{Another interesting area for improvement is the region localization generalization capability, where our current model
seems to struggle with out-of-distribution union partialities. We are optimistic that a more diverse and extensive training set would boost the generalization performance.
Moreover, we did not consider unprocessed partial single-view or depth scans of physical objects. We expect a drop in performance on such data comparable to other spectral methods. We consider improving the robustness of spectral methods on natural non-pre-processed data as an essential and challenging research direction.
}
%
%
%


\paragraph*{Acknowledgements}
Parts of this work were supported by the ERC Starting Grants No. 802554 (SPECGEO) and No. 758800 (EXPROTEA), the ANR AI Chair AIGRETTE, the SAPIENZA BE-FOR-ERC 2020 Grant (NONLINFMAPS), and the GALILEO 2022 Fellowship $G22\_4$.

\printbibliography                


\clearpage

\appendix
\section{Supplementary Material}

We report additional results that were not included in the main manuscript.
Further, we describe with more details the architectures involved in the proposed method. 



%
    
\subsection{Additional results}\label{sec:results}
In this section, we collect additional results for the experiments and applications described in the main manuscript.

\begin{figure}[b]
\centering
    \begin{overpic}[trim=0cm 0cm 0cm -0.0cm,clip,width=0.99\linewidth]{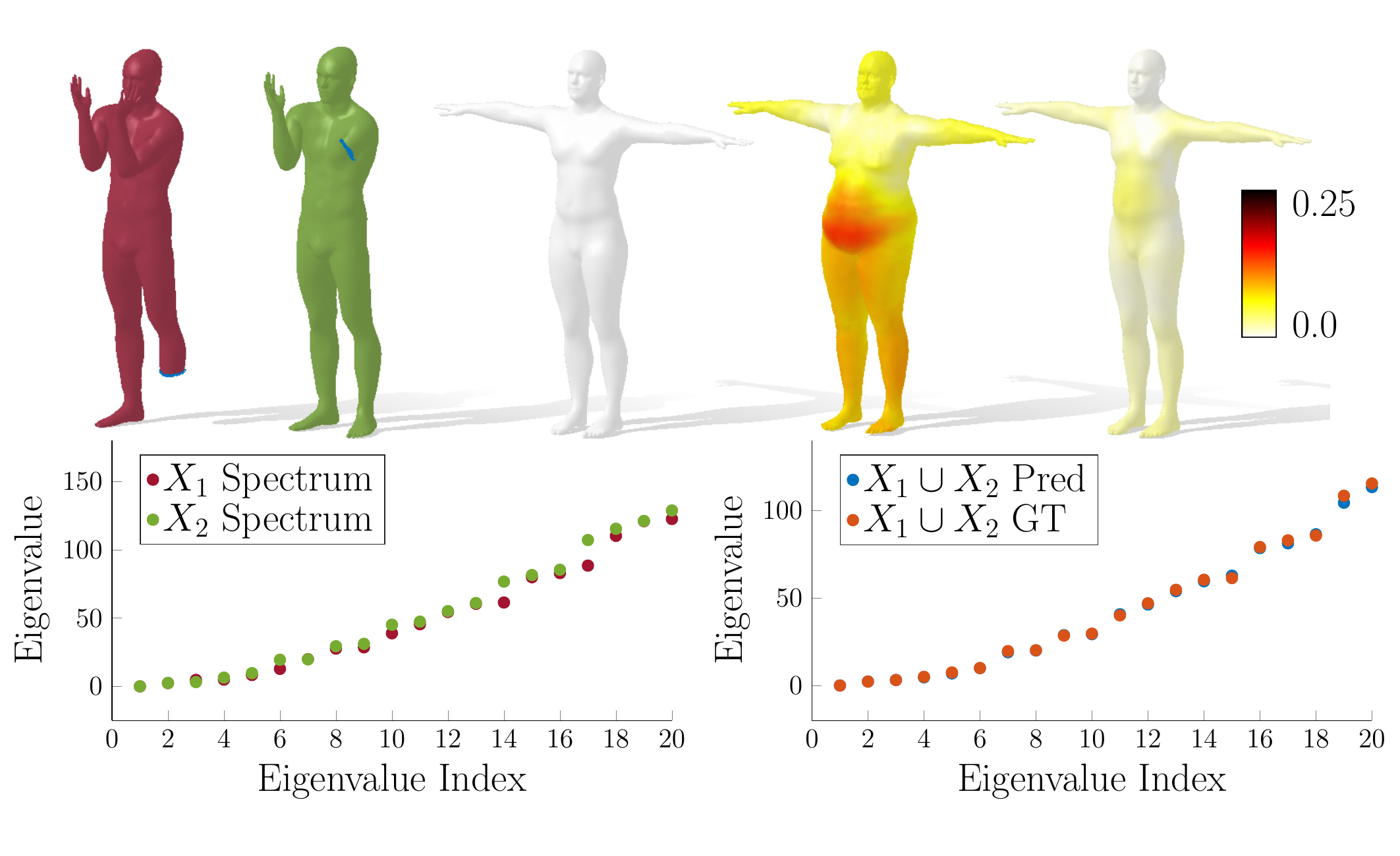}
        \put(8,60){\footnotesize $\M_1$ }
        \put(17,60){$\cup$ }
        \put(23,60){\footnotesize $\M_2$ }
        \put(40,60){\footnotesize GT}
        \put(56,60){\footnotesize GT($\M_2$)}
        \put(78,60){\footnotesize Ours}
    \end{overpic}\\
    \begin{overpic}[trim=0cm 0.7cm 0cm -0.3cm,clip,width=0.99\linewidth]{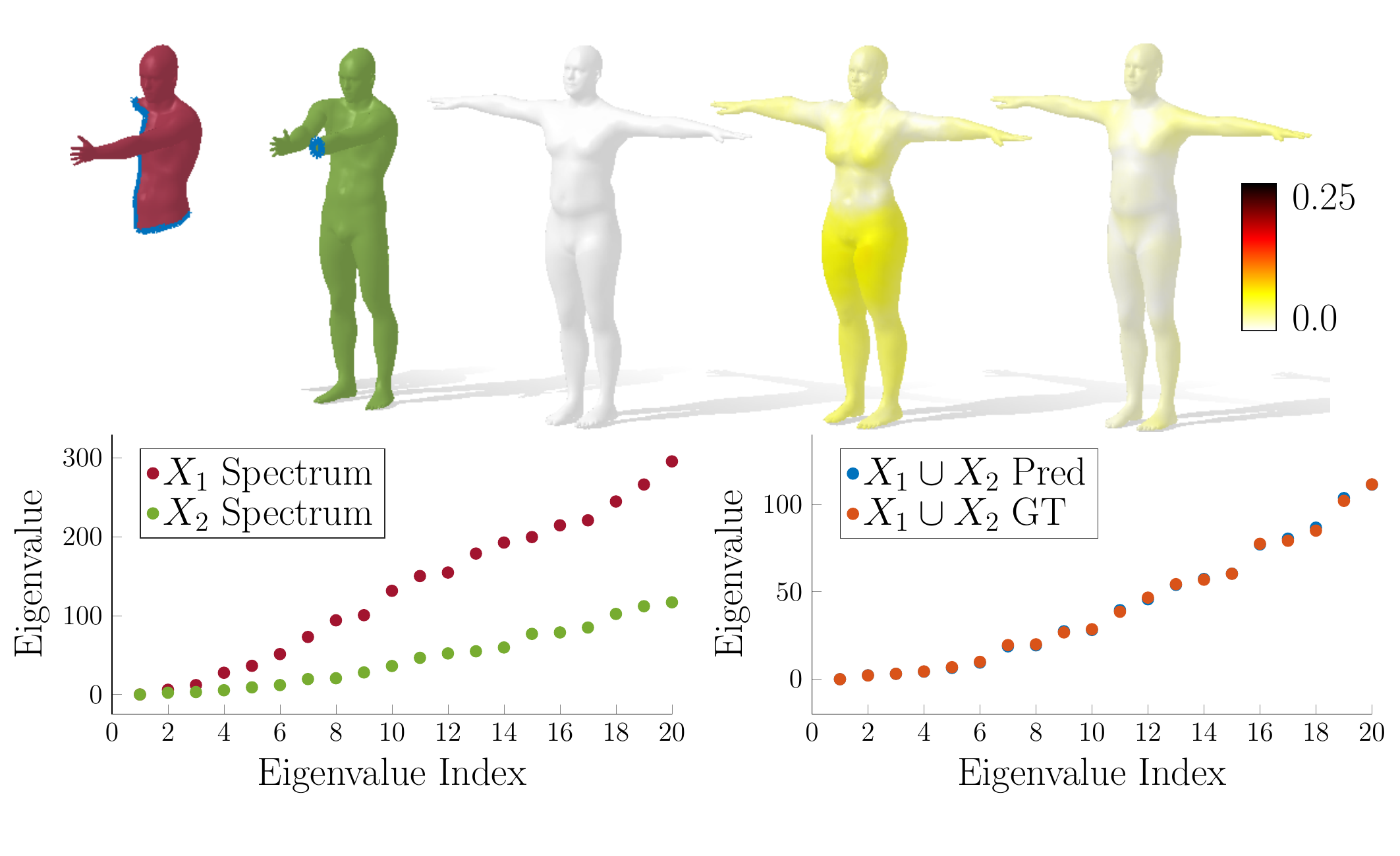}
        \put(8,57){\footnotesize $\M_1$ }
        \put(17,57){$\cup$ }
        \put(23,57){\footnotesize $\M_2$ }
        \put(40,57){\footnotesize GT}
        \put(56,57){\footnotesize GT($\M_2$)}
        \put(78,57){\footnotesize Ours}
    \end{overpic}
    \caption{\label{fig:instant}  
    Comparison of the reconstruction obtained by running the state-of-the-art method of~\cite{Instant2020} on the green shape, yielding the fourth shape, and the reconstruction obtained from our predicted full spectrum, yielding the last shape.}
\end{figure}
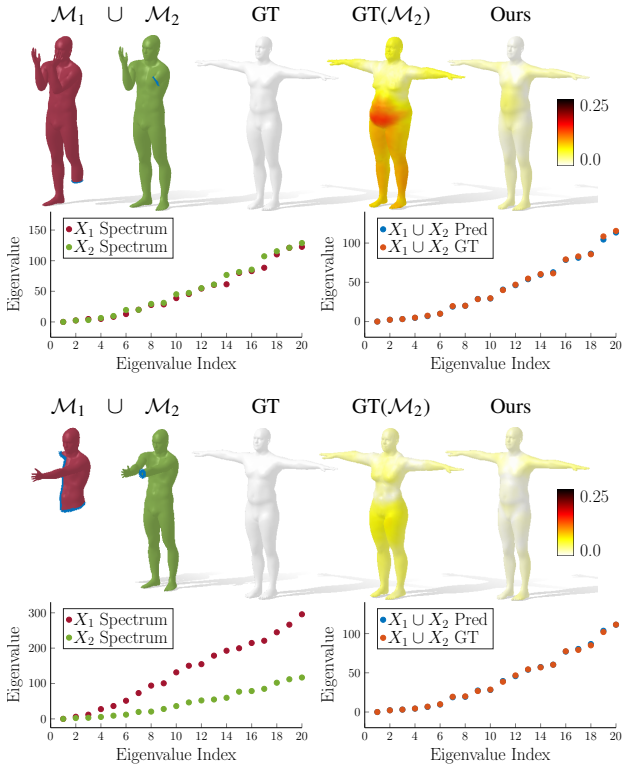

\paragraph*{Additional results on geometry reconstruction}
In Fig.~\ref{fig:instant}, we report additional examples of shape-from-spectrum recovery. These qualitative results confirm that we outperform \cite{Instant2020} when applied to partial shapes.

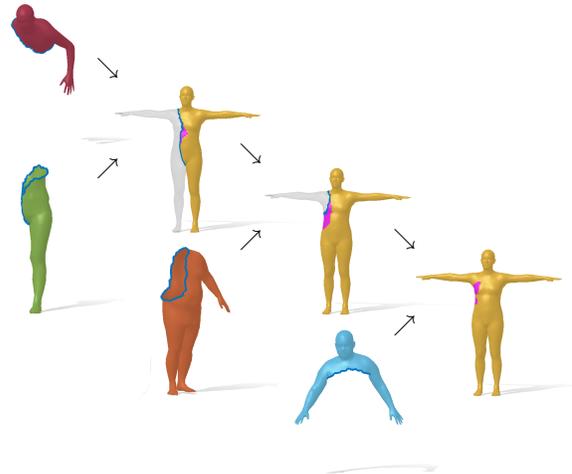
\begin{figure}[b]
\centering
    \begin{overpic}[trim=0cm 0cm 0cm 0cm,clip,width=0.9\linewidth]{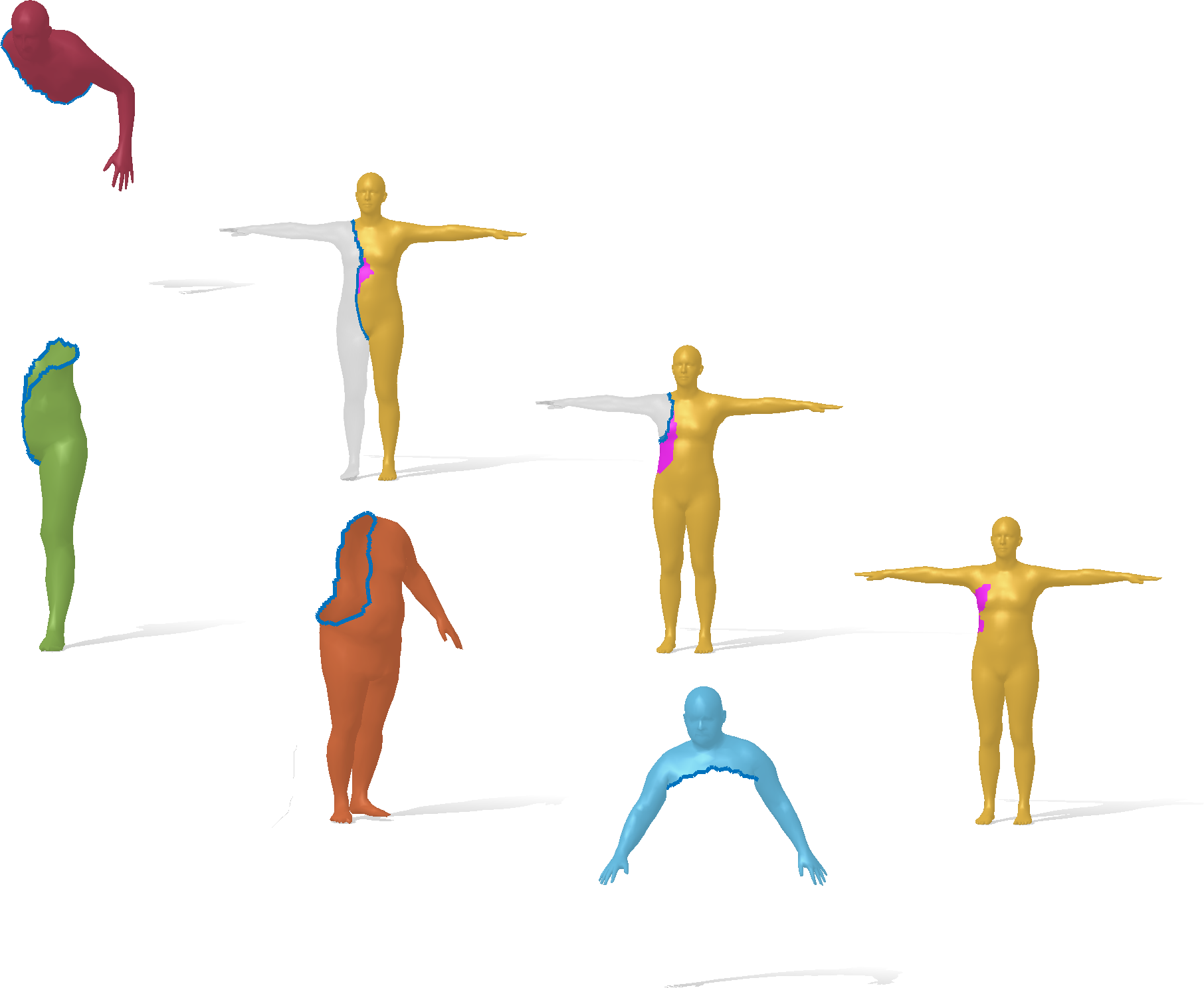}
        \put(15,70){\footnotesize $\searrow$}
        \put(15,52.5){\footnotesize $\nearrow$}

        \put(40,55){\footnotesize $\searrow$}
        \put(40,40){\footnotesize $\nearrow$}
        
        \put(67,40){\footnotesize $\searrow$}
        \put(67,25){\footnotesize $\nearrow$}
        
    \end{overpic}
    \caption{\label{fig:comp4} Example of associativity with 4 partial shapes.}
\end{figure}

\paragraph*{Associativity with more than 3 partial shapes}
In Fig.~\ref{fig:comp4} we show an example of iterative spectral union, with four different partial shapes.
Deeper iterative unions are more difficult since the prediction error in each step is amplified by the subsequent steps.

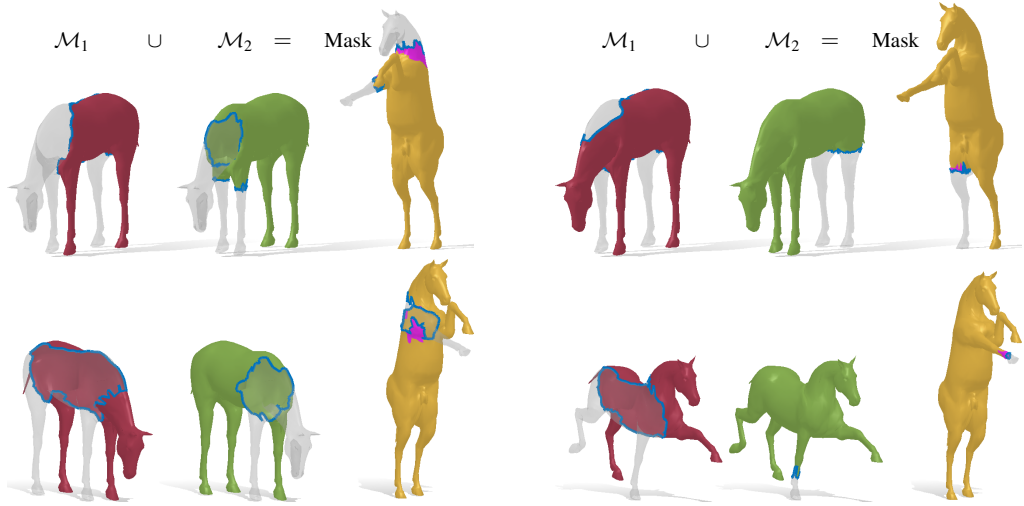
\begin{figure*}[t!]
\centering
    \begin{overpic}[trim=0cm 0cm 0cm 0cm,clip,width=0.35\linewidth]{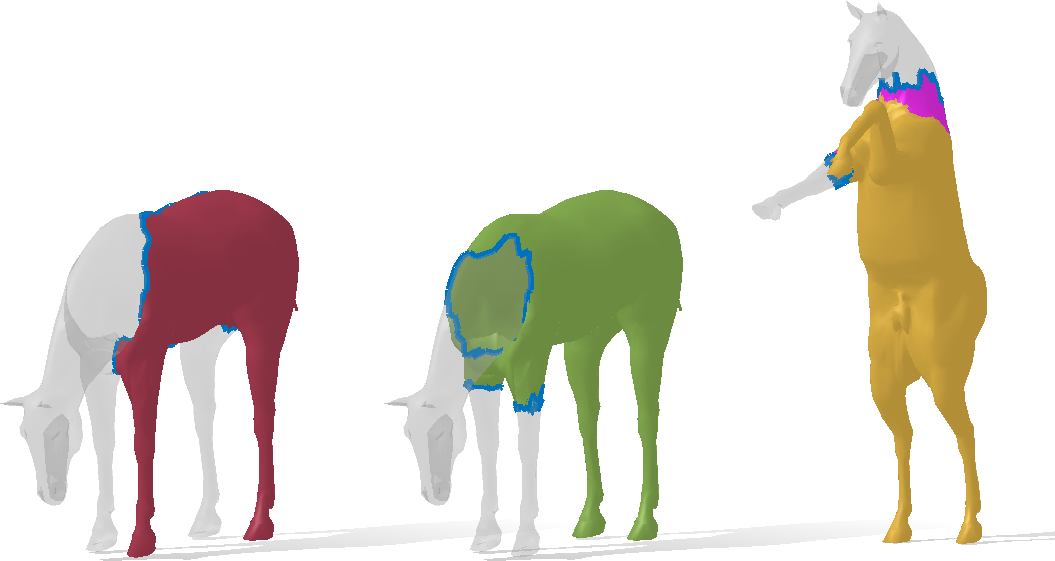}
        \put(10,45){\footnotesize $\M_1$ }
        \put(30,45){\footnotesize $\cup$ }
        \put(45,45){\footnotesize $\M_2$ }
        \put(57,45){$=$}
        \put(68,45){\footnotesize Mask}
    \end{overpic}\hspace{1cm}
    \begin{overpic}[trim=0cm 0cm 0cm 0cm,clip,width=0.35\linewidth]{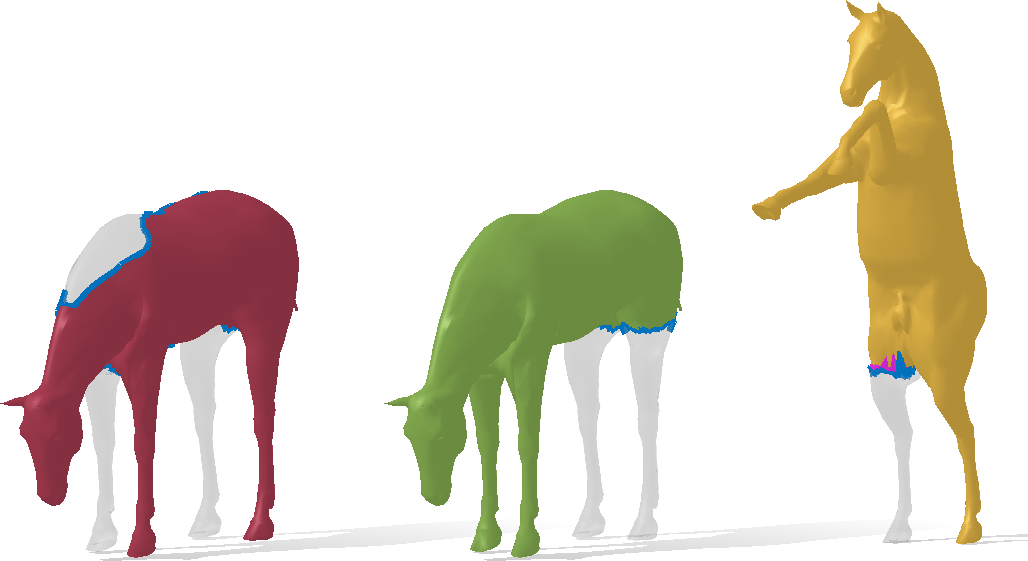}
        \put(10,45){\footnotesize $\M_1$ }
        \put(30,45){\footnotesize $\cup$ }
        \put(45,45){\footnotesize $\M_2$ }
        \put(57,45){$=$}
        \put(68,45){\footnotesize Mask}
    \end{overpic}
    \includegraphics[width=0.35\linewidth]{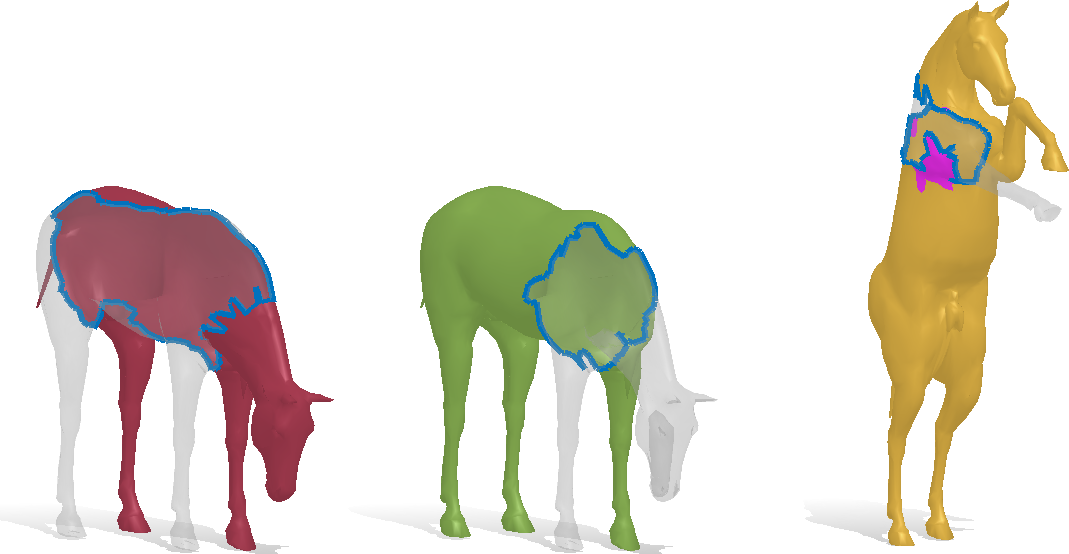}\hspace{1cm}
    \includegraphics[width=0.35\linewidth]{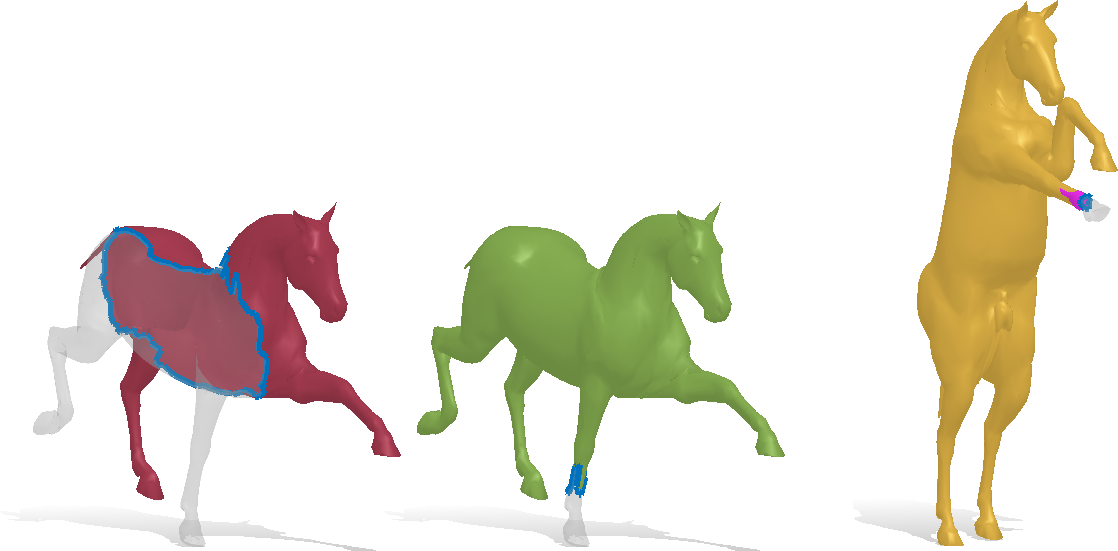}
    \caption{\label{fig:horse_with_gt} 
    Region localization on horses. We predict the indicator function that describes the union of two partial shapes, given their eigenvalues.}
\end{figure*}

\paragraph*{Additional results on the {\em horse} class}
We report additional results on the horse class. 
In these experiments, the spectral union operator is pre-trained on humans and fine-tuned to horses from the TOSCA dataset. Then, a region localization model is trained specifically for horses, slightly modified to account for the different number of vertices in the template.

\begin{figure}[b]
\centering
    \begin{overpic}[trim=0cm 0cm 0cm 0cm,clip,width=0.75\linewidth]{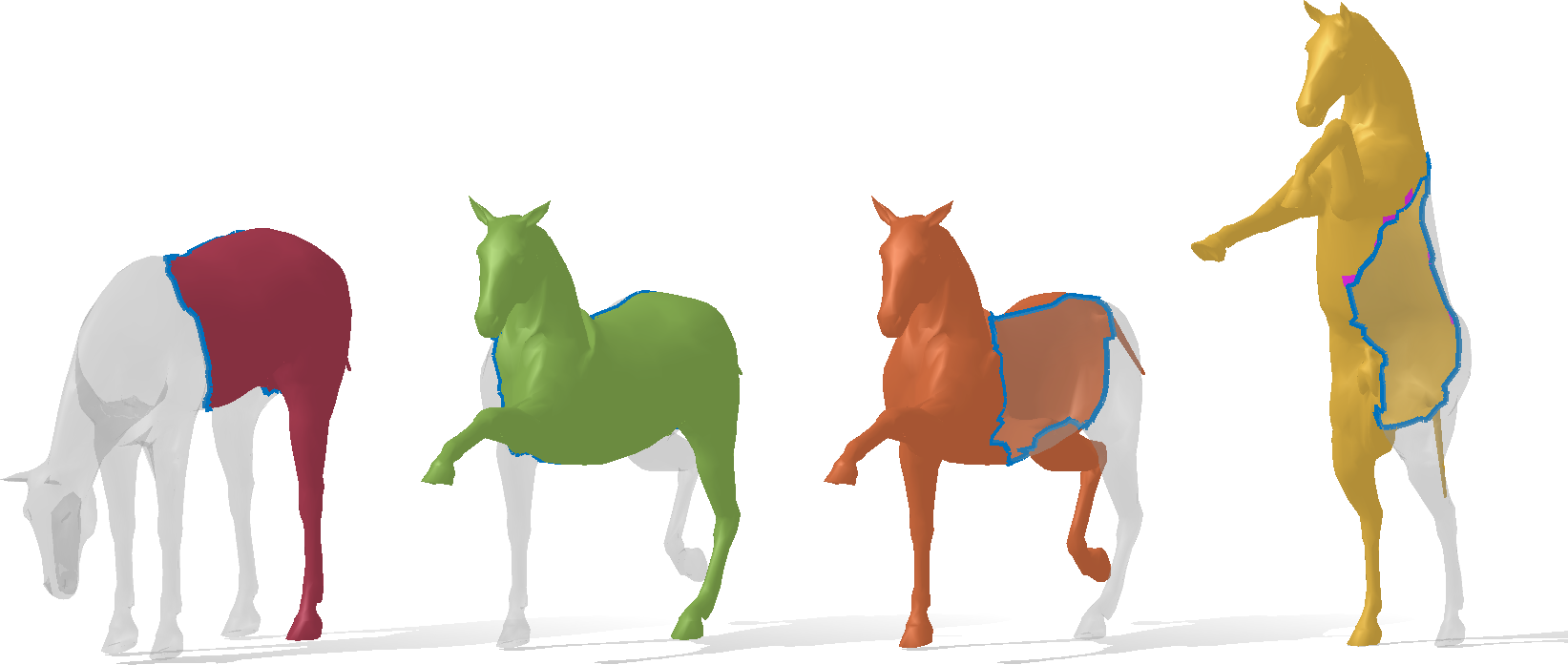}
        \put(5, 40){\footnotesize$\M_1$}
        \put(19, 40){\footnotesize$\cup$}
        \put(29, 40){\footnotesize$\M_2$}
        \put(44, 40){\footnotesize$\cup$}
        \put(55, 40){\footnotesize$\M_3$}
        \put(65, 40){\footnotesize$=$}
        \put(71, 40){\footnotesize Mask}
    \end{overpic}
    \includegraphics[width=0.75\linewidth]{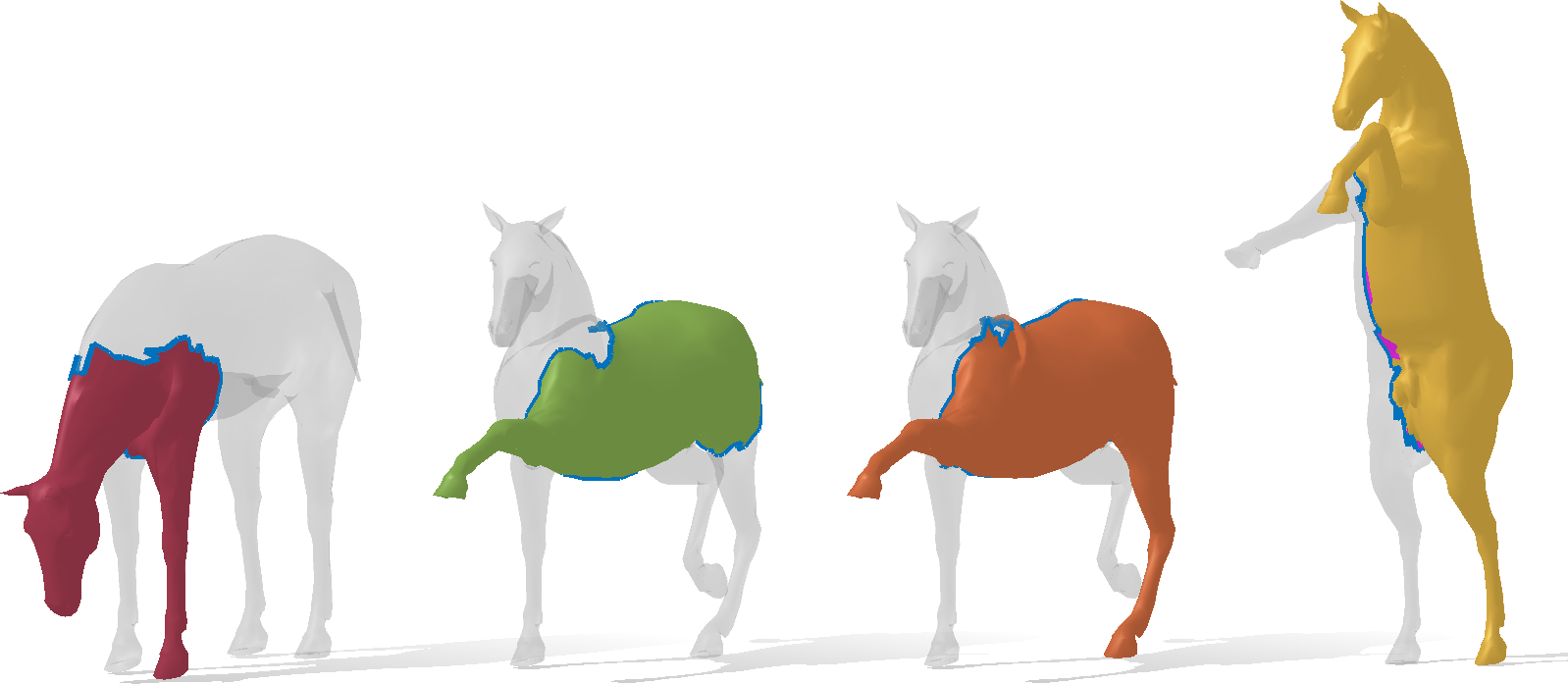}
    \caption{\label{fig:horse_composition} 
    Example of associativity.}
\end{figure}

\begin{figure}[b!]
\centering
    \begin{overpic}[trim=0cm 0cm 0cm 0cm,clip,width=0.65\linewidth]{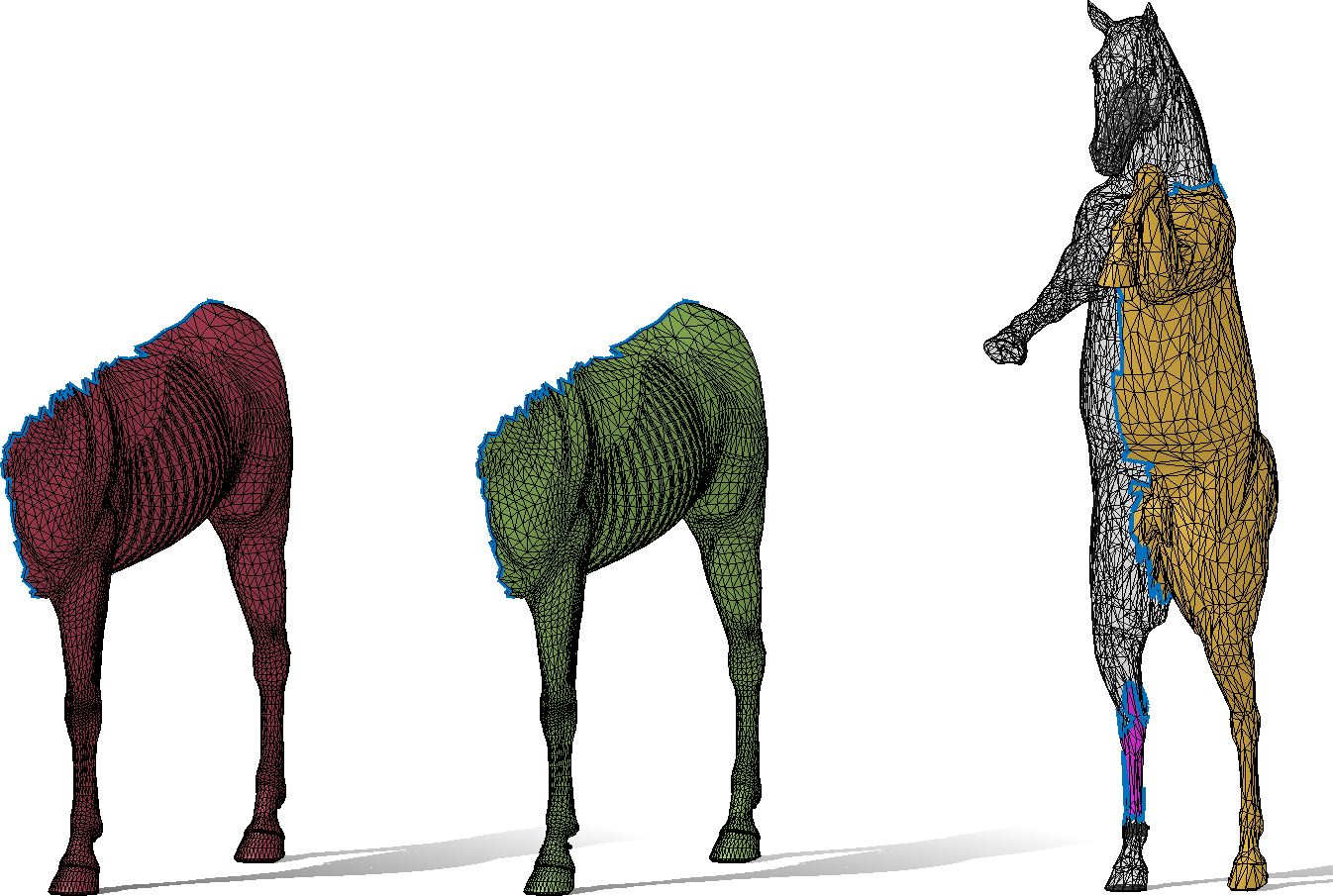}
        \put(5,60){\footnotesize $\M_1$ }
        \put(26,60){\footnotesize $\cup$ }
        \put(42,60){\footnotesize $\M_2$ }
        \put(57,60){$=$}
        \put(68,60){\footnotesize Mask}
    \end{overpic}
    \includegraphics[width=0.65\linewidth]{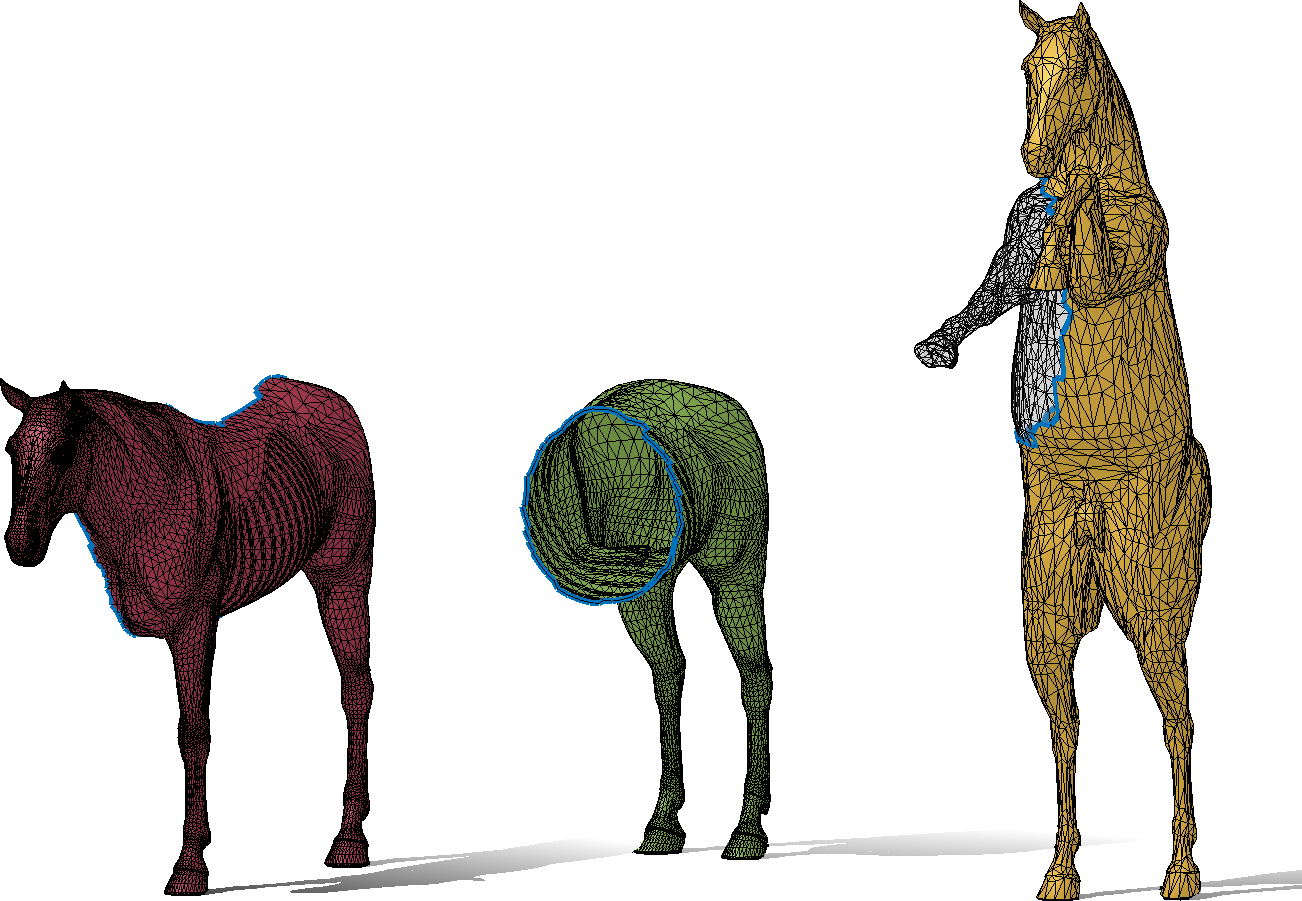}
    \caption{\label{fig:other_horses} 
    Region localization on different horses. The partial shapes have a different triangulation, vertex density and style.}
\end{figure}

\begin{figure*}[h]
\centering
    \begin{overpic}[trim=0cm 0cm 0cm 0cm,clip,width=0.305\linewidth]{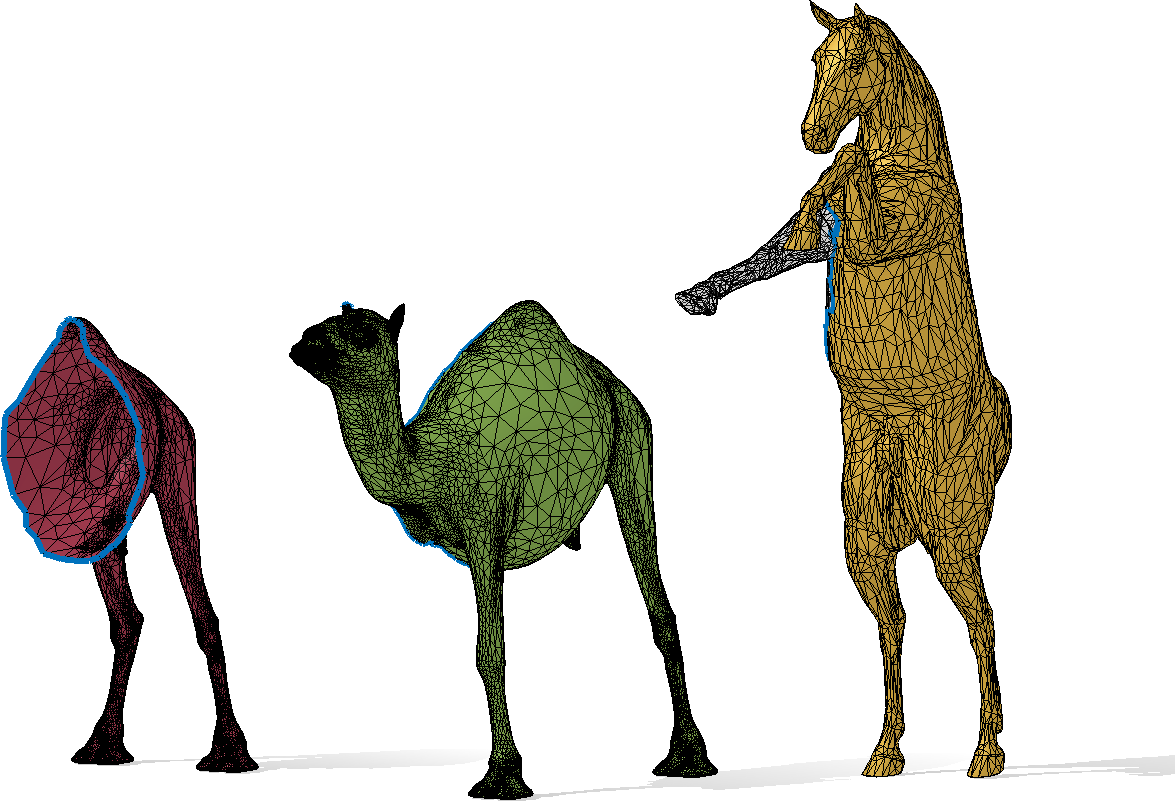}
        \put(3,60){\footnotesize $\M_1$ }
        \put(20,60){\footnotesize $\cup$ }
        \put(32,60){\footnotesize $\M_2$ }
        \put(45,60){$=$}
        \put(55,60){\footnotesize Mask}
    \end{overpic}\hfill
    \begin{overpic}[trim=0cm 0cm 0cm 0cm,clip,width=0.305\linewidth]{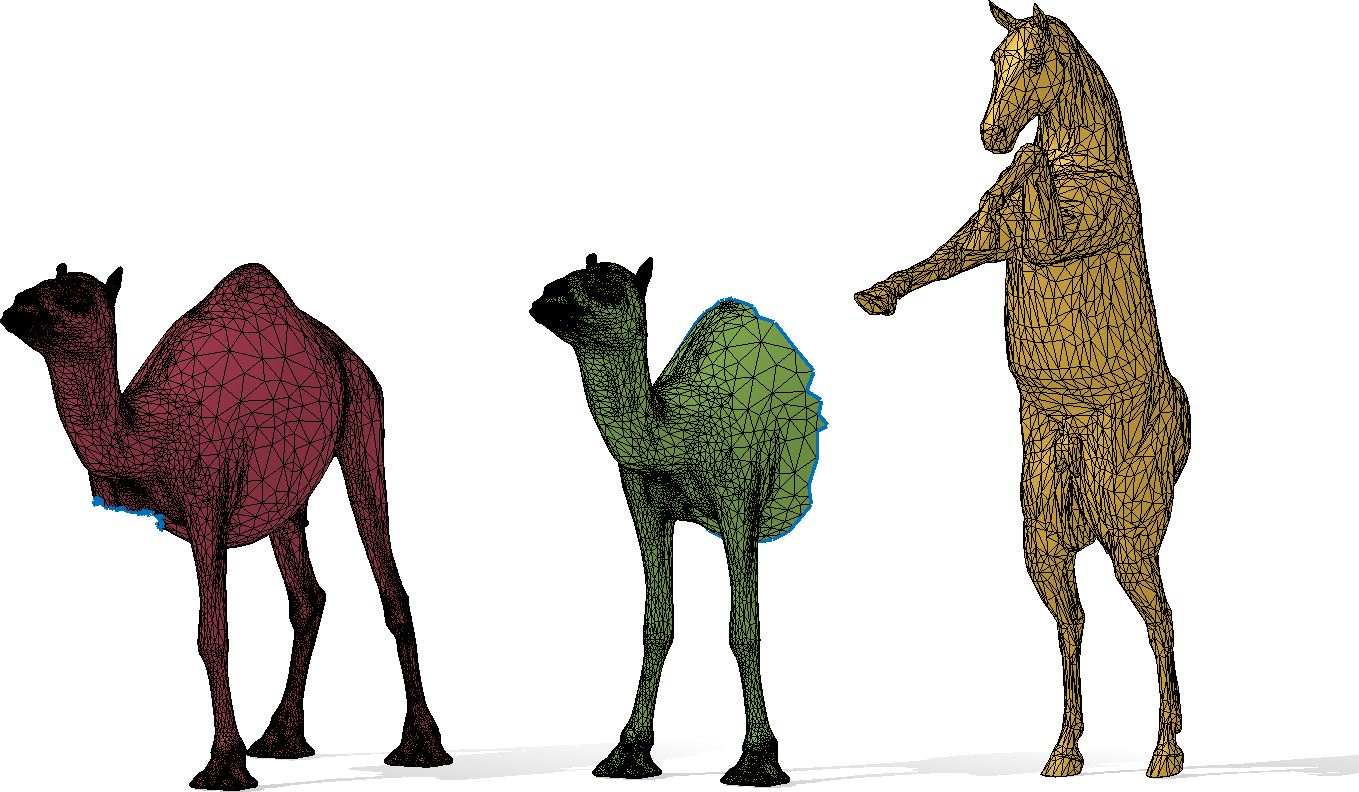}
        \put(3,60){\footnotesize $\M_1$ }
        \put(20,60){\footnotesize $\cup$ }
        \put(32,60){\footnotesize $\M_2$ }
        \put(45,60){$=$}
        \put(55,60){\footnotesize Mask}
    \end{overpic}\hfill
    \begin{overpic}[trim=0cm 0cm 0cm 0cm,clip,width=0.305\linewidth]{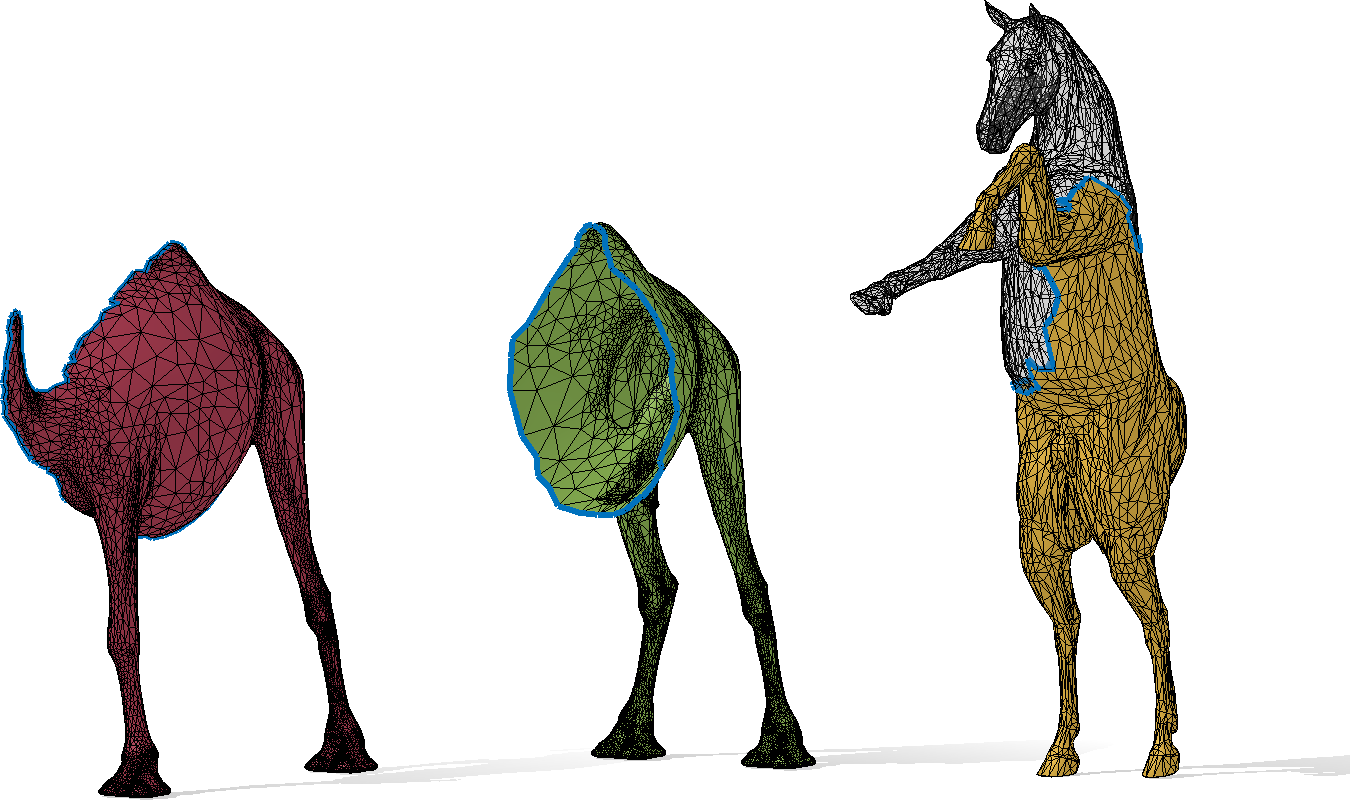}
        \put(3,60){\footnotesize $\M_1$ }
        \put(20,60){\footnotesize $\cup$ }
        \put(32,60){\footnotesize $\M_2$ }
        \put(45,60){$=$}
        \put(55,60){\footnotesize Mask}
    \end{overpic}\\
    \includegraphics[width=0.305\linewidth]{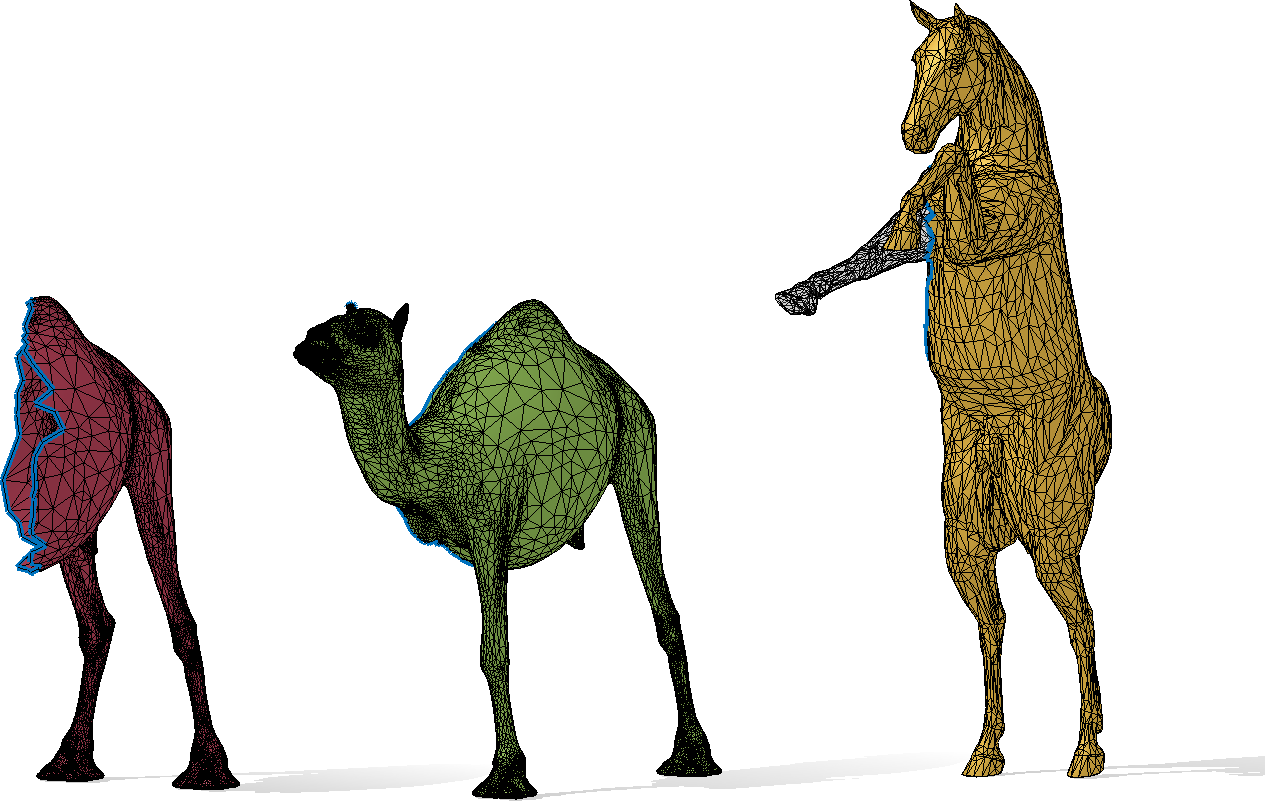}\hfill
    \includegraphics[width=0.305\linewidth]{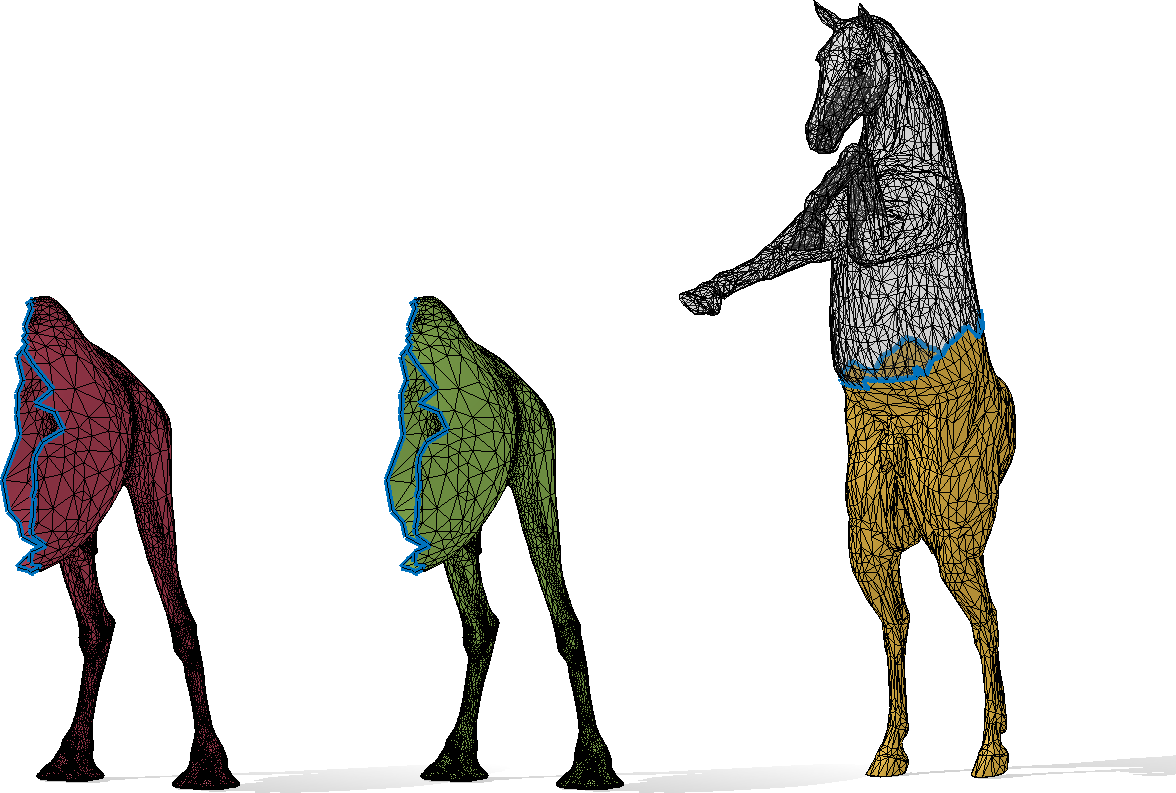}\hfill
    \includegraphics[width=0.305\linewidth]{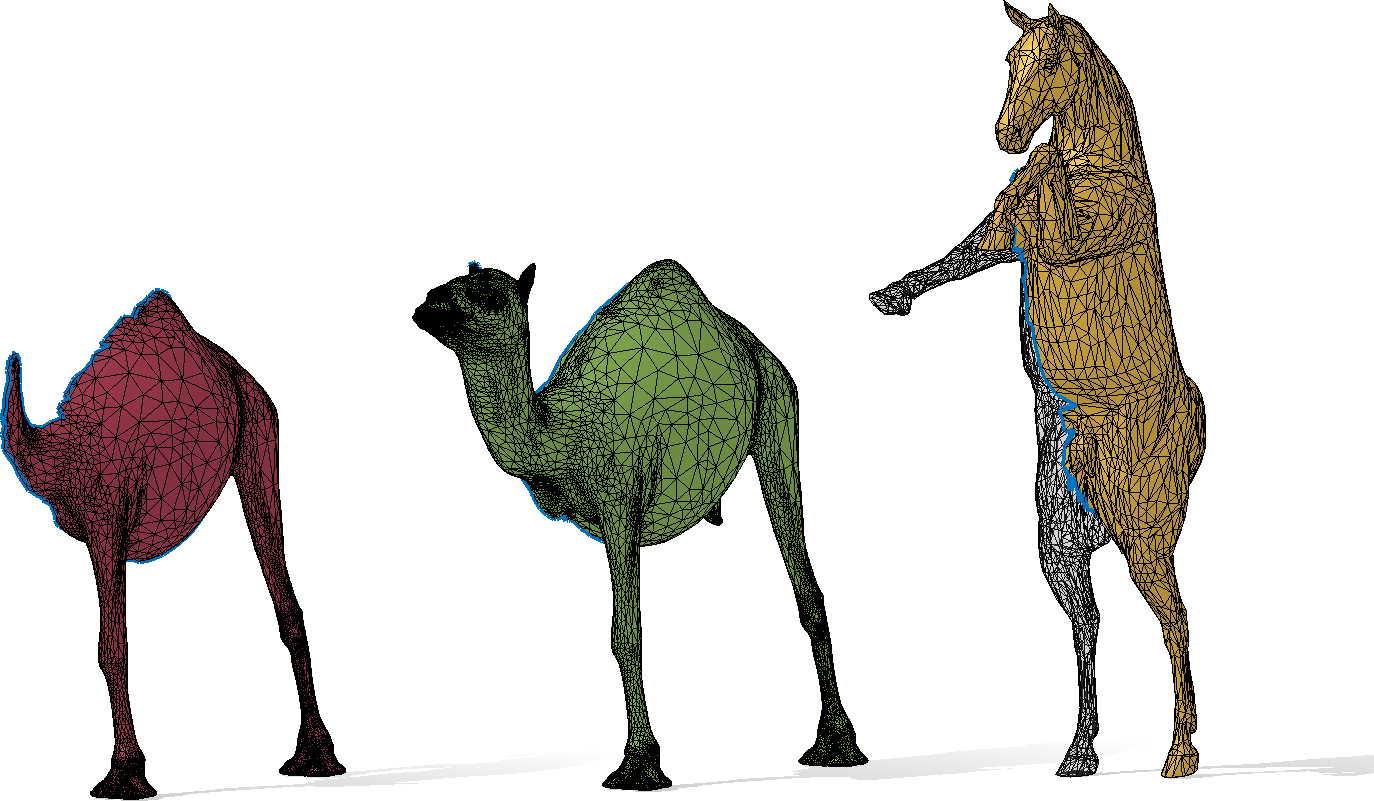}
    \caption{\label{fig:camel} 
    Region localization on a camel, the model is trained on horses.}
\end{figure*}

In Figure~\ref{fig:horse_with_gt} we report qualitative results of region localization on horses;
in Figure~\ref{fig:horse_composition} we show associativity examples;
in Figure~\ref{fig:other_horses} we present qualitative results on horses with different triangulation, vertex density and style with respect to the horses used in the training phase;
in Figure~\ref{fig:camel} we show that our method is able to generalize to non-isometric but similar enough deformations.

\begin{figure*}[ht]
\centering
    \includegraphics[width=0.45\linewidth]{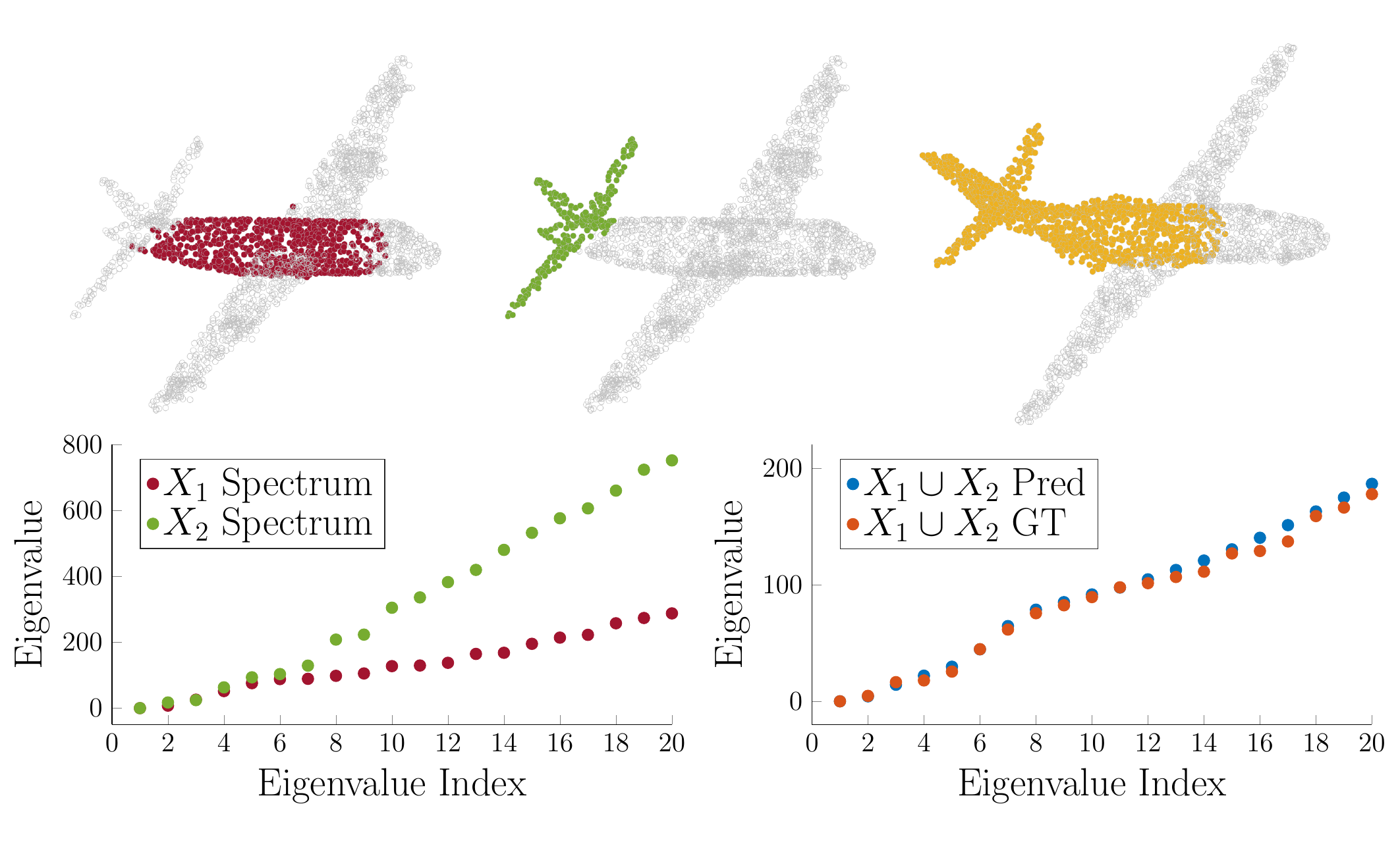}\hspace{1cm}
    \includegraphics[width=0.45\linewidth]{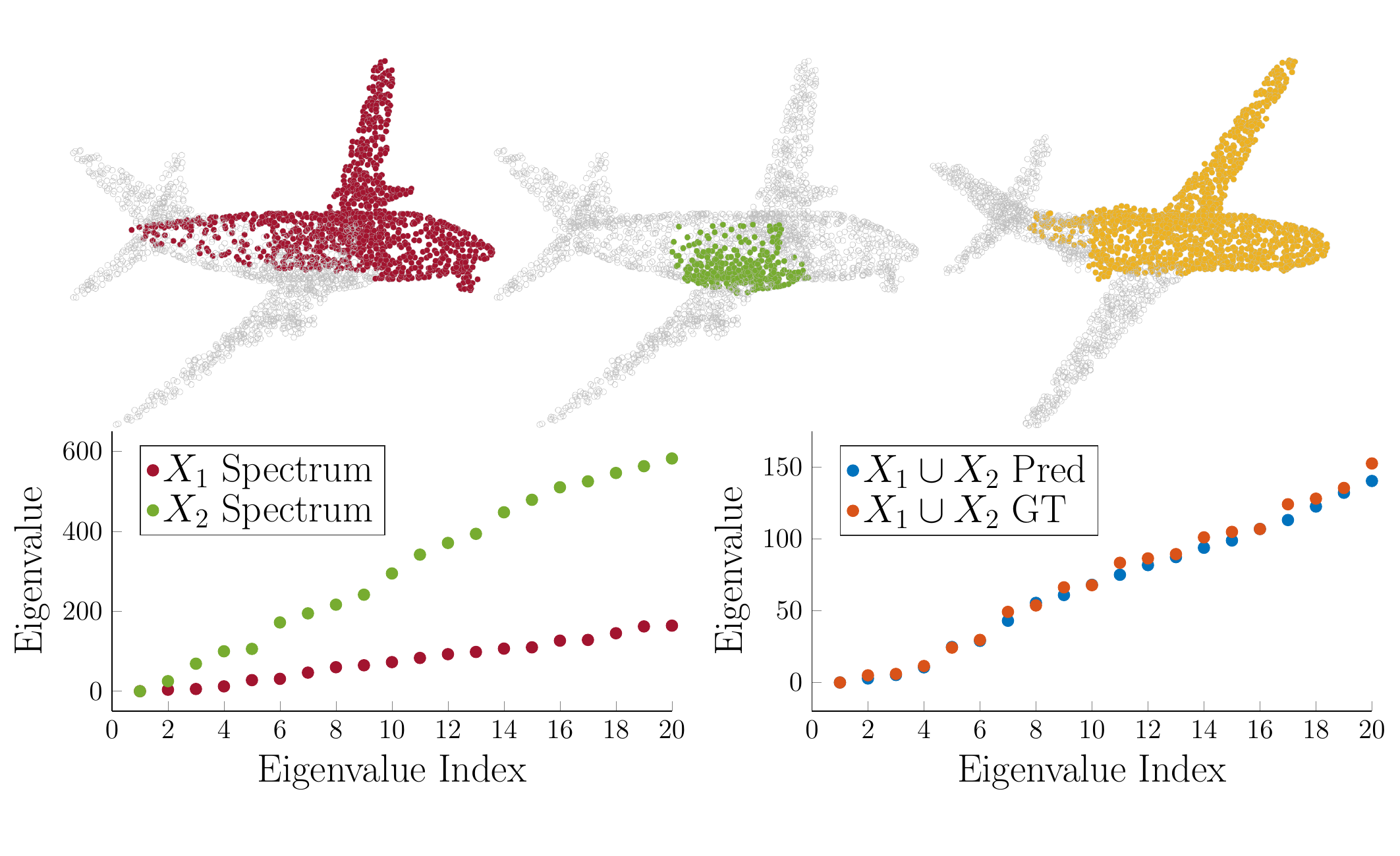}\\[0.5cm]
    \includegraphics[width=0.45\linewidth]{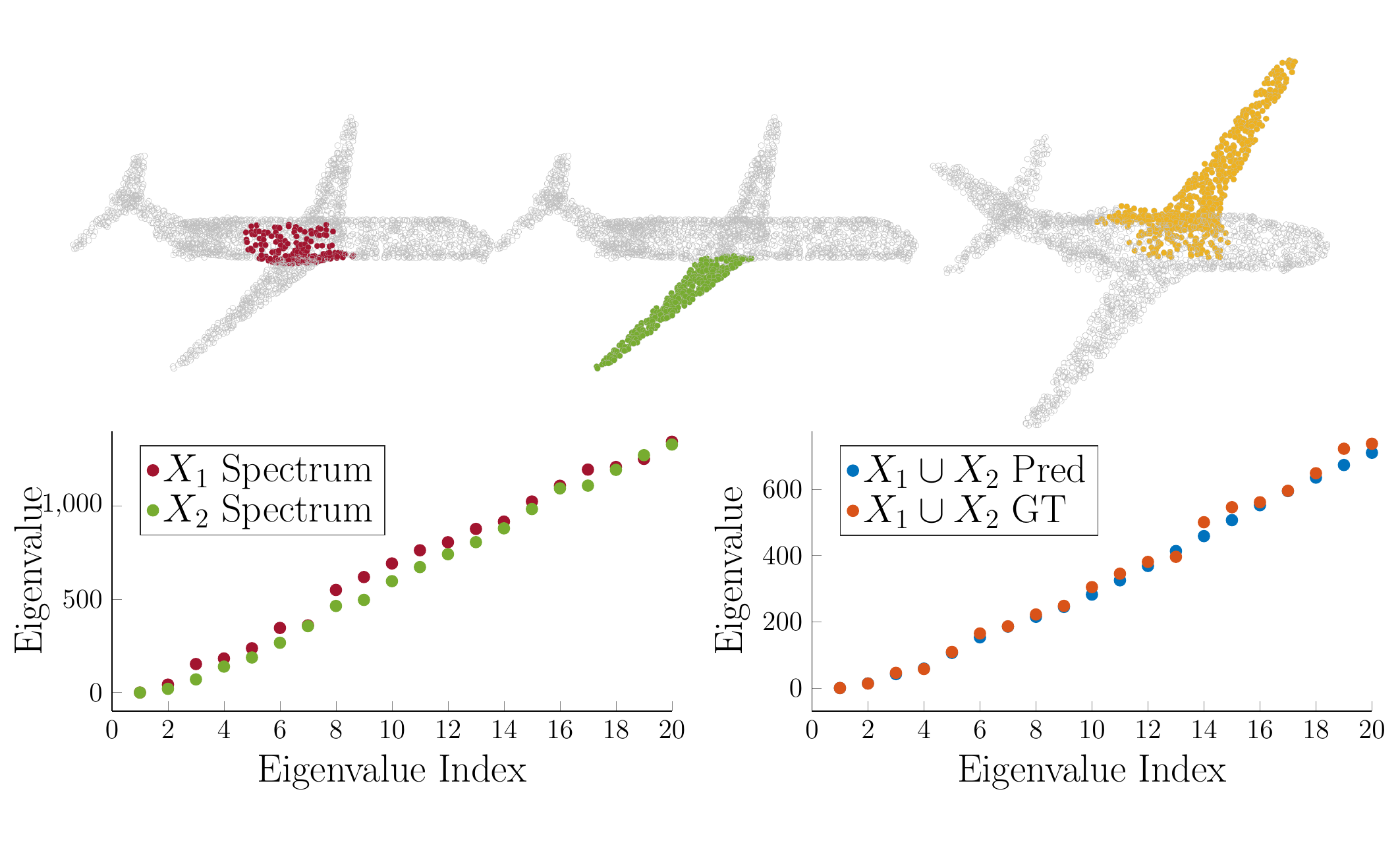}\hspace{1cm}
    \includegraphics[width=0.45\linewidth]{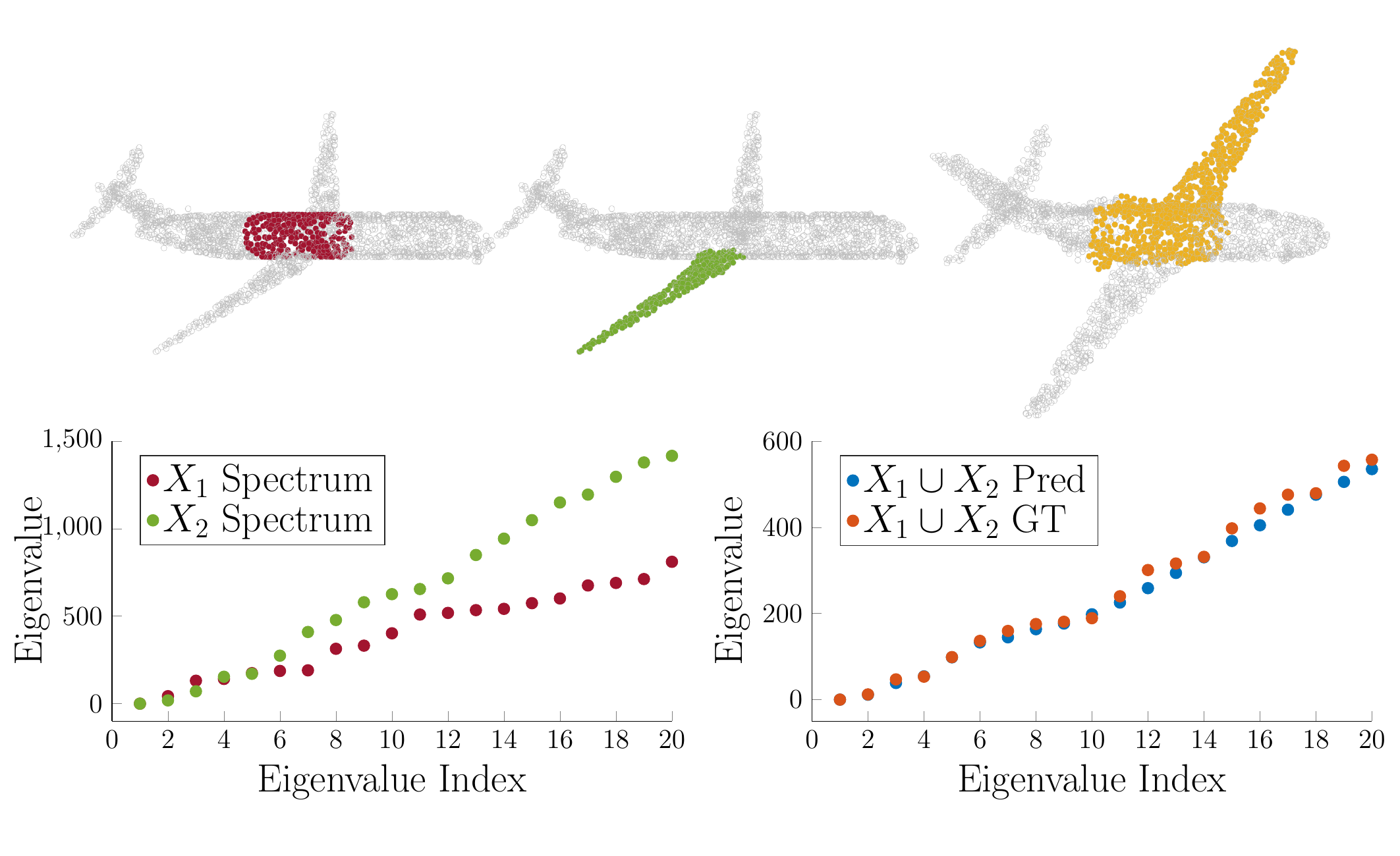}\\[0.5cm]
    \includegraphics[width=0.45\linewidth]{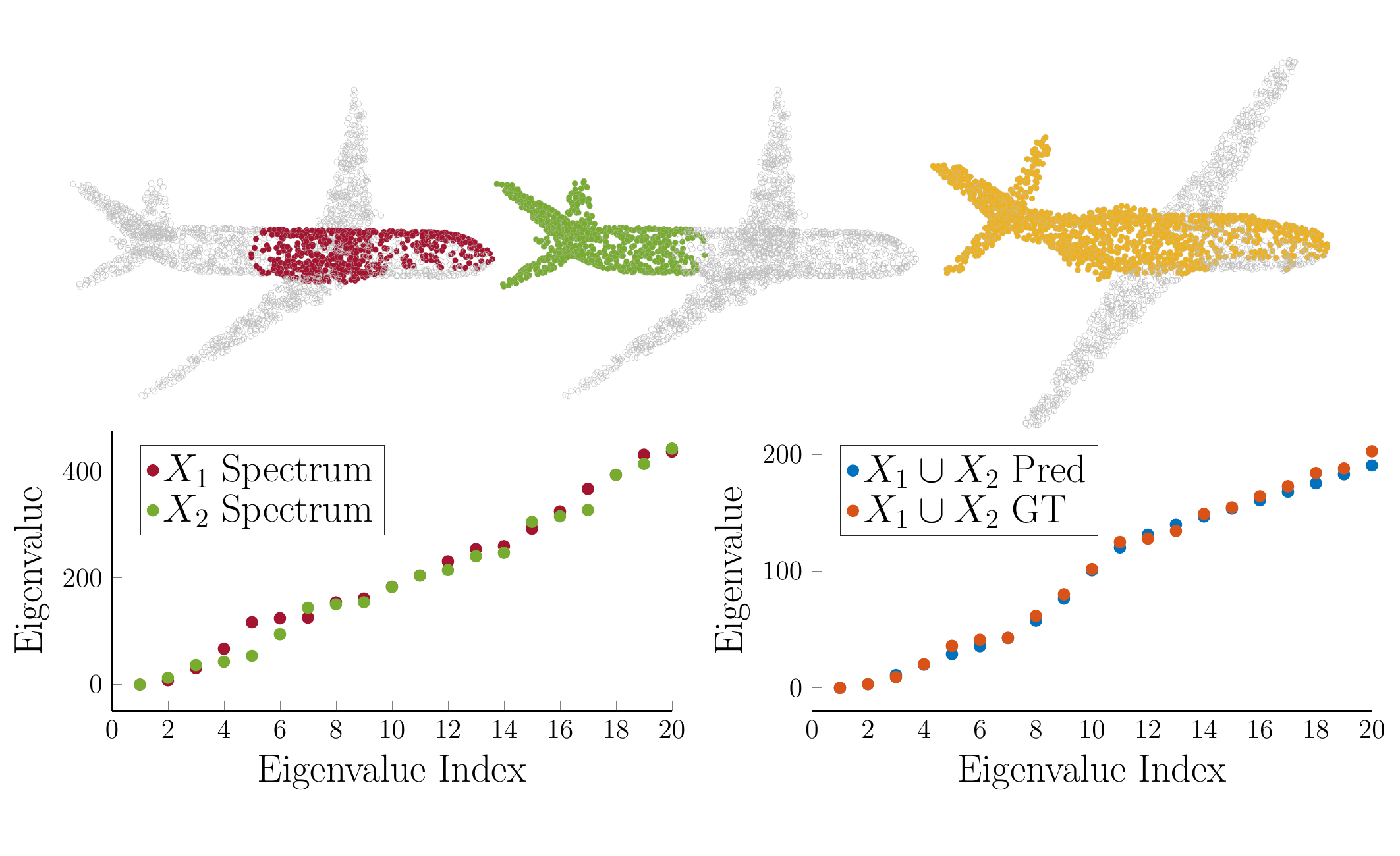}\hspace{1cm}
    \includegraphics[width=0.45\linewidth]{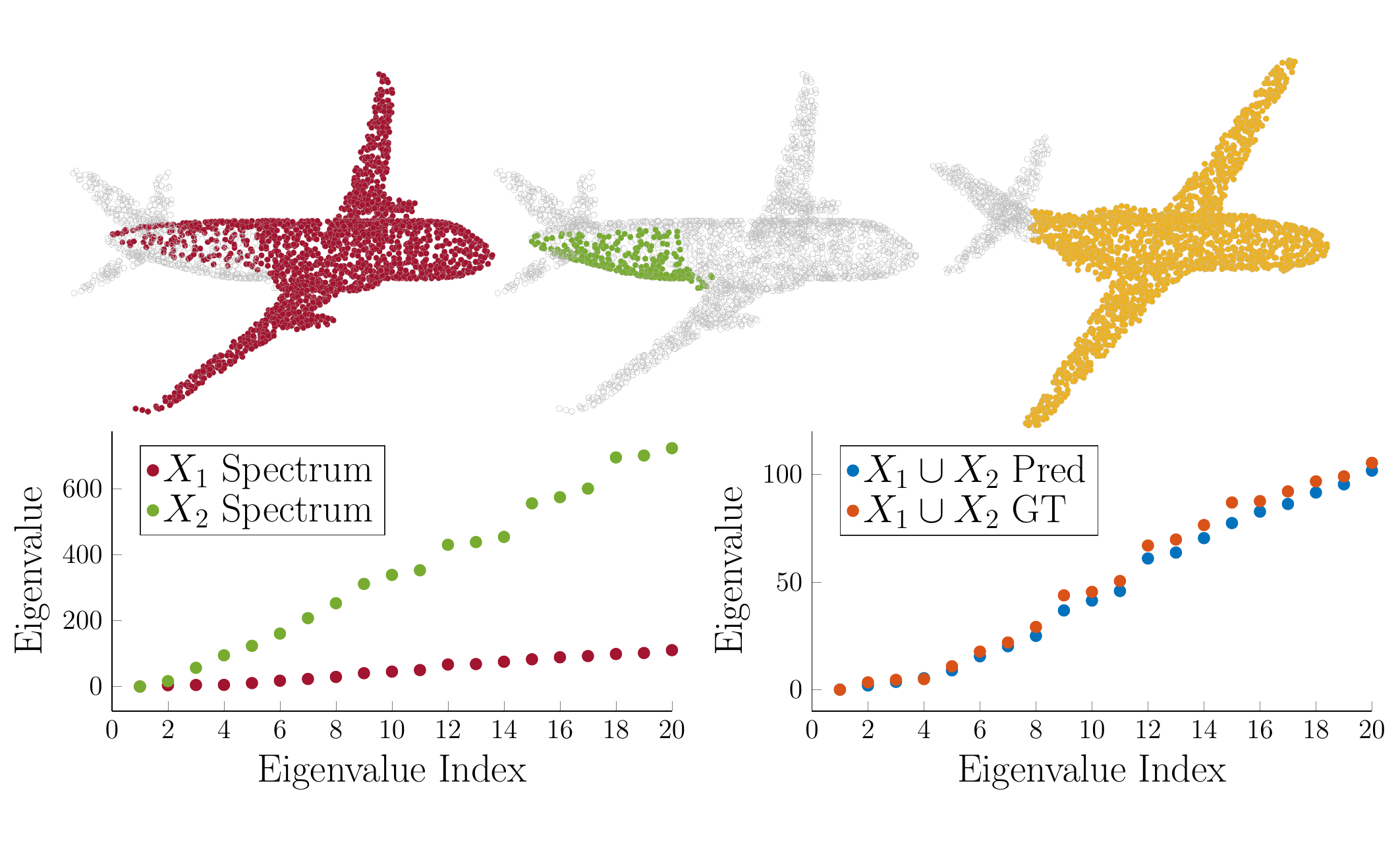}\\[0.5cm]
    \includegraphics[width=0.45\linewidth]{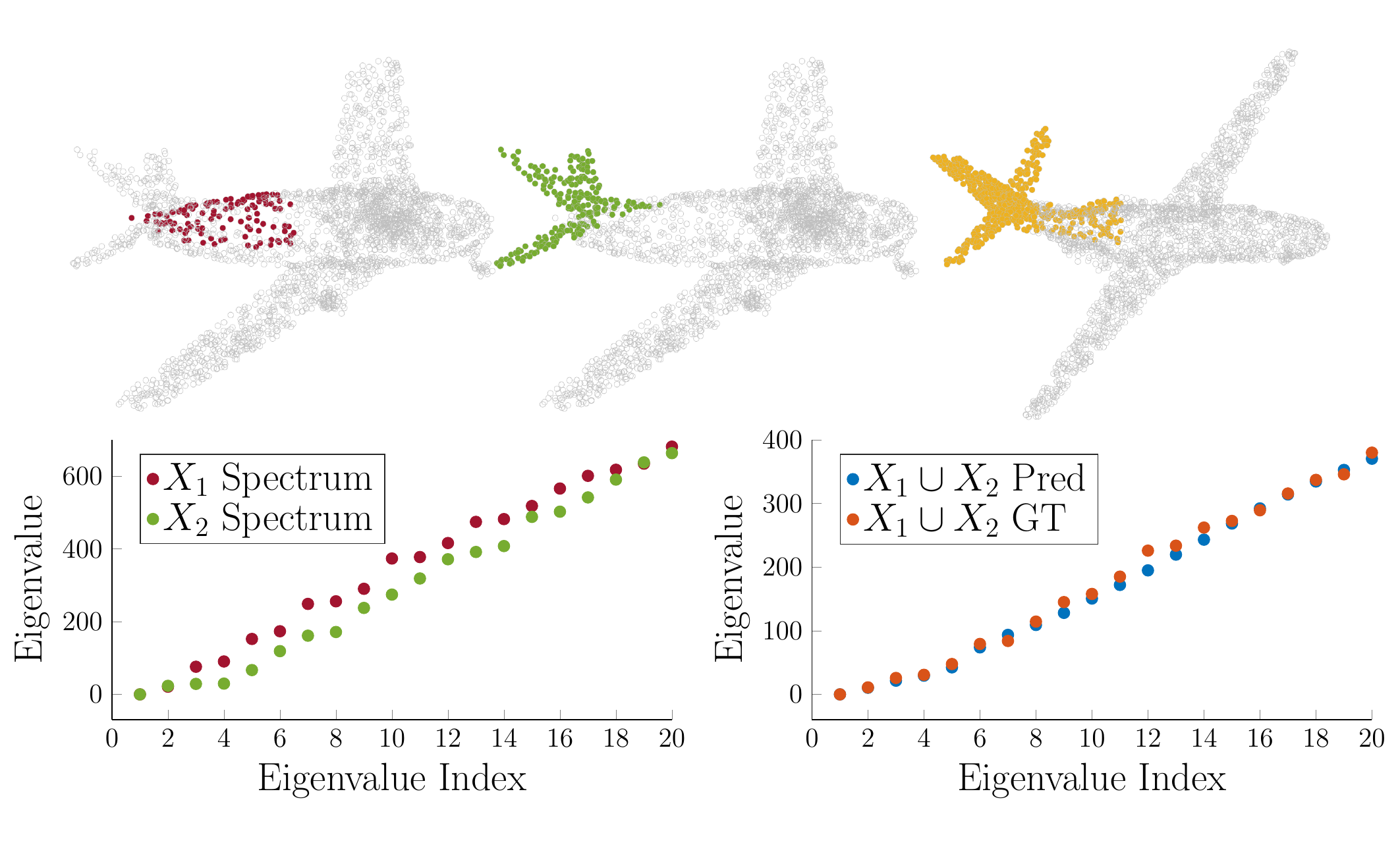}\hspace{1cm}
    \includegraphics[width=0.45\linewidth]{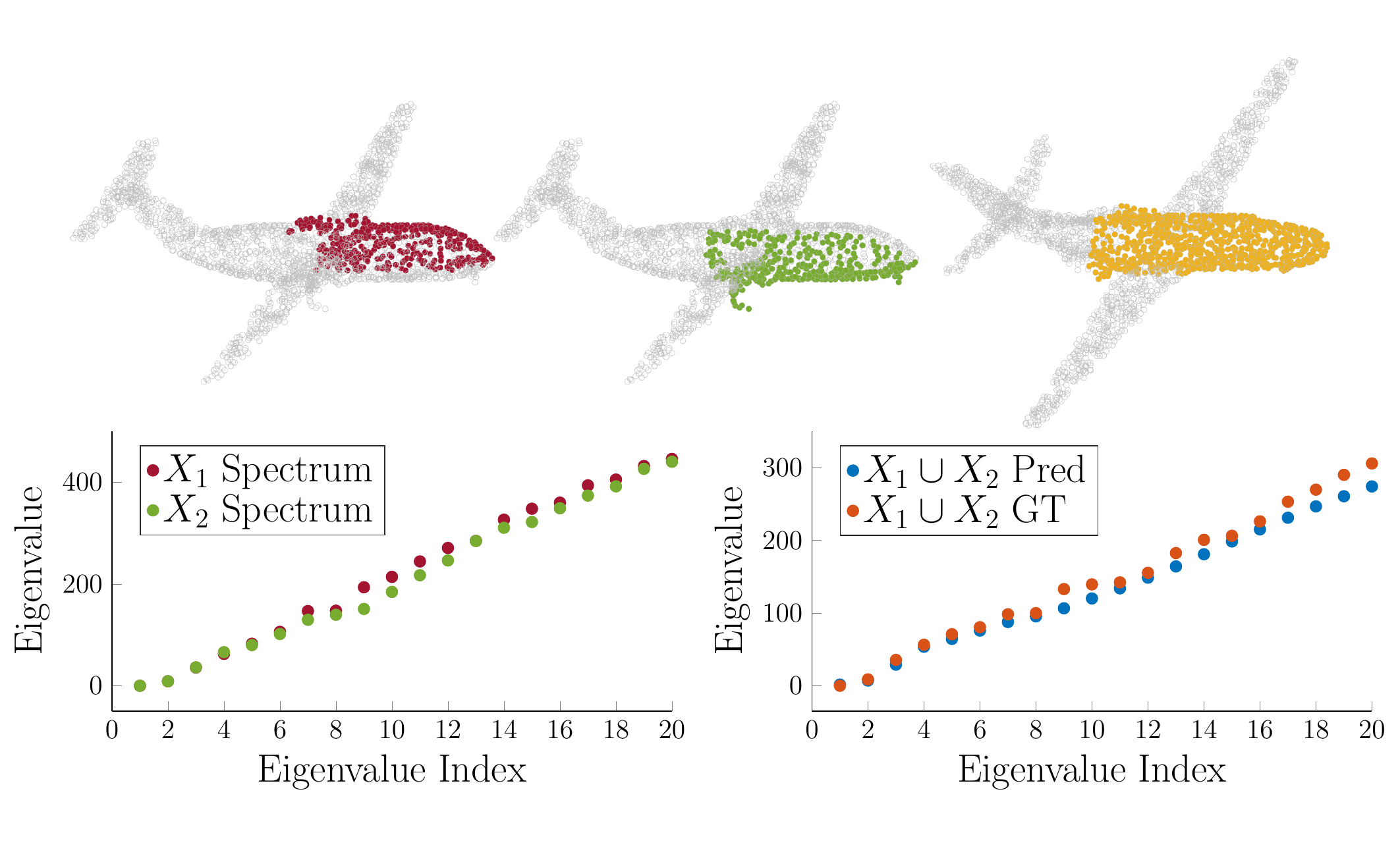}

    \caption{\label{fig:aereo_with_evals} 
    Region localization on aereoplanes. The model is trained and tested on point clouds.}
\end{figure*}

\begin{figure*}[ht]
\vspace{3cm}
    \centering

    \includegraphics[width=0.45\linewidth]{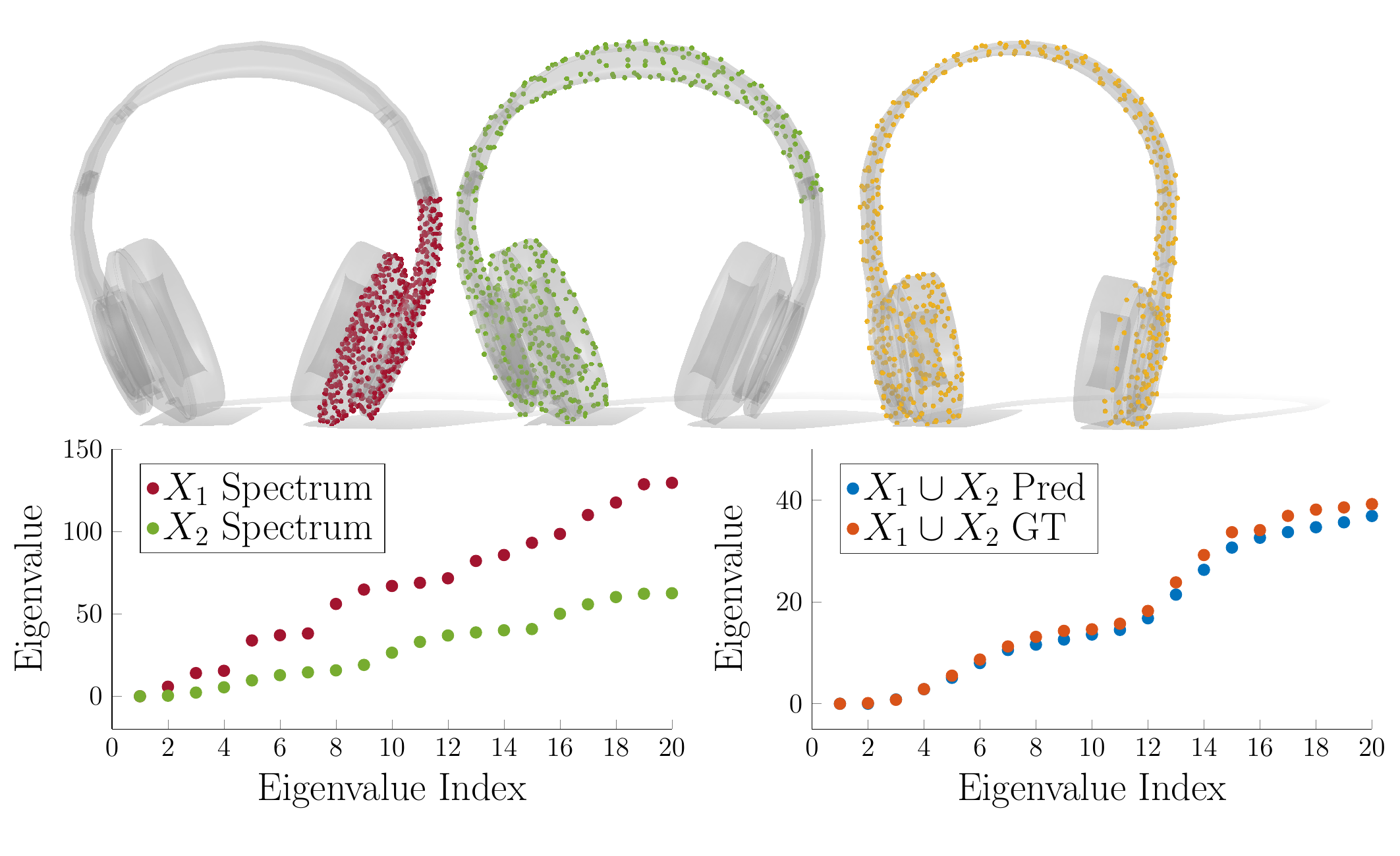}\hspace{1cm}
    \includegraphics[width=0.45\linewidth]{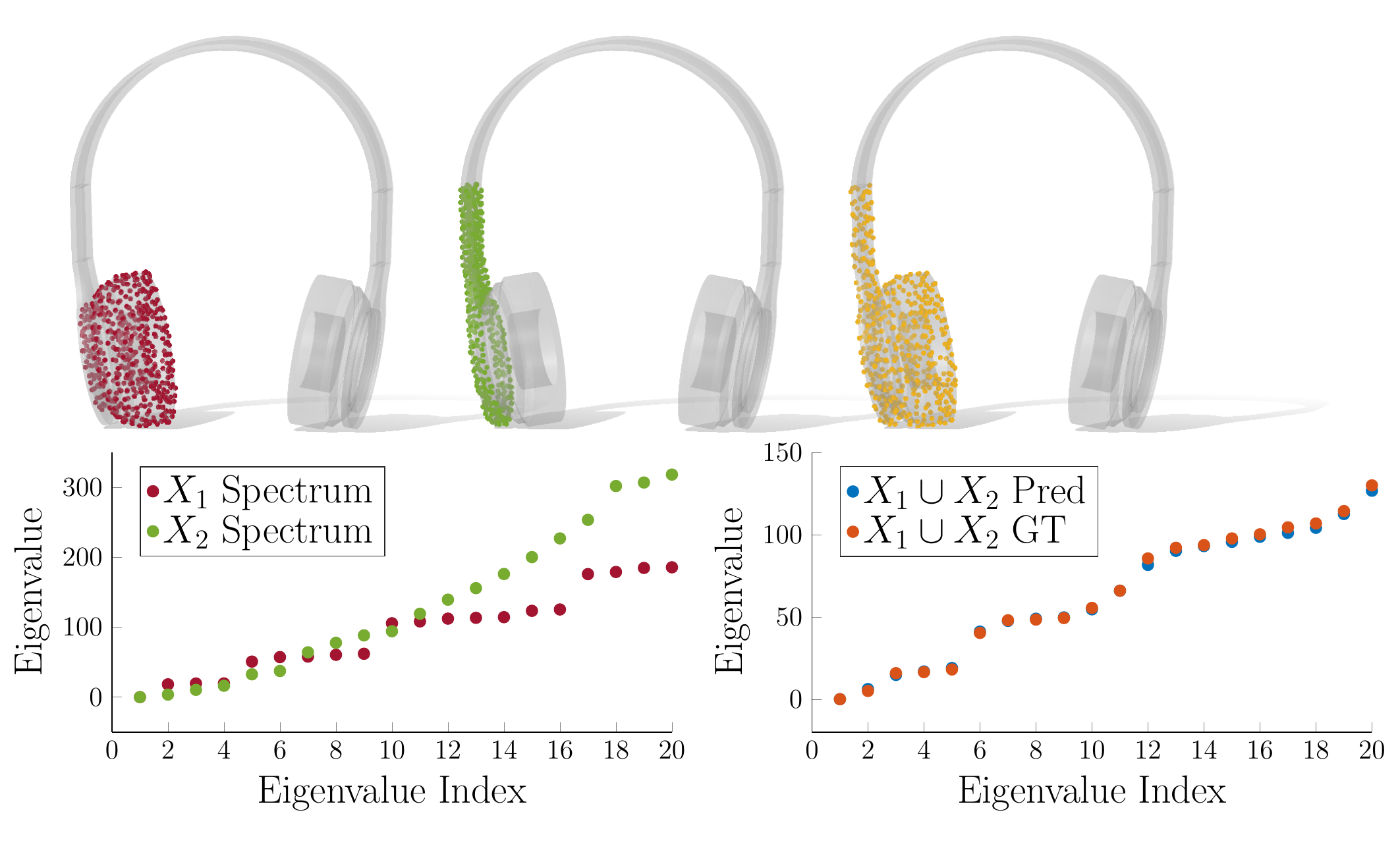}\\ 
    \includegraphics[width=0.45\linewidth]{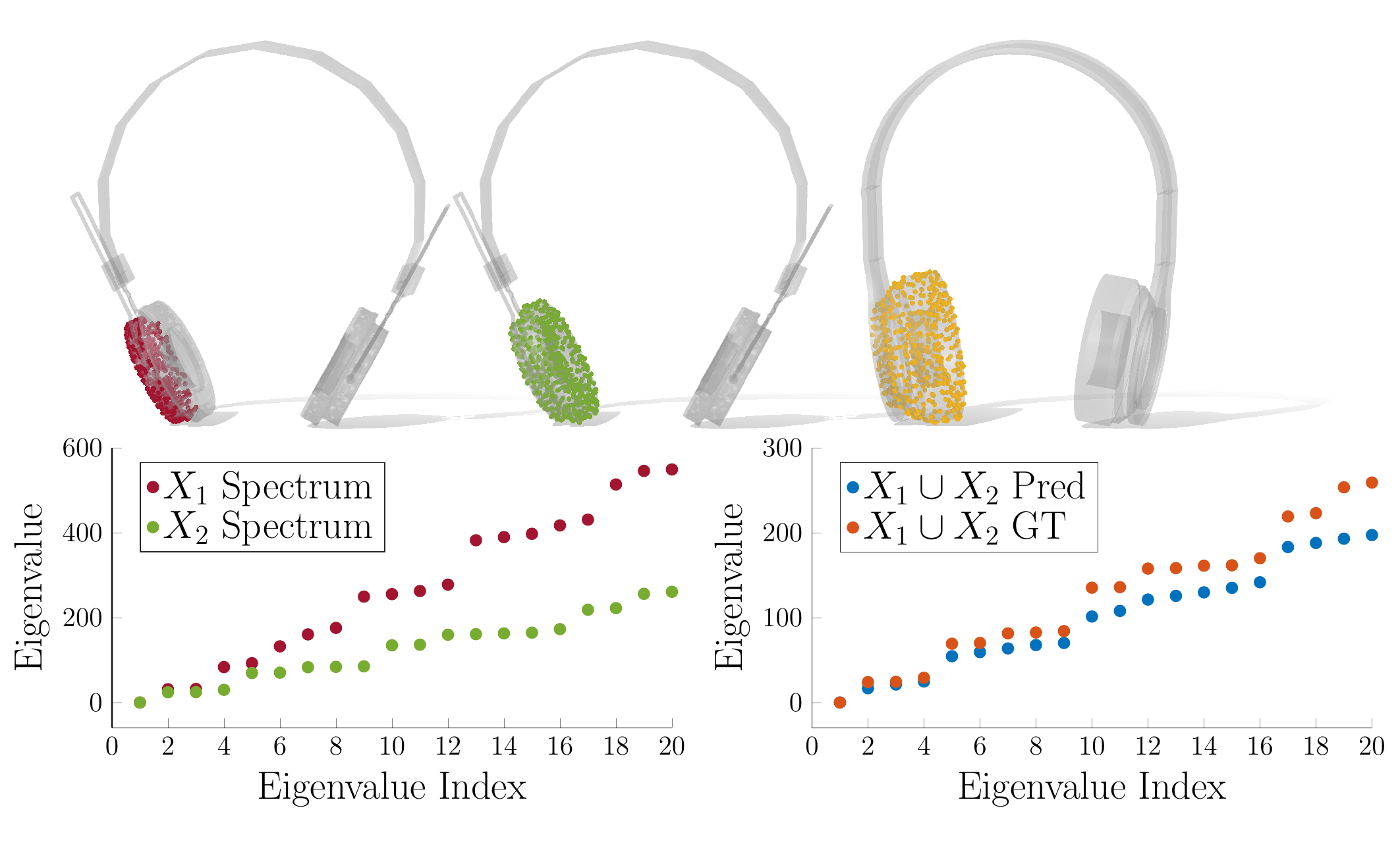}\hspace{1cm}
    \includegraphics[width=0.45\linewidth]{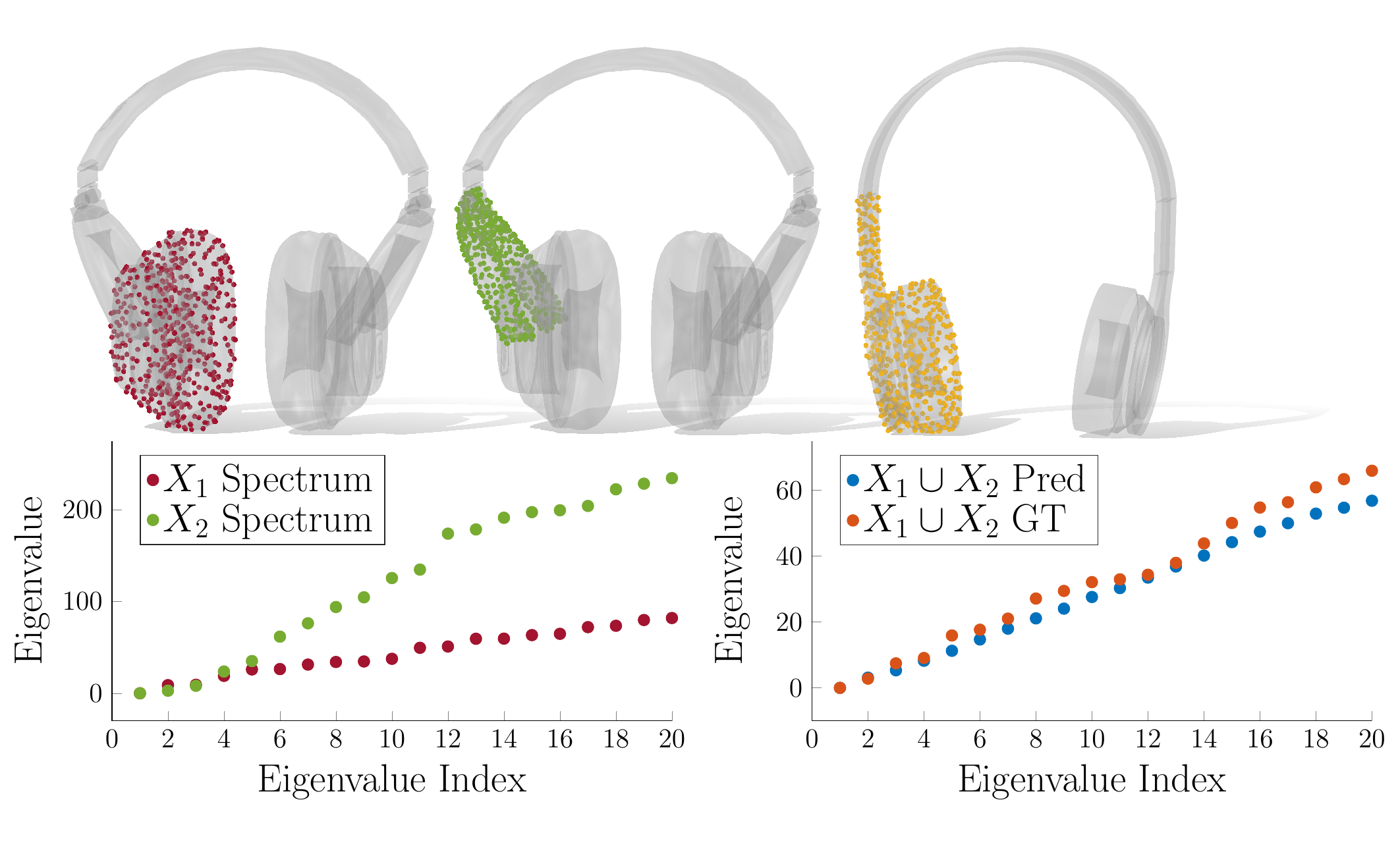}\\ 
    \includegraphics[width=0.45\linewidth]{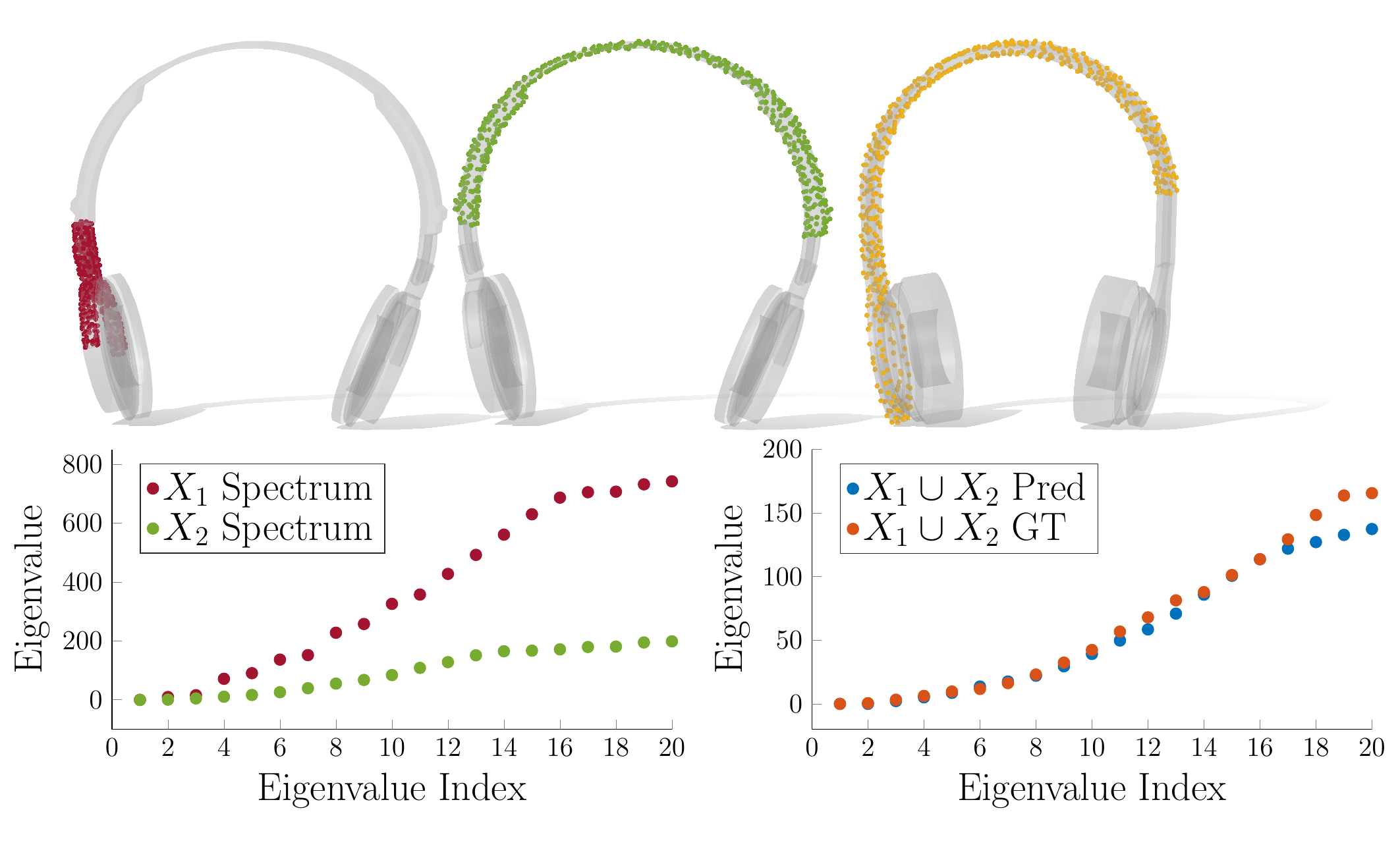}\hspace{1cm}
    \includegraphics[width=0.45\linewidth]{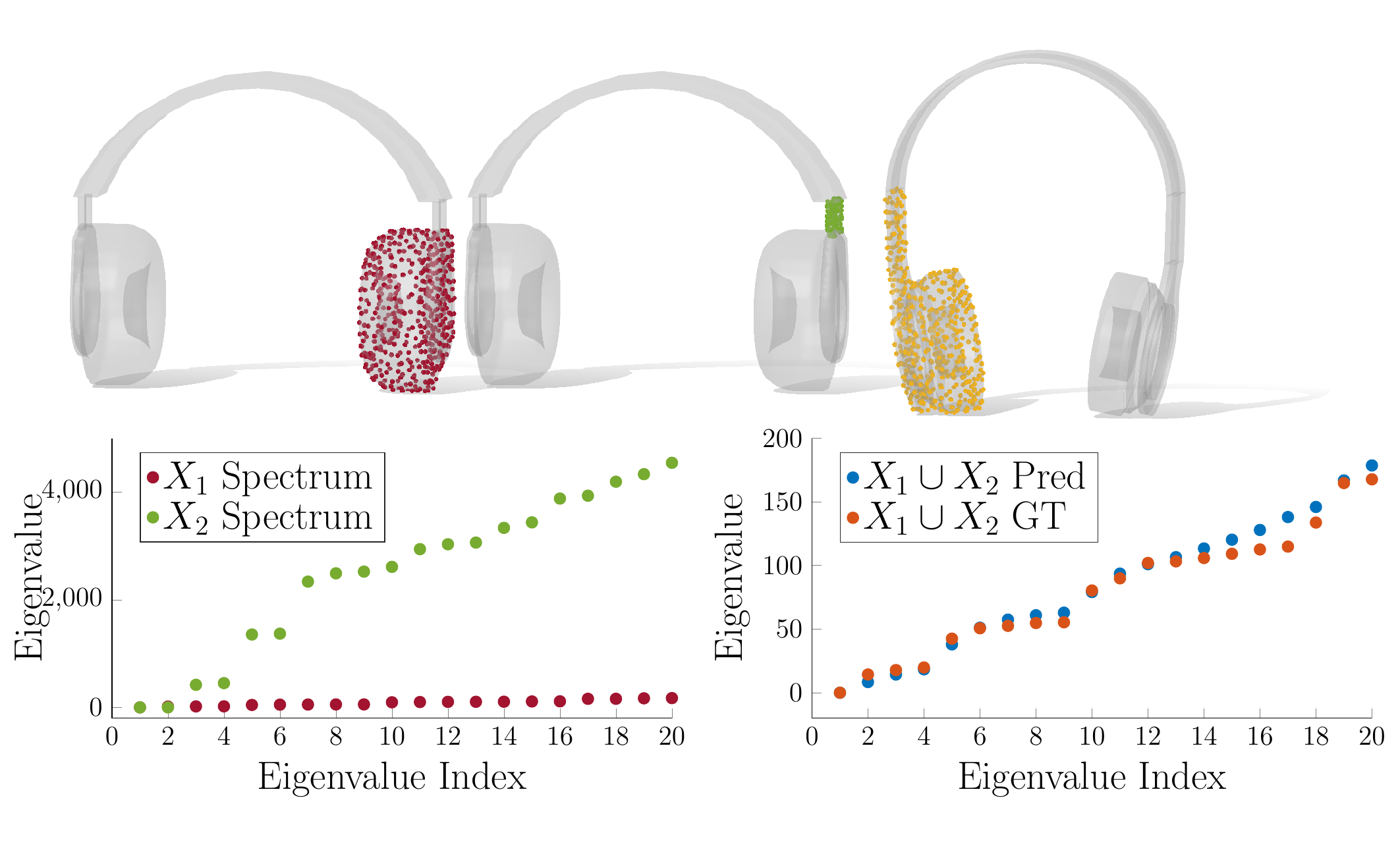}\\
    \caption{\label{fig:earphones_with_evals}
    Region localization on headphones, trained and tested on point clouds. 
    In the examples on the right column, despite significant changes in the geometry of the partialities, the model localizes the same correct region.\\
    \vspace{3cm}}
    
\end{figure*}

\paragraph*{Additional results on {\em Point clouds}}
To show the flexibility of our approach we consider a further class of shapes composed by airplanes from \cite{shapenet2015} represented as point clouds. We report in Figure \ref{fig:aereo_with_evals} some qualitative results on the region localization task with this class.
We also report additional results on headphones \cite{Mo_2019_CVPR} in Figure \ref{fig:earphones_with_evals}. These results show that our model generalizes to different source geometries, as long as the class shape does not change.





%

\subsection{Architecture}\label{sec:architecture}
In this section, we describe in detail the proposed neural architecture. 
%
%
Note that since surface area directly affects the magnitude of the eigenvalues, at test time the shapes are normalized to have the same area of the shapes seen at training time.

\subsubsection{Spectral union model}
\begin{figure*}[ht]
\centering
    \begin{overpic}[trim=0cm 0cm 0cm 0cm,clip, tics=2, height=0.95\textheight]{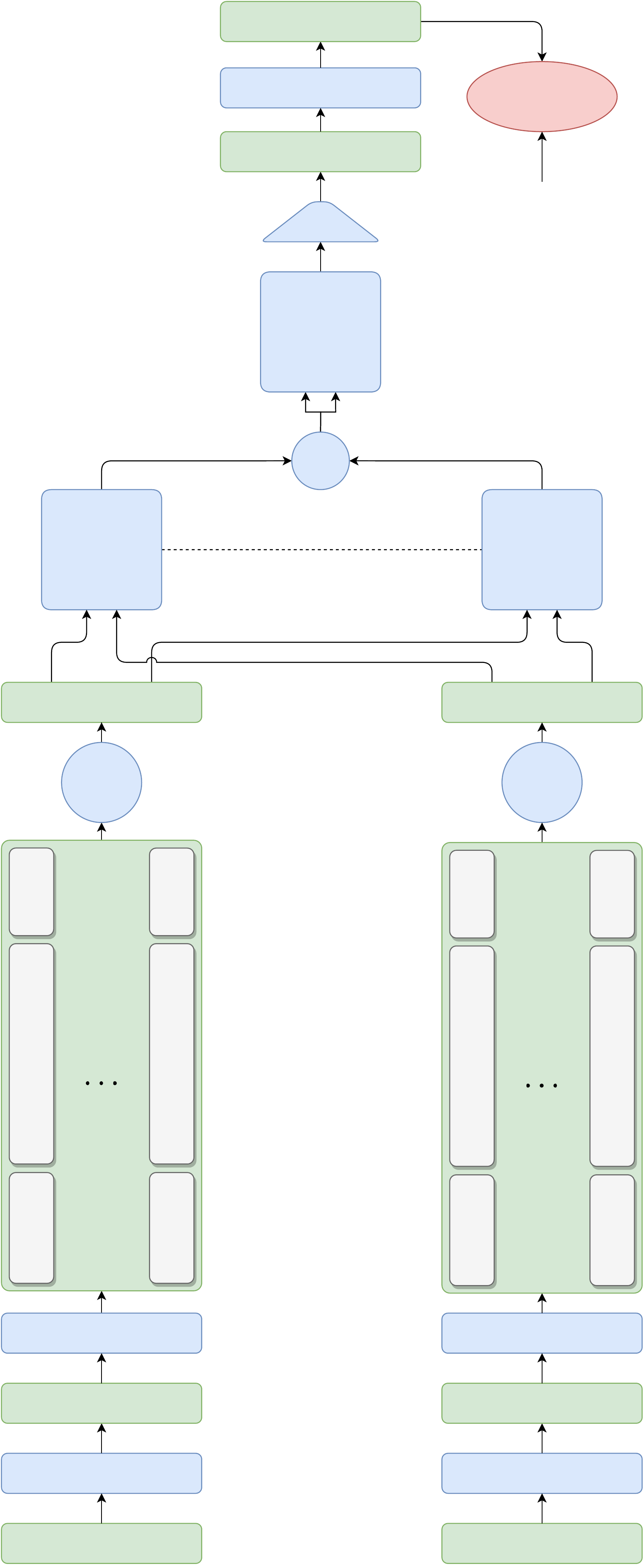}

    \put(6, 1){\footnotesize $\bm{\Lambda}_{1}$}
    \put(2, 5.5){\footnotesize $\mathrm{offset\ encoding}$}
    \put(4.25, 10){\footnotesize $\mathrm{off}(\bm{\Lambda}_{1})$}
    \put(6, 14.5){\footnotesize $\bm{E}$}

    \put(34, 1){\footnotesize $\bm{\Lambda}_{2}$}
    \put(30, 5.5){\footnotesize $\mathrm{offset\ encoding}$}
    \put(32.25, 10){\footnotesize $\mathrm{off}(\bm{\Lambda}_{2})$}
    \put(34, 14.5){\footnotesize $\bm{E}$}

    \put(1.25, 18.5){\footnotesize \rotatebox{90}{${\color{darkblue}x}_1^1\cdots {\color{darkblue}x}_{\ell/2}^1$}}
    \put(1.25, 27){\footnotesize \rotatebox{90}{${\color{darkblue}x}_{\ell/2}^{}\cdots{\color{darkblue}x}_{\ell}\times\mathrm{off}(\lambda_{1})$}}
    \put(1.25, 41){\footnotesize \rotatebox{90}{$\mathrm{off}(\lambda_{1})^{}$}}
    
    \put(10.25, 18.5){\footnotesize \rotatebox{90}{${\color{darkblue}x}_1^k\cdots {\color{darkblue}x}_{\ell/2}^k$}}
    \put(10.25, 27){\footnotesize \rotatebox{90}{${\color{darkblue}x}_{\ell/2}^{}\cdots{\color{darkblue}x}_{\ell}\times\mathrm{off}(\lambda_{k})$}}
    \put(10.25, 41){\footnotesize \rotatebox{90}{$\mathrm{off}(\lambda_{k})^{}$}}
    
    \put(29.35, 18.5){\footnotesize \rotatebox{90}{${\color{darkblue}x}_1^1\cdots {\color{darkblue}x}_{\ell/2}^1$}}
    \put(29.35, 27){\footnotesize \rotatebox{90}{${\color{darkblue}x}_{\ell/2}^{}\cdots{\color{darkblue}x}_{\ell}\times\mathrm{off}(\lambda_{1})$}}
    \put(29.35, 41){\footnotesize \rotatebox{90}{$\mathrm{off}(\lambda_{1})^{}$}}
    
    \put(38.35, 18.5){\footnotesize \rotatebox{90}{${\color{darkblue}x}_1^k\cdots {\color{darkblue}x}_{\ell/2}^k$}}
    \put(38.35, 27){\footnotesize \rotatebox{90}{${\color{darkblue}x}_{\ell/2}^{}\cdots{\color{darkblue}x}_{\ell}\times\mathrm{off}(\lambda_{k})$}}
    \put(38.35, 41){\footnotesize \rotatebox{90}{$\mathrm{off}(\lambda_{k})^{}$}}
    
    \put(5.75, 50.5){\scriptsize pos.}
    \put(4.5, 49.25){\scriptsize encoding}
    
    \put(33.75, 50.5){\scriptsize pos.}
    \put(32.5, 49.25){\scriptsize encoding}
    
    \put(3, 54.75){\footnotesize $\bm{\Lambda}_{1}$ embedding}
    \put(31, 54.75){\footnotesize $\bm{\Lambda}_{2}$ embedding}

    \put(19.9, 63.5){\footnotesize \color{gray}{$\bm{\Theta}$}}

    \put(5.5, 64.5){\footnotesize $\bm{T_A}$}
    \put(33.5, 64.5){\footnotesize $\bm{T_A}$}

    \put(5,61.5){\tiny $D$}
    \put(7,61.5){\tiny $E$}
    \put(33.25,61.5){\tiny $E$}
    \put(35.25,61.5){\tiny $D$}
    
    \put(19.8, 70.3){\footnotesize $\bm{+}$}

    \put(19.5, 78.5){\footnotesize $\bm{T_B}$}
    
    \put(20, 85.5){\footnotesize $\bm{\rho}$}
    
    \put(16.35, 89.85){\footnotesize $\mathrm{off}(\bm{\widetilde{\Lambda}}_{\M_1 \cup \M_2}) $}
    
    \put(16.35, 94){\footnotesize $\mathrm{offset\ decoding}$}

    \put(17.25, 98.25){\footnotesize $\bm{\widetilde{\Lambda}}_{\M_1 \cup \M_2}$}

    \put(31.5, 87){\footnotesize $\bm{{\Lambda}}_{\M_1 \cup \M_2}$}

    \put(33, 93.5){\footnotesize $\mathbf{mse}$}

    \end{overpic}
    \caption{\label{fig:architecture_union}  Detailed architecture of the spectral union operator. }
\end{figure*}

In Fig.~\ref{fig:architecture_union} we show the detailed architecture of the spectral union model.

\paragraph*{Hyperparameters}
The dimensionality of each embedding is $32$. 
$\bm{T_A}$ has $8$ heads, $6$ layers, the dimensionality of the internal feed-forward layer is $64$ and the dropout is $0.1$. $\bm{T_B}$ has $8$ heads, $3$ layers, the dimensionality of the feed-forward is $32$ and the dropout is $0.1$.
Thus, $\bm{\rho}$ reduces the embedding  dimensionality from $32$ to $1$.

\paragraph*{Training}
The model is trained until convergence.
The training randomly augments online each input independently. 
The batch size is $32$. The optimizer used is Adam with learning rate of $2e{-4}$ and weight decay $1e{-5}$.
The learning rate changes according to cosine annealing with warm restarts scheduler and it restarts every 10 epochs, doubling the number of epochs between restarts at each restart.

\subsubsection{Region Localization model}
\begin{figure*}[ht]
\centering
    \begin{overpic}[trim=0cm 0cm 0cm 0cm, clip, tics=2, width=0.975\textwidth]{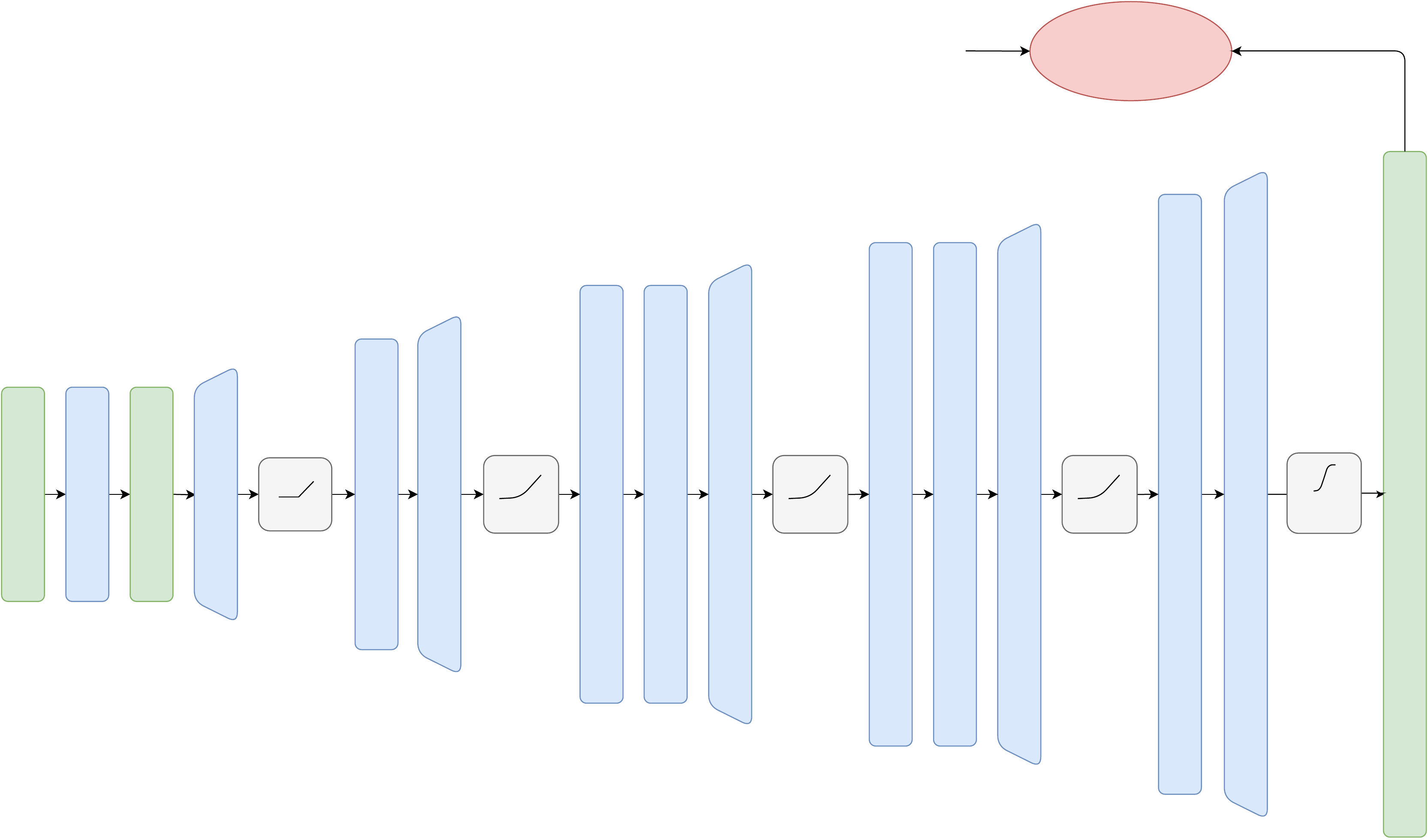}
    \put(1, 23.5){\footnotesize \rotatebox{90}{$\bm{\Lambda}$}}
    \put(5.5, 18.5){\footnotesize \rotatebox{90}{$\mathrm{offset\ encoding}$}}
    \put(9.75, 22){\footnotesize \rotatebox{90}{$\mathrm{off}(\bm{\Lambda})$}}
    
    \put(14.5, 21.75){\footnotesize \rotatebox{90}{$\mathrm{Linear}$}}

    \put(19, 22){\tiny $\mathrm{ReLU}$}

    \put(25.5, 19.5){\footnotesize \rotatebox{90}{$\mathrm{Layer\ Norm}$}}

    \put(30, 21.75){\footnotesize \rotatebox{90}{$\mathrm{Linear}$}}

    \put(35, 22){\tiny $\mathrm{ELU}$}

    \put(41.5, 20.75){\footnotesize \rotatebox{90}{$\mathrm{Dropout}$}}

    \put(46, 19.5){\footnotesize \rotatebox{90}{$\mathrm{Layer\ Norm}$}}

    \put(50.25, 21.75){\footnotesize \rotatebox{90}{$\mathrm{Linear}$}}

    \put(55.5, 22){\tiny $\mathrm{ELU}$}

    \put(61.5, 20.75){\footnotesize \rotatebox{90}{$\mathrm{Dropout}$}}

    \put(66.25, 19.5){\footnotesize \rotatebox{90}{$\mathrm{Layer\ Norm}$}}

    \put(70.75, 21.75){\footnotesize \rotatebox{90}{$\mathrm{Linear}$}}

    \put(75.75, 22){\tiny $\mathrm{ELU}$}

    \put(81.75, 20.75){\footnotesize \rotatebox{90}{$\mathrm{Dropout}$}}

    \put(86.5, 21.75){\footnotesize \rotatebox{90}{$\mathrm{Linear}$}}

    \put(90.45, 22){\tiny $\mathrm{sigmoid}$}

    \put(97.5, 22.5){\footnotesize \rotatebox{90}{$\bm{\widetilde{R}_{\M}}$}}

    \put(64, 55){\footnotesize $\bm{{R}_{\M}}$}

    \put(77.5, 54.75){\footnotesize $\mathbf{mse}$}

    \end{overpic}
    \caption{\label{fig:architecture_region} Detailed architecture of the region localization MLP for humans.}
\end{figure*}
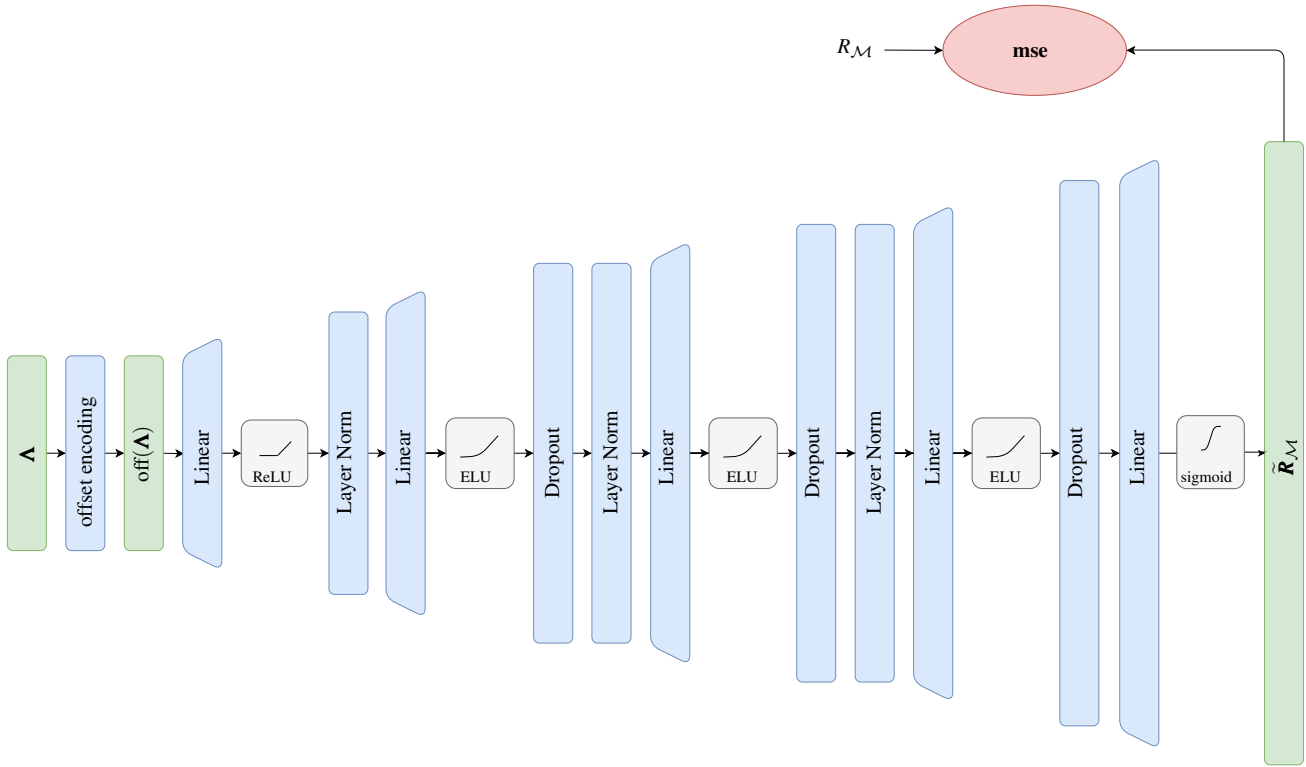

In Fig.~\ref{fig:architecture_region} we show the detailed architecture of the region localization model for humans.

\paragraph*{Hyperparameters}
The dense layers increase the dimensionality of the input sequence from $20$ to $6890$, for humans, i.e. the number of vertices in the fixed template. 
In particular, the layers apply the following transformations $20 \to 1300 \to 2600 \to 3900 \to 5200 \to 6890$. The dropout is always set to $p=0.5$.

\paragraph*{Training}
The model is trained to localize the region from both the predicted union eigenvalues and all the ground-truth eigenvalues, to which we add random noise.
The model is early stopped, monitoring the IoU metric on a validation set.
The batch size is $32$. The optimizer used is Adam with learning rate of $5e{-5}$ and weight decay $1e{-6}$. 
The scheduler adopted is again the cosine annealing with warm restarts, with the same hyperparameters.

\subsubsection{Data processing for point clouds}
The aereoplanes from \cite{shapenet2015} and headphones from \cite{Mo_2019_CVPR} are point clouds with \emph{semantic} segmentation.

We performed some data processing to: (1) extract shapes with only given segments (e.g., discarding earphones or strange headphones), (2) extract random partialities from each shape and (3) find the segment-level matching between each shape and a fixed template for the region localization task.
We did this by defining the graph of the segments for each shape, then searching for sub-graphs with determined properties for (2) and solving the graph isomorphism against the template for (3).

We obtained 75 headphones and 964 aereoplanes for the training set, we extract random pairs of partialities from each shape.

\end{document}